\documentclass[11pt,fleqn,twoside]{article}
\usepackage{amsmath}
\usepackage{amssymb}
\usepackage{xcolor}
\usepackage{colortbl}
\usepackage{booktabs}
\usepackage{pstricks,multido,pst-plot}
\usepackage{tikz}	
\usepackage{pgfplots}
\usepackage{paralist}
\usepackage{wrapfig}
\usepackage{graphicx,epsfig}
\usepackage{marginnote}
\usepackage{todonotes}
\usepackage[margin=2cm,top=1.4cm,bottom=2.1cm,centering]{geometry}   
\usepackage{tcolorbox,pgfkeys}
\usepackage{hyperref}                      
\usepackage{microtype}
\definecolor{lightgray}{gray}{0.95}
\definecolor{lightyellow}{rgb}{1.00,.98,.50}
\definecolor{myblue}{rgb}{.39,.54,0.82}
\definecolor{midblue}{rgb}{.7,.7,1}
\definecolor{vertbleu}{rgb}{.73,.88,.93}
\definecolor{lightblue}{rgb}{0.71,0.88,0.96}
\definecolor{mygray}{rgb}{.80,.80,.80}
\definecolor{lightred}{rgb}{0.96,0.84,0.84}
\definecolor{mygreen}{rgb}{0.40,0.75,0.16}
\definecolor{mybrown}{rgb}{0.69,0.49,0.30}
\definecolor{mypink}{rgb}{1.00,0.22,0.57}
\definecolor{mygrey}{rgb}{.90,.90,.90}
\definecolor{cerulean}{cmyk}{0.94,0.11,0,0}

\newtcolorbox{mybox}{colback=yellow!20!white,colframe=gray,boxrule=0.3mm,arc=3mm,outer arc=1mm}

\usepackage{enumitem}

\catcode`\@=11
\def\numberbysection{\@addtoreset{equation}{section}
     \def\theequation{\thesection.\arabic{equation}}}
\numberbysection
\def\be{\begin{equation}}
\def\ee{\end{equation}}
\newcommand\bea{\begin{eqnarray}}
\newcommand\eea{\end{eqnarray}}
\renewcommand\phi{\varphi}

\newcommand\egal{&\!\!\!=\!\!\!&}

\renewcommand{\ge}{\geqslant}
\renewcommand{\le}{\leqslant}
\def\benn{\begin{eqnarray*}}
\def\eenn{\end{eqnarray*}}

\def\Z{{\mathbb Z}}

\def\N{{\mathbb N}}

\def\P{{\mathbb P}}

\def\vdimerb{\pspolygon[linewidth=0.4pt,linecolor=black,fillstyle=solid,fillcolor=myblue](0,0)(0,2)(1,2)(1,0)}
\def\vdimery{\pspolygon[linewidth=0.4pt,linecolor=black,fillstyle=solid,fillcolor=yellow](0,0)(0,2)(1,2)(1,0)}
\def\hdimerg{\pspolygon[linewidth=0.4pt,linecolor=black,fillstyle=solid,fillcolor=green](0,0)(2,0)(2,1)(0,1)}

\def\hdimerr{\pspolygon[linewidth=0.4pt,linecolor=black,fillstyle=solid,fillcolor=red](0,0)(2,0)(2,1)(0,1)}
\def\hdimerw{\pspolygon[linewidth=0.2pt,linecolor=black](0,0)(2,0)(2,1)(0,1)}
\def\vdimerw{\pspolygon[linewidth=0.2pt,linecolor=black](0,0)(0,2)(1,2)(1,0)}

\newcommand*\xbar[1]{%
  \hbox{%
    \vbox{%
      \hrule height 0.5pt % The actual bar
      \kern0.3ex%         % Distance between bar and symbol
      \hbox{%
        \kern-0.2em%      % Shortening on the left side
        \ensuremath{#1}%
        \kern-0.0em%      % Shortening on the right side
      }%
    }%
  }%
} 

\hypersetup{
colorlinks=true,
citecolor=red,
linkcolor=blue,
urlcolor=black
}

\begin{document}
\renewcommand{\baselinestretch}{1.1}

\title{\vspace{-7truemm}\textbf{\Large Double tangent method for two-periodic Aztec diamonds}}

\date{}
\maketitle

\begin{center}
{\vspace{-14mm}\large \textsc{Philippe Ruelle}}
\\[.5cm]
{\em Institut de Recherche en Math\'ematique et Physique\\ 
Universit\'e catholique de Louvain, Louvain-la-Neuve, B-1348, Belgium}
\\[.2cm]

%{\tt bryan.debin\,@\,uclouvain.be,}
%\qquad
%{\tt philippe.ruelle\,@\,uclouvain.be}
%
\end{center}

\vspace{0.1cm} 

\begin{abstract}
We use the octahedron recurrence, which generalizes the quadratic recurrence found by Kuo for standard Aztec diamonds, in order to compute boundary one-refined and two-refined partition functions for two-periodic Aztec diamonds. In a first approach, the geometric tangent method allows to derive the parametric form of the arctic curve, separating the solid and liquid phases. This is done by using the recent reformulation of the tangent method and the one-refined partition functions without extension of the domain. In a second part, we use the two-refined tangent method to rederive the arctic curve from the boundary two-refined partition functions, which we compute exactly on the lattice. The curve satisfies the known algebraic equation of degree 8, of which either tangent method gives an explicit parametrization.
\end{abstract}

%\centerline{{\bf \Large Two-periodic Aztec diamonds with two tangent methods}}
%\vskip 0.7truecm
%\centerline{(Working notes)}
%\bigskip
%\centerline{\large Philippe Ruelle}
%\bigskip
%\centerline{Universit\'e catholique de Louvain}
%\centerline{Institut de Physique Th\'eorique}
%\centerline{Chemin du Cyclotron, 2}
%\centerline{B-1348 \hskip 0.25truecm Louvain-la-Neuve, Belgium}
%\medskip
%\centerline{\it \today}
%\vskip 0.8truecm
%\hrule \medskip
%\noindent \small
%We use Kuo's non-linear recurrence, renamed octahedron recurrence by Speyer in a more general context, in order to compute boundary one-refined and two-refined partition functions for two-periodic Aztec diamonds. In a first approach, the recently reformulated tangent method allows to derive the parametric form of the arctic curve, separating the solid and liquid phases. This is done by using the one-refined partition functions without extension of the domain. In the second part, we use the two-refined tangent method to rederive the arctic curve from the boundary two-refined partition function, which we compute exactly on the lattice. The curve satisfies the known algebraic equation of degree 8, of which either tangent method gives an explicit parametrization.
\vspace{4mm} 
\hrule
\normalsize

%%%%%%%%%%%%%%%%%%%%%%%%%%%%%%%%%%%%%%%%%%%%%%%%%%%%%%%%%%%%%%%%%%%%%%%%%%

%\noindent
%{\bf \red Changes to be made when revising:}  
%
%\noindent
%(1) Below (2,2), change summation for product.\\
%(2) Check (5.10) vs previously established formula in the multirefined paper. OK, not quite the same !\\
%(3) Acknowledgements: remove the `a' before Senior.
%\vskip 2mm \noindent  
%{\bf \red v1 :} The one- and two-refined partition functions are computed by using edge variables. These are replaced in the next version by boundary face variables, accounted for by the octahedral weighting.
%
%\vspace{4mm} 
%\hrule
%
%%%%%%%%%%%%%%%%%%%%%%%%%%%%%%%%%%%%%%%%%%%%%%%%%%%%%%%%%%%%%%%%%%%%%%%%%%%

\baselineskip=4.5mm

\tableofcontents 

\renewcommand{\baselinestretch}{1.1}
\selectfont

%%%%%%%%%%%%%%%%%%%%%%%%%%%%%%%%%%%%%%%%%%%%%%%%%%%%%%%%%%%%%%%%%%%%%%%%%%

\vskip 1truecm
\noindent
{\bf \large 1. Introduction}
\addcontentsline{toc}{subsection}{1. Introduction}
\setcounter{section}{2}
\setcounter{equation}{0}

\medskip
\noindent
Some statistical models exhibit the arctic phenomenon: they show genuinely different behaviours in different regions of the domain on which they are defined. Random tilings of Aztec diamonds by dominos are the most famous example \cite{EKLP92a}. Aztec diamonds are finite squarelike domains, with staircase boundaries (an essential ingredient \cite{St21}). When the size of the domain gets large, one observes a solid, frozen phase in the four corners, and a liquid, entropic phase in the central region. In the limit of infinite size, equivalently of zero mesh after an appropriate rescaling, the interface between the two phases becomes sharp and converges to the inscribed circle, called the arctic circle \cite{JPS98}.

Similar arctic phenomena occur in other random tiling models, for instance tilings of polygons by lozenges, see \cite{CLP98} and the recent review \cite{Go21}, but also in interacting vertex models \cite{CP10,CPZ10,DDFG20,dGKW21} and lattice path models \cite{DFL18}. These models show the same arctic phenomenon, namely a solid phase and a liquid phase separated by an interface which condenses in the scaling limit onto a curve, called the arctic curve. The presence of two phases is generically due to specific boundary conditions, though the probability measure used may also have spectacular consequences.

This is the case for two-periodic Aztec diamonds, for which the measure on the domino tilings is no longer uniform but depends on two alternating weights \cite{DFSG14}. The effect of the new measure is that a third phase, called gaseous, appears inside the liquid phase, so that there are now two interfaces separating the three phases. Two-periodic Aztec diamonds and generalizations thereof have been subsequently studied by several authors \cite{CY14,CJ16,DK21,Be19,JM21,BD22}.

The most basic problem, solved for all models quoted above, is to determine the shape of the arctic curves in the scaling limit. In many cases, this has been done by looking at an appropriate observable which serves as an order parameter and probes the different phases. Recently an alternative but heuristic method, called the tangent method, has been proposed to compute the arctic curves separating the outer, solid phase from the liquid phase \cite{CS16} (it does not give access to the interfaces between inner phases). The tangent method has been successfully used in many models \cite{CS16,DFL18,DR19,CPS19,DFG19,DDFG20,dKDR22}. 

Our objective here is to apply the tangent method (in two different versions) to compute the (outer) arctic curve of two-periodic Aztec diamonds. As mentioned above, this curve is known and has actually been computed in three different ways: \cite{DFSG14} used the octahedron recurrence (as in the present work), while \cite{CJ16} and \cite{DK21} computed the inverse of the Kasteleyn matrix, using two different approaches. Whereas \cite{DFSG14} considered a bulk observable, namely the average occupancy rate around a face, we consider boundary correlation functions, or refined partition functions, more appropriate to apply the tangent method. In both cases, the approach allows to compute the outer arctic curve only; the inner component can however be obtained from the real section of the algebraic curve describing the outer component. In contrast, the routes followed in \cite{CJ16} and \cite{DK21} are technically more complex and sophisticated but they also yield more, and in particular, they yield both the outer and the inner components.

The plan of the paper is as follows. Section 2 presents the basis of the lattice calculations, namely the non-linear recurrence satisfied by the lattice partition function for tilings of Aztec diamonds with a fairly general probability measure (octahedron recurrence). Section 3 specializes the general setting to the two-periodic measure and contains first (known) results. A first application of the (geometric) tangent method is carried out in Section 4 and yields the parametric form of the arctic curve separating the solid and liquid phases. This first method is based on one-refined lattice partition functions, which concern the tilings with a fixed number of vertical dominos along the northwest (NW) boundary, as well as the required information related to the lattice path description of the tilings. The second application, which uses the two-refined tangent method, is worked out in Section 5 and relies on two-refined partition functions, for which the numbers of vertical dominos on the NW and northeast (NE) boundaries are fixed. Section 6 makes some observations pointing to the general expectation that multirefinements of partition functions can all be obtained, in the scaling limit, from one-refinements. It also clarifies the relation between the tangent methods used with and without an extension of the domain. Section 7 discusses the possible application of the tangent methods used here to Aztec diamonds with weights of higher periodicity. An appendix contains the technical proof of some statements made in the text.

%%%%%%%%%%%%%%%%%%%%%%%%%%%%%%%%%%%%%%%%%%%%%%%%%%%%%%%%%%%%%%%%%%%%%%%%%%

\vskip 0.5truecm
\noindent
{\bf \large 2. Octahedron recurrence}
\addcontentsline{toc}{subsection}{2. Octahedron recurrence}
\setcounter{section}{2}
\setcounter{equation}{0}

\medskip
\noindent
The number of ways an Aztec diamond of order $n$ can be tiled by dominos is equal to $T_n = 2^{n(n+1)/2}$. This surprising result has received different proofs \cite{EKLP92a,EKLP92b,EF05}, but the proof by Kuo \cite{Ku04} plays an essential role in the present work. By using a method called graphical condensation, he shows that the concentric superposition of two Aztec diamond tilings, of order $n$ and of order $n-2$, can be uniquely decomposed into two Aztec diamond tilings of order $n-1$, superposed with a slight shift, either horizontal or vertical. The isomorphism can be graphically represented as follows,

\begin{figure}[h]
\begin{center}
\psset{unit=.4cm}
\hspace{1.5cm}
\pspicture(0,1)(5,6)
\pspolygon[linewidth=0.2pt,linecolor=black,](0,0)(-3,3)(0,6)(3,3)
\pspolygon[linewidth=0.2pt,linecolor=black,](0,0.8)(-2.2,3)(0,5.2)(2.2,3)
\rput(5,3){$=$}
\endpspicture
\hspace{1.8cm}
\pspicture(0,1)(5,6)
\pspolygon[linewidth=0.2pt,linecolor=black,](0,0.25)(-2.75,3)(0,5.75)(2.75,3)
\rput(0.5,0){\pspolygon[linewidth=0.2pt,linecolor=black,](0,0.25)(-2.75,3)(0,5.75)(2.75,3)}
\psline[linewidth=0.2pt,linecolor=black](0.1,5.8)(0.4,5.8)
\psline[linewidth=0.2pt,linecolor=black](0.1,0.2)(0.4,0.2)
\rput(5,3){$\cup$}
\endpspicture
\hspace{1.7cm}
\pspicture(0,1)(5,6)
\rput(0,-0.25){\pspolygon[linewidth=0.2pt,linecolor=black,](0,0.25)(-2.75,3)(0,5.75)(2.75,3)}
\rput(0,0.25){\pspolygon[linewidth=0.2pt,linecolor=black,](0,0.25)(-2.75,3)(0,5.75)(2.75,3)}
\psline[linewidth=0.2pt,linecolor=black](-2.8,2.85)(-2.8,3.15)
\psline[linewidth=0.2pt,linecolor=black](2.8,2.85)(2.8,3.15)
\endpspicture
\end{center}
\end{figure}

\noindent
and leads to the following non-linear recurrence relation,
\be
T_n \, T_{n-2} = 2 \, T_{n-1}^2.
\label{kuo}
\ee
It linearizes for the ratios $S_n=T_n/T_{n-1}$, giving the simple relation $S_n = 2 S_{n-1}$. With the initial conditions $T_0=1$ and $T_1=2$, it easily yields the result quoted above. 

\begin{figure}[t]
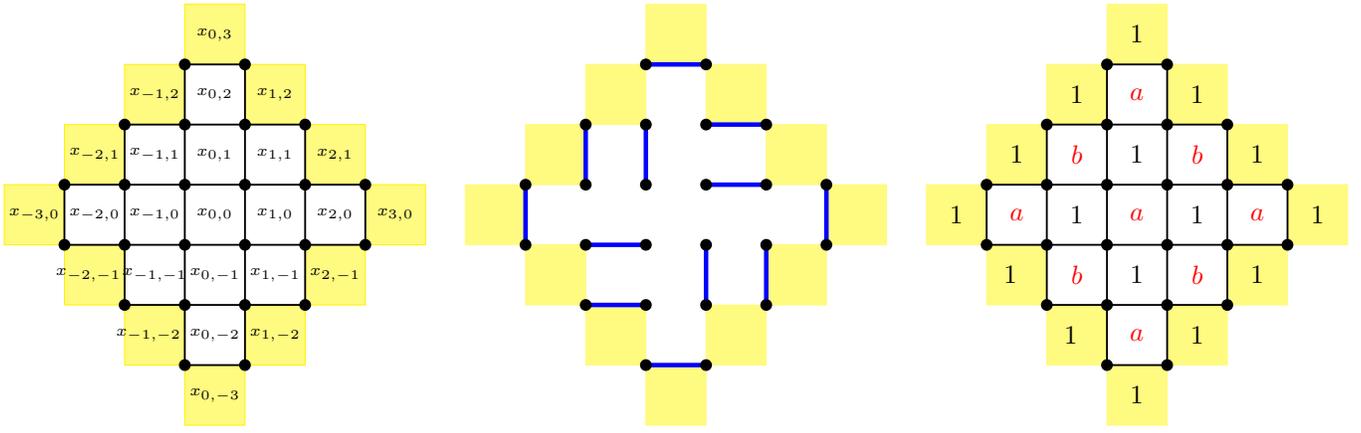

\begin{center}
\psset{unit=.8cm}
\pspicture(0,2)(6,8)
\rput(-0.5,4.5){\multido{\nx=0+1}{4}{\rput(\nx,\nx){\pspolygon[linewidth=0pt,linecolor=yellow,fillstyle=solid,fillcolor=lightyellow](0,0)(0,1)(1,1)(1,0)}}}
\rput(2.5,1.5){\multido{\nx=0+1}{4}{\rput(\nx,\nx){\pspolygon[linewidth=0pt,linecolor=yellow,fillstyle=solid,fillcolor=lightyellow](0,0)(0,1)(1,1)(1,0)}}}
\rput(0.5,3.5){\multido{\nx=0+1}{2}{\rput(\nx,-\nx){\pspolygon[linewidth=0pt,linecolor=yellow,fillstyle=solid,fillcolor=lightyellow](0,0)(0,1)(1,1)(1,0)}}}
\rput(3.5,6.5){\multido{\nx=0+1}{2}{\rput(\nx,-\nx){\pspolygon[linewidth=0pt,linecolor=yellow,fillstyle=solid,fillcolor=lightyellow](0,0)(0,1)(1,1)(1,0)}}}
\multido{\nt=4+1}{2}{\rput(0,\nt){\pscircle[linecolor=black,fillstyle=solid,fillcolor=black](0.5,0.5){0.08}}}
\multido{\nt=3+1}{4}{\rput(1,\nt){\pscircle[linecolor=black,fillstyle=solid,fillcolor=black](0.5,0.5){0.08}}}
\multido{\nt=2+1}{6}{\rput(2,\nt){\pscircle[linecolor=black,fillstyle=solid,fillcolor=black](0.5,0.5){0.08}}}
%\multido{\nt=1+1}{8}{\rput(3,\nt){\pscircle[linecolor=black,fillstyle=solid,fillcolor=black](0.5,0.5){0.08}}}
%\multido{\nt=0+1}{10}{\rput(4,\nt){\pscircle[linecolor=black,fillstyle=solid,fillcolor=black](0.5,0.5){0.08}}}
%\multido{\nt=-1+1}{12}{\rput(5,\nt){\pscircle[linecolor=black,fillstyle=solid,fillcolor=black](0.5,0.5){0.12}}}
%\multido{\nt=-1+1}{12}{\rput(6,\nt){\pscircle[linecolor=black,fillstyle=solid,fillcolor=black](0.5,0.5){0.12}}}
%\multido{\nt=0+1}{10}{\rput(5,\nt){\pscircle[linecolor=black,fillstyle=solid,fillcolor=black](0.5,0.5){0.08}}}
%\multido{\nt=1+1}{8}{\rput(4,\nt){\pscircle[linecolor=black,fillstyle=solid,fillcolor=black](0.5,0.5){0.08}}}
\multido{\nt=2+1}{6}{\rput(3,\nt){\pscircle[linecolor=black,fillstyle=solid,fillcolor=black](0.5,0.5){0.08}}}
\multido{\nt=3+1}{4}{\rput(4,\nt){\pscircle[linecolor=black,fillstyle=solid,fillcolor=black](0.5,0.5){0.08}}}
\multido{\nt=4+1}{2}{\rput(5,\nt){\pscircle[linecolor=black,fillstyle=solid,fillcolor=black](0.5,0.5){0.08}}}
\psline[linewidth=.7pt,linecolor=black](2.5,7.5)(3.5,7.5)
\psline[linewidth=.7pt,linecolor=black](1.5,6.5)(4.5,6.5)
\psline[linewidth=.7pt,linecolor=black](0.5,5.5)(5.5,5.5)
\psline[linewidth=.7pt,linecolor=black](0.5,4.5)(5.5,4.5)
\psline[linewidth=.7pt,linecolor=black](1.5,3.5)(4.5,3.5)
\psline[linewidth=.7pt,linecolor=black](2.5,2.5)(3.5,2.5)
\psline[linewidth=.7pt,linecolor=black](0.5,5.5)(0.5,4.5)
\psline[linewidth=.7pt,linecolor=black](1.5,6.5)(1.5,3.5)
\psline[linewidth=.7pt,linecolor=black](2.5,7.5)(2.5,2.5)
\psline[linewidth=.7pt,linecolor=black](3.5,7.5)(3.5,2.5)
\psline[linewidth=.7pt,linecolor=black](4.5,6.5)(4.5,3.5)
\psline[linewidth=.7pt,linecolor=black](5.5,5.5)(5.5,4.5)
\rput(3,5){\tiny $x_{0,0}$}
\rput(4,5){\tiny $x_{1,0}$}
\rput(5,5){\tiny $x_{2,0}$}
\rput(6,5){\tiny $x_{3,0}$}
\rput(2,5){\tiny $x_{-1,0}$}
\rput(1,5){\tiny $x_{-2,0}$}
\rput(0,5){\tiny $x_{-3,0}$}
\rput(1,6){\tiny $x_{-2,1}$}
\rput(2,6){\tiny $x_{-1,1}$}
\rput(3,6){\tiny $x_{0,1}$}
\rput(4,6){\tiny $x_{1,1}$}
\rput(5,6){\tiny $x_{2,1}$}
\rput(2,7){\tiny $x_{-1,2}$}
\rput(3,7){\tiny $x_{0,2}$}
\rput(4,7){\tiny $x_{1,2}$}
\rput(3,8){\tiny $x_{0,3}$}
\rput(0.9,4){\tiny $x_{-2,-1}$}
\rput(2,4){\tiny $x_{-1,-1}$}
\rput(3,4){\tiny $x_{0,-1}$}
\rput(4,4){\tiny $x_{1,-1}$}
\rput(5,4){\tiny $x_{2,-1}$}
\rput(1.9,3){\tiny $x_{-1,-2}$}
\rput(3,3){\tiny $x_{0,-2}$}
\rput(4,3){\tiny $x_{1,-2}$}
\rput(3,2){\tiny $x_{0,-3}$}
\endpspicture
\hspace{1.2cm}
\pspicture(0,2)(6,8)
\rput(-0.5,4.5){\multido{\nx=0+1}{4}{\rput(\nx,\nx){\pspolygon[linewidth=0pt,linecolor=lightyellow,fillstyle=solid,fillcolor=lightyellow](0,0)(0,1)(1,1)(1,0)}}}
\rput(2.5,1.5){\multido{\nx=0+1}{4}{\rput(\nx,\nx){\pspolygon[linewidth=0pt,linecolor=lightyellow,fillstyle=solid,fillcolor=lightyellow](0,0)(0,1)(1,1)(1,0)}}}
\rput(0.5,3.5){\multido{\nx=0+1}{2}{\rput(\nx,-\nx){\pspolygon[linewidth=0pt,linecolor=lightyellow,fillstyle=solid,fillcolor=lightyellow](0,0)(0,1)(1,1)(1,0)}}}
\rput(3.5,6.5){\multido{\nx=0+1}{2}{\rput(\nx,-\nx){\pspolygon[linewidth=0pt,linecolor=lightyellow,fillstyle=solid,fillcolor=lightyellow](0,0)(0,1)(1,1)(1,0)}}}
\psline[linewidth=1.7pt,linecolor=blue](5.5,5.5)(5.5,4.5)
\psline[linewidth=1.7pt,linecolor=blue](3.5,5.5)(4.5,5.5)
\psline[linewidth=1.7pt,linecolor=blue](3.5,6.5)(4.5,6.5)
\psline[linewidth=1.7pt,linecolor=blue](3.5,7.5)(2.5,7.5)
\psline[linewidth=1.7pt,linecolor=blue](2.5,6.5)(2.5,5.5)
\psline[linewidth=1.7pt,linecolor=blue](1.5,6.5)(1.5,5.5)
\psline[linewidth=1.7pt,linecolor=blue](0.5,5.5)(0.5,4.5)
\psline[linewidth=1.7pt,linecolor=blue](2.5,4.5)(1.5,4.5)
\psline[linewidth=1.7pt,linecolor=blue](2.5,3.5)(1.5,3.5)
\psline[linewidth=1.7pt,linecolor=blue](2.5,2.5)(3.5,2.5)
\psline[linewidth=1.7pt,linecolor=blue](3.5,3.5)(3.5,4.5)
\psline[linewidth=1.7pt,linecolor=blue](4.5,3.5)(4.5,4.5)
\multido{\nt=4+1}{2}{\rput(0,\nt){\pscircle[linecolor=black,fillstyle=solid,fillcolor=black](0.5,0.5){0.08}}}
\multido{\nt=3+1}{4}{\rput(1,\nt){\pscircle[linecolor=black,fillstyle=solid,fillcolor=black](0.5,0.5){0.08}}}
\multido{\nt=2+1}{6}{\rput(2,\nt){\pscircle[linecolor=black,fillstyle=solid,fillcolor=black](0.5,0.5){0.08}}}
%\multido{\nt=1+1}{8}{\rput(3,\nt){\pscircle[linecolor=black,fillstyle=solid,fillcolor=black](0.5,0.5){0.08}}}
%\multido{\nt=0+1}{10}{\rput(4,\nt){\pscircle[linecolor=black,fillstyle=solid,fillcolor=black](0.5,0.5){0.08}}}
%\multido{\nt=-1+1}{12}{\rput(5,\nt){\pscircle[linecolor=black,fillstyle=solid,fillcolor=black](0.5,0.5){0.12}}}
%\multido{\nt=-1+1}{12}{\rput(6,\nt){\pscircle[linecolor=black,fillstyle=solid,fillcolor=black](0.5,0.5){0.12}}}
%\multido{\nt=0+1}{10}{\rput(5,\nt){\pscircle[linecolor=black,fillstyle=solid,fillcolor=black](0.5,0.5){0.08}}}
%\multido{\nt=1+1}{8}{\rput(4,\nt){\pscircle[linecolor=black,fillstyle=solid,fillcolor=black](0.5,0.5){0.08}}}
\multido{\nt=2+1}{6}{\rput(3,\nt){\pscircle[linecolor=black,fillstyle=solid,fillcolor=black](0.5,0.5){0.08}}}
\multido{\nt=3+1}{4}{\rput(4,\nt){\pscircle[linecolor=black,fillstyle=solid,fillcolor=black](0.5,0.5){0.08}}}
\multido{\nt=4+1}{2}{\rput(5,\nt){\pscircle[linecolor=black,fillstyle=solid,fillcolor=black](0.5,0.5){0.08}}}
\endpspicture
\hspace{1.2cm}
\pspicture(0,2)(6,8)
\rput(-0.5,4.5){\multido{\nx=0+1}{4}{\rput(\nx,\nx){\pspolygon[linewidth=0pt,linecolor=lightyellow,fillstyle=solid,fillcolor=lightyellow](0,0)(0,1)(1,1)(1,0)}}}
\rput(2.5,1.5){\multido{\nx=0+1}{4}{\rput(\nx,\nx){\pspolygon[linewidth=0pt,linecolor=lightyellow,fillstyle=solid,fillcolor=lightyellow](0,0)(0,1)(1,1)(1,0)}}}
\rput(0.5,3.5){\multido{\nx=0+1}{2}{\rput(\nx,-\nx){\pspolygon[linewidth=0pt,linecolor=lightyellow,fillstyle=solid,fillcolor=lightyellow](0,0)(0,1)(1,1)(1,0)}}}
\rput(3.5,6.5){\multido{\nx=0+1}{2}{\rput(\nx,-\nx){\pspolygon[linewidth=0pt,linecolor=lightyellow,fillstyle=solid,fillcolor=lightyellow](0,0)(0,1)(1,1)(1,0)}}}
\multido{\nt=4+1}{2}{\rput(0,\nt){\pscircle[linecolor=black,fillstyle=solid,fillcolor=black](0.5,0.5){0.08}}}
\multido{\nt=3+1}{4}{\rput(1,\nt){\pscircle[linecolor=black,fillstyle=solid,fillcolor=black](0.5,0.5){0.08}}}
\multido{\nt=2+1}{6}{\rput(2,\nt){\pscircle[linecolor=black,fillstyle=solid,fillcolor=black](0.5,0.5){0.08}}}
%\multido{\nt=1+1}{8}{\rput(3,\nt){\pscircle[linecolor=black,fillstyle=solid,fillcolor=black](0.5,0.5){0.08}}}
%\multido{\nt=0+1}{10}{\rput(4,\nt){\pscircle[linecolor=black,fillstyle=solid,fillcolor=black](0.5,0.5){0.08}}}
%\multido{\nt=-1+1}{12}{\rput(5,\nt){\pscircle[linecolor=black,fillstyle=solid,fillcolor=black](0.5,0.5){0.12}}}
%\multido{\nt=-1+1}{12}{\rput(6,\nt){\pscircle[linecolor=black,fillstyle=solid,fillcolor=black](0.5,0.5){0.12}}}
%\multido{\nt=0+1}{10}{\rput(5,\nt){\pscircle[linecolor=black,fillstyle=solid,fillcolor=black](0.5,0.5){0.08}}}
%\multido{\nt=1+1}{8}{\rput(4,\nt){\pscircle[linecolor=black,fillstyle=solid,fillcolor=black](0.5,0.5){0.08}}}
\multido{\nt=2+1}{6}{\rput(3,\nt){\pscircle[linecolor=black,fillstyle=solid,fillcolor=black](0.5,0.5){0.08}}}
\multido{\nt=3+1}{4}{\rput(4,\nt){\pscircle[linecolor=black,fillstyle=solid,fillcolor=black](0.5,0.5){0.08}}}
\multido{\nt=4+1}{2}{\rput(5,\nt){\pscircle[linecolor=black,fillstyle=solid,fillcolor=black](0.5,0.5){0.08}}}
\psline[linewidth=.7pt,linecolor=black](2.5,7.5)(3.5,7.5)
\psline[linewidth=.7pt,linecolor=black](1.5,6.5)(4.5,6.5)
\psline[linewidth=.7pt,linecolor=black](0.5,5.5)(5.5,5.5)
\psline[linewidth=.7pt,linecolor=black](0.5,4.5)(5.5,4.5)
\psline[linewidth=.7pt,linecolor=black](1.5,3.5)(4.5,3.5)
\psline[linewidth=.7pt,linecolor=black](2.5,2.5)(3.5,2.5)
\psline[linewidth=.7pt,linecolor=black](0.5,5.5)(0.5,4.5)
\psline[linewidth=.7pt,linecolor=black](1.5,6.5)(1.5,3.5)
\psline[linewidth=.7pt,linecolor=black](2.5,7.5)(2.5,2.5)
\psline[linewidth=.7pt,linecolor=black](3.5,7.5)(3.5,2.5)
\psline[linewidth=.7pt,linecolor=black](4.5,6.5)(4.5,3.5)
\psline[linewidth=.7pt,linecolor=black](5.5,5.5)(5.5,4.5)
\rput(3,5){\red \small $a$}
\rput(4,5){\small $1$}
\rput(5,5){\red \small $a$}
\rput(6,5){\small $1$}
\rput(2,5){\small $1$}
\rput(1,5){\red \small $a$}
\rput(0,5){\small $1$}
\rput(1,6){\small $1$}
\rput(2,6){\red \small $b$}
\rput(3,6){\small $1$}
\rput(4,6){\red \small $b$}
\rput(5,6){\small $1$}
\rput(2,7){\small $1$}
\rput(3,7){\red \small $a$}
\rput(4,7){\small $1$}
\rput(3,8){\small $1$}
\rput(0.9,4){\small $1$}
\rput(2,4){\red \small $b$}
\rput(3,4){\small $1$}
\rput(4,4){\red \small $b$}
\rput(5,4){\small $1$}
\rput(1.9,3){\small $1$}
\rput(3,3){\red \small $a$}
\rput(4,3){\small $1$}
\rput(3,2){\small $1$}
\endpspicture
\end{center}
\caption{{\it Left}: the graph $\hat{\cal A}_3$, the extended Aztec graph of order 3. {\it Center}: a typical perfect matching of ${\cal A}_3$. {\it Right}: the weights used to define the two-periodic Aztec diamond, for which the lowest and highest faces of ${\cal A}_n$ have weight $a$, whatever the parity of $n$.}
\label{fig1}
\end{figure}

A few years later, Speyer made a very neat observation \cite{Sp07}: the above recurrence not only holds for uniformly weighted Aztec diamonds but also for much more general measures. To describe it, we use the language of perfect matchings of the dual graph of the Aztec diamond, or simply the Aztec graph. The Figure \ref{fig1} shows the Aztec graph of order 3, extended by the outer faces, in yellow. The Aztec graph of order $n$ will be denoted by ${\cal A}_n$, while $\hat{\cal A}_n$ will denote its extension by the outer faces  (the set of vertices and the set of edges are identical for ${\cal A}_n$ and $\hat{\cal A}_n$). The faces of $\hat{\cal A}_n$ are labelled a pair $(k,\ell)$ of integer coordinates, with $-n \le k,\ell \le n$ and $|k|+|\ell| \le n$; a weight $x_{k,\ell}$ is attached to the face $(k,\ell)$. A typical perfect matching is shown in the central panel of Figure 1.

The measure is defined by giving a perfect matching $M$ of ${\cal A}_n$ a weight equal to 
\be
\prod_{(k,\ell) \in \hat{\cal A}_{n}} x_{k,\ell}^{1-N_{k,\ell}},
\label{weight}
\ee
where $N_{k,\ell}$ is the number of edges of $M$ which are around the face $(k,\ell)$. The product extends to all faces in $\hat{\cal A}_n$, which therefore includes inner faces, namely those of ${\cal A}_n$, as well as boundary faces all around it. Thus $N_{k,\ell}$ can be equal to 0,1 or 2 for an inner face, and 0 or 1 for a boundary face. We refer to this as the {\it octahedral measure}.

The observation made in \cite{Sp07} (see also \cite{DF14}) is that the sum of the weights of all perfect matchings of ${\cal A}_n$, equivalently the partition function with respect to the octahedral measure, namely
\be
T_{n;0,0} = \sum_{M\, {\rm of}{\cal A}_{n}} \; \prod_{(k,\ell) \in \hat{\cal A}_{n}} x_{k,\ell}^{1-N_{k,\ell}},
\ee
satisfies a recurrence relation generalizing (\ref{kuo}),
\be
T_{n;0,0} \, T_{n-2;0,0} = T_{n-1;1,0} \, T_{n-1;-1,0} + T_{n-1;0,1} \, T_{n-1;0,-1},
\label{octa}
\ee
where $T_{k;i,j}$ is the partition function for the perfect matchings of the Aztec subgraph ${\cal A}_k \subset {\cal A}_n$ of order $k$ centered at the face $(i,j)$, computed with respect to the face weights of $\hat{\cal A}_n$ (thus $T_{k;i,j}$ depends on a subset of the weights used to computed $T_{n;0,0}$). The four terms in the r.h.s. of (\ref{octa}) correspond to the four slightly shifted Aztec diamonds shown in the picture before (\ref{kuo}). The recurrence relation (\ref{octa}) is subjected to the initial conditions
\be
T_{0;i,j}=1, \quad {\rm and} \quad T_{1;i,j} = x_{i,j}^{-1} \, (x^{}_{i,j-1} \, x^{}_{i,j+1} + x^{}_{i-1,j} \, x^{}_{i+1,j}),
\ee
from which all higher order partition functions can be computed. For uniform weights (all $x_{i,j}=1$), $T_{n;i,j} \equiv T_n$ does not depend on $i,j$ and reduces to the number of domino tilings of the Aztec diamond of order $n$. 

One can generalize the above set-up by assigning weights $x_{k,\ell}$ to all faces $(k,\ell)$ of $\Z^2$, and defining $T_{n;i,j}$ to be the partition function for the perfect matchings of the graph ${\cal A}_n$ centered at the face $(i,j)$. Then these partitions functions satisfy a set of non-linear recurrence relations, called the octahedron recurrence \cite{Sp07},
\be
T_{n;i,j} \, T_{n-2;i,j} = T_{n-1;i+1,j}\, T_{n-1;i-1,j} + T_{n-1;i,j+1}\,T_{n-1;i,j-1},
\ee
for $i,j \in \Z$, subjected to appropriate initial conditions. The structure of these relations is reminiscent of (and actually related to) Dodgson's condensation formula and appears in other contexts. We refer to \cite{Sp07} for more background and applications of $T$-systems. There is a further generalization where the edges get extra weights contributing to the total weight of a perfect matching \cite{Sp07}; this will be used in Section 7.

The measure that is more often used in the literature gives a perfect matching the weight
\be
\prod_{(k,\ell) \in {\cal A}_{n}} x_{k,\ell}^{N_{k,\ell}}.
\label{conv}
\ee
Apart from a global factor and a change of parameters $x^{}_{k,\ell} \to x_{k,\ell}^{-1}$, a more essential difference is that the product over the faces does not include the boundary faces of $\hat{\cal A}_n$. Keeping the general weights on the faces of ${\cal A}_n$ as above, the resulting partition functions, which we will generically denote by $Z_n$, do not quite satisfy the octahedron recurrence. In the case of two-periodic Aztec diamonds, they do satisfy it but the resulting recurrence is not as simple. For this reason, we will mostly adopt the octahedral weighting (\ref{weight}) used to write the $T$-system, and revert to the conventional measure at the end of the calculations. For the path description however, the conventional measure is easier.

\vskip 0.5truecm
\noindent
{\bf \large 3. Two-periodic Aztec diamonds}
\addcontentsline{toc}{subsection}{3. Two-periodic Aztec diamonds}
\setcounter{section}{3}
\setcounter{equation}{0}

\medskip
\noindent
Two-periodic Aztec diamonds are a particularly simple example of the previous set-up. We use here the convention that the two-periodic Aztec diamonds with parameters $(a,b)$ are such that the lowest and highest faces of the Aztec graph have weight $a$ irrespective of the parity of $n$, see the right picture in Figure \ref{fig1} (this convention somewhat simplifies the path description given later). As a consequence, the leftmost and rightmost faces of the graph have weight $a$ if $n$ is odd, $b$ if $n$ is even. It also implies that the number of $a$--faces and the number of $b$--faces are the same if $n$ is even, and equal to $\frac{n^2}2$; for $n$ odd, there is one more $a$--face than $b$--faces, and so the two numbers are respectively $\frac{n^2+1}2$ and $\frac{n^2-1}2$. Although we could set $ab=1$ without loss of generality, we keep $a$ and $b$ throughout. The ratio $\frac ab$, or any function of it, is the only essential parameter; in the following we will mostly use
\be
\beta = \frac {a^2}{b^2}.
\ee

It is easy to make contact with the previous section. For $n$ odd, we choose the weight of all odd faces to be $x_{k,\ell}=1$ ($k+\ell$ odd), and the weight of the even faces to be $x_{k,\ell}=a$ if $k,\ell$ are both even, and $x_{k,\ell}=b$ if $k,\ell$ are both odd, as shown on the right of Figure \ref{fig1}; for $n$ even, we choose the weight of all even faces to be $x_{k,\ell}=1$ ($k+\ell$ even), and the weight of the odd faces to be $x_{k,\ell}=a$ if $k$ is even, and $x_{k,\ell}=b$ if $k$ odd. We note that the faces surrounding ${\cal A}_n$, namely the boundary faces of $\hat{\cal A}_n$, have all weight 1; the same is true of the subgraphs ${\cal A}_{n-1}$ and ${\cal A}_{n-2}$ involved in the recurrence. 

Let $T_n(a,b)$ be the weighted sum over the domino tilings of the two-periodic Aztec diamond of order $n$ and parameters $(a,b)$, each tiling being weighted according to the octahedral measure (\ref{weight}). We note that if the Aztec graph of order $n$ centered at $(0,0)$ has parameters $(a,b)$, those of order $n-1$ centered at $(0,\pm 1)$ and $(\pm 1,0)$ have parameters $(a,b)$ and $(b,a)$ respectively. Therefore the recurrence (\ref{octa}) yields
\be
T_n(a,b)\,T_{n-2}(a,b) = T_{n-1}^2(a,b) + T_{n-1}^2(b,a).
\ee
The boundary conditions $T_0(a,b)=1$ and $T_1(a,b)=\frac 2a$ allow to compute $T_n(a,b)$ for any $n$, with the following result \cite{DFSG14},
\be
T_n(a,b) = \Big(\frac2{ab}\Big)^{\lfloor \!\frac{(n+1)^2}4\!\rfloor} \big(a^2 + b^2 \big)^{\lfloor\!\frac{n^2}4\!\rfloor} \times \begin{cases}
\vspace{-1.5mm}
1 & {\rm if\ }n = 0 \bmod 2,\\
\vspace{-1.5mm}
b & {\rm if\ }n = 1 \bmod 4,\\
a & {\rm if\ }n = 3 \bmod 4.\\
\end{cases}
\label{Fn}
\ee

As explained in the previous section, the partition function $Z_n(a,b)$ with respect to the standard measure (\ref{conv}) is easily related to $T_n(a,b)$ by a global factor and a change of parameters. We obtain, using the numbers of $a$-- and $b$--faces given above, 
\be
Z_n(a,b) = (ab)^{\frac{n^2}2} \begin{Bmatrix} 1 \\ \sqrt{\frac ab} \end{Bmatrix} T_n\Big(\frac 1a,\frac 1b\Big) = 
\big(2ab\big)^{\lfloor\!\frac{(n+1)^2}4\!\rfloor} \big(a^2 + b^2 \big)^{\lfloor\!\frac{n^2}4\!\rfloor} \times \begin{cases}
\vspace{-1.5mm}
1 & {\rm if\ }n \neq 1 \bmod 4,\\
\noalign{\medskip}
\frac ab & {\rm if\ }n = 1 \bmod 4,\\
\end{cases}
\label{ZF}
\ee 
where the two numbers within the curly brackets refer to the two cases $n$ even or $n$ odd. The first few partition functions read
\be
Z_2(a,b) = \big(2ab\big)^2\,(a^2+b^2), \quad Z_3(a,b) = \big(2ab\big)^4\,(a^2+b^2)^2, \quad Z_4(a,b) = \big(2ab\big)^6\,(a^2+b^2)^4.
\label{Zn}
\ee

%%%%%%%%%%%%%%%%%%%%%%%%%%%%%%%%%%%%%%%%%%%%%%%%%%%%%%%%%%%%%%%%%%%%%%%%%%

\vskip 0.5truecm
\noindent
{\bf \large 4. One-refined partition functions}
\addcontentsline{toc}{subsection}{4. One-refined partition functions}
\setcounter{section}{4}
\setcounter{equation}{0}

\begin{figure}[h]
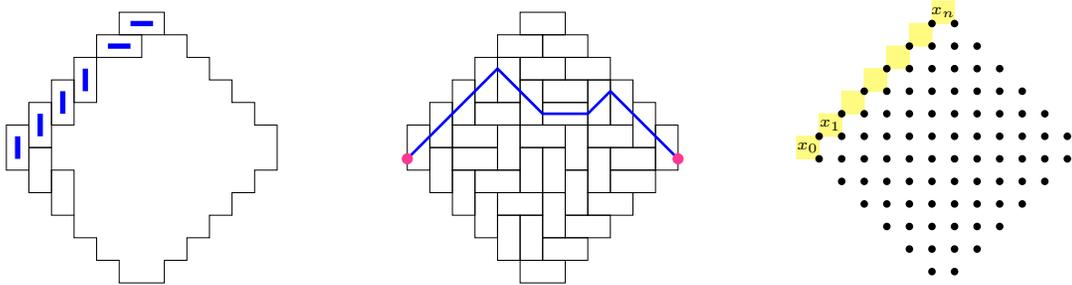

\begin{center}
\psset{unit=.3cm}
\pspicture(0,0)(14,12)
\rput(5,10){\hdimerw}
\psline[linewidth=2.0pt,linecolor=blue](5.5,10.5)(6.5,10.5)
\rput(4,9){\hdimerw}
\psline[linewidth=2.0pt,linecolor=blue](4.5,9.5)(5.5,9.5)
%\rput(6,9){\hdimerw}
%\rput(4,7){\vdimer}
%\rput(5,8){\hdimerw}
%\rput(7,8){\hdimerw}
\rput(2,6){\vdimerw}
\psline[linewidth=2.0pt,linecolor=blue](2.5,6.5)(2.5,7.5)
\rput(1,5){\vdimerw}
\psline[linewidth=2.0pt,linecolor=blue](1.5,5.5)(1.5,6.5)
\rput(0,4){\vdimerw}
\psline[linewidth=2.0pt,linecolor=blue](0.5,4.5)(0.5,5.5)
\rput(1,3){\vdimerw}
\rput(2,2){\vdimerw}
%\rput(3,1){\vdimerw}
%\rput(4,0){\vdimerw}
%\rput(5,-1){\hdimerw}
\rput(3,7){\vdimerw}
\psline[linewidth=2.0pt,linecolor=blue](3.5,7.5)(3.5,8.5)
\psline[linewidth=0.2pt,linecolor=black](7,10)(8,10)(8,9)(9,9)(9,8)(10,8)(10,7)(11,7)(11,6)(12,6)(12,4)(11,4)(11,3)(10,3)(10,2)(9,2)(9,1)(8,1)(8,0)(7,0)(7,-1)(5,-1)(5,0)(4,0)(4,1)(3,1)(3,2)
\endpspicture
\hspace{1cm}
\pspicture(0,0)(14,12)
\rput(5,10){\hdimerw}
\rput(4,9){\hdimerw}
\rput(6,9){\hdimerw}
\rput(4,7){\vdimerw}
\rput(5,8){\hdimerw}
\rput(7,8){\hdimerw}
\rput(2,6){\vdimerw}
\rput(1,5){\vdimerw}
\rput(0,4){\vdimerw}
\rput(1,3){\vdimerw}
\rput(2,2){\vdimerw}
\rput(3,1){\vdimerw}
\rput(4,0){\vdimerw}
\rput(5,-1){\hdimerw}
\rput(6,0){\hdimerw}
\rput(7,1){\hdimerw}
\rput(8,2){\hdimerw}
\rput(9,3){\hdimerw}
\rput(11,4){\vdimerw}
\rput(10,5){\vdimerw}
\rput(9,6){\vdimerw}
\rput(8,6){\vdimerw}
\rput(6,7){\hdimerw}
\rput(5,6){\vdimerw}
\rput(3,7){\vdimerw}
\rput(8,5){\hdimerw}
\rput(9,4){\hdimerw}
\rput(7,4){\vdimerw}
\rput(8,3){\vdimerw}
\rput(6,6){\hdimerw}
\rput(5,0){\vdimerw}
\rput(6,1){\vdimerw}
\rput(7,2){\vdimerw}
\rput(6,3){\vdimerw}
\rput(5,5){\hdimerw}
\rput(2,5){\hdimerw}
\rput(2,4){\hdimerw}
\rput(3,6){\hdimerw}
\rput(4,4){\vdimerw}
\rput(3,3){\hdimerw}
\rput(4,2){\hdimerw}
\rput(5,3){\vdimerw}
%\psline[linewidth=1.0pt,linecolor=mypink](5,-0.5)(7,-0.5)
%\psline[linewidth=1.0pt,linecolor=mypink](4,0.5)(5,1.5)(6,0.5)(8,0.5)
%\psline[linewidth=1.0pt,linecolor=mypink](3,1.5)(4,2.5)(6,2.5)(7,1.5)(9,1.5)
%\psline[linewidth=1.0pt,linecolor=mypink](2,2.5)(3,3.5)(5,3.5)(6,4.5)(8,2.5)(10,2.5)
%\psline[linewidth=1.0pt,linecolor=mypink](1,3.5)(2,4.5)(4,4.5)(5,5.5)(7,5.5)(9,3.5)(11,3.5)
\psline[linewidth=1.0pt,linecolor=blue](0,4.5)(4,8.5)(5,7.5)(6,6.5)(8,6.5)(9,7.5)(12,4.5)
%\psline[linewidth=1.0pt,linestyle=dashed,dash=.30 .20,linecolor=mypink](-4,8.5)(6,-1.5)(16,8.5)
%\pscircle[linecolor=mypink,fillstyle=solid,fillcolor=mypink](5,-0.5){0.2}
%\pscircle[linecolor=mypink,fillstyle=solid,fillcolor=mypink](4,0.5){0.2}
%\pscircle[linecolor=mypink,fillstyle=solid,fillcolor=mypink](3,1.5){0.2}
%\pscircle[linecolor=mypink,fillstyle=solid,fillcolor=mypink](2,2.5){0.2}
%\pscircle[linecolor=mypink,fillstyle=solid,fillcolor=mypink](1,3.5){0.2}
\pscircle[linecolor=mypink,fillstyle=solid,fillcolor=mypink](0,4.5){0.2}
%\pscircle[linecolor=mypink,fillstyle=solid,fillcolor=mypink](7,-0.5){0.2}
%\pscircle[linecolor=mypink,fillstyle=solid,fillcolor=mypink](8,0.5){0.2}
%\pscircle[linecolor=mypink,fillstyle=solid,fillcolor=mypink](9,1.5){0.2}
%\pscircle[linecolor=mypink,fillstyle=solid,fillcolor=mypink](10,2.5){0.2}
%\pscircle[linecolor=mypink,fillstyle=solid,fillcolor=mypink](11,3.5){0.2}
\pscircle[linecolor=mypink,fillstyle=solid,fillcolor=mypink](12,4.5){0.2}
%\rput(-0.7,1.5){\footnotesize $(-i,i-n)$}
%\rput(12.5,1.5){\footnotesize $(j,j-n)$}
\endpspicture
\hspace{1cm}
\pspicture(0,0)(10,12)
\rput(-0.5,4.5){\multido{\nx=0+1}{7}{\rput(\nx,\nx){\pspolygon[linewidth=0pt,linecolor=lightyellow,fillstyle=solid,fillcolor=lightyellow](0,0)(0,1)(1,1)(1,0)}}}
\multido{\nt=4+1}{2}{\rput(0,\nt){\pscircle[linecolor=black,fillstyle=solid,fillcolor=black](0.5,0.5){0.12}}}
\multido{\nt=3+1}{4}{\rput(1,\nt){\pscircle[linecolor=black,fillstyle=solid,fillcolor=black](0.5,0.5){0.12}}}
\multido{\nt=2+1}{6}{\rput(2,\nt){\pscircle[linecolor=black,fillstyle=solid,fillcolor=black](0.5,0.5){0.12}}}
\multido{\nt=1+1}{8}{\rput(3,\nt){\pscircle[linecolor=black,fillstyle=solid,fillcolor=black](0.5,0.5){0.12}}}
\multido{\nt=0+1}{10}{\rput(4,\nt){\pscircle[linecolor=black,fillstyle=solid,fillcolor=black](0.5,0.5){0.12}}}
\multido{\nt=-1+1}{12}{\rput(5,\nt){\pscircle[linecolor=black,fillstyle=solid,fillcolor=black](0.5,0.5){0.12}}}
\multido{\nt=-1+1}{12}{\rput(6,\nt){\pscircle[linecolor=black,fillstyle=solid,fillcolor=black](0.5,0.5){0.12}}}
\multido{\nt=0+1}{10}{\rput(7,\nt){\pscircle[linecolor=black,fillstyle=solid,fillcolor=black](0.5,0.5){0.12}}}
\multido{\nt=1+1}{8}{\rput(8,\nt){\pscircle[linecolor=black,fillstyle=solid,fillcolor=black](0.5,0.5){0.12}}}
\multido{\nt=2+1}{6}{\rput(9,\nt){\pscircle[linecolor=black,fillstyle=solid,fillcolor=black](0.5,0.5){0.12}}}
\multido{\nt=3+1}{4}{\rput(10,\nt){\pscircle[linecolor=black,fillstyle=solid,fillcolor=black](0.5,0.5){0.12}}}
\multido{\nt=4+1}{2}{\rput(11,\nt){\pscircle[linecolor=black,fillstyle=solid,fillcolor=black](0.5,0.5){0.12}}}
\rput(0,5){\tiny $x_0$}
\rput(1,6){\tiny $x_1$}
\rput(6,11){\tiny $x_n$}
\endpspicture
\end{center}
\caption{Illustration of the refinement given by the number of vertical dominos along the NW boundary, here four in the Aztec graph of order 6.}
\label{fig2}
\end{figure}

\medskip
\noindent
The refinement we want to compute concerns the perfect matchings which have a fixed number of vertical edges (or dominos) along the NW boundary of the Aztec diamond, as shown in Figure \ref{fig2}. In the path description that we will use later on, the presence of vertical edges forces the uppermost path to follow the NW boundary over a fixed distance, which is precisely the right setting to apply the tangent method.

%The dual graph of an Aztec diamond of order $n$ comprises $2n$ edges along the NW boundary, $n$ of which are horizontal and $n$ are vertical, forming a staircase starting with a vertical step and ending with a horizontal step. We assign the $2n$ steps (edges) the weights $\alpha_1,\delta_1,\alpha_2,\delta_2,\ldots,\alpha_n,\delta_n$. All other edges of the dual graph keep a weight equal to 1. Let $T_n(a,b|\alpha_1,\delta_1,\ldots,\alpha_n,\delta_n)$ be the corresponding weighted sum over all perfect matchings. It can be computed from the generalized octahedron equation (\ref{mocta}).

To do this, we take advantage of the fact that the octahedral measure includes contributions coming from the outer faces of $\hat{\cal A}_n$. So we introduce face variables $x_0,x_1,\ldots,x_n$ on the faces along the NW boundary, as shown in Figure \ref{fig2}, and denote by $T_n(a,b|x_0,x_1,\ldots,x_n)$ the corresponding octahedral partition function.

The octahedron recurrence now implies the following relation
\bea
T_n(a,b|x_0,x_1,\ldots,x_n)\,T_{n-2}(a,b) \egal T_{n-1}(a,b)\,T_{n-1}(a,b|x_1,x_2\ldots,x_n) \nonumber\\
&& \hspace{1cm} + \; T_{n-1}(b,a)\,T_{n-1}(b,a|x_0,x_1,\ldots,x_{n-1}),
\label{recurF}
\eea
and boundary conditions $T_0=1$ and $T_1(a,b|x_0,x_1) = (x_0 + x_1)/a$. 

For $n=3$ for instance, we get, using the recurrence, the following result,
\be
T_3(a,b|x_0,x_1,x_2,x_3) = \frac a4 \Big(\frac2{ab}\Big)^4 \, (a^2+b^2) \,\Big\{x_0 \, b^2 + x_1 \, (2a^2+b^2) + x_2 \, (2a^2+b^2) + x_3 \, b^2\Big\}. 
\ee
The four terms in the brackets correspond to the four possible types of perfect matchings, namely those with 0,\,1,\,2 or 3 vertical edges along the NW boundary, as depicted below.

\begin{figure}[h]
\begin{center}
\psset{unit=.5cm}
\pspicture(0,0)(4,3.5)
\pspolygon[linewidth=0pt,linecolor=lightyellow,fillstyle=solid,fillcolor=lightyellow](0,0)(0,1)(1,1)(1,0)
\rput(0.5,0.5){\small $x_0$}
\psline[linewidth=2.0pt,linecolor=blue](1,1)(2,1)
\psline[linewidth=2.0pt,linecolor=blue](2,2)(3,2)
\psline[linewidth=2.0pt,linecolor=blue](3,3)(4,3)
\pscircle[linecolor=black,fillstyle=solid,fillcolor=black](1,0){0.12}
\pscircle[linecolor=black,fillstyle=solid,fillcolor=black](1,1){0.12}
\pscircle[linecolor=black,fillstyle=solid,fillcolor=black](2,1){0.12}
\pscircle[linecolor=black,fillstyle=solid,fillcolor=black](2,2){0.12}
\pscircle[linecolor=black,fillstyle=solid,fillcolor=black](3,2){0.12}
\pscircle[linecolor=black,fillstyle=solid,fillcolor=black](3,3){0.12}
\pscircle[linecolor=black,fillstyle=solid,fillcolor=black](4,3){0.12}
\endpspicture
\hspace{1.5cm}
\pspicture(0,0)(4,3.5)
\rput(1,1){\pspolygon[linewidth=0pt,linecolor=lightyellow,fillstyle=solid,fillcolor=lightyellow](0,0)(0,1)(1,1)(1,0)}
\rput(1.5,1.5){\small $x_1$}
\psline[linewidth=2.0pt,linecolor=blue](1,0)(1,1)
\psline[linewidth=2.0pt,linecolor=blue](2,2)(3,2)
\psline[linewidth=2.0pt,linecolor=blue](3,3)(4,3)
\pscircle[linecolor=black,fillstyle=solid,fillcolor=black](1,0){0.12}
\pscircle[linecolor=black,fillstyle=solid,fillcolor=black](1,1){0.12}
\pscircle[linecolor=black,fillstyle=solid,fillcolor=black](2,1){0.12}
\pscircle[linecolor=black,fillstyle=solid,fillcolor=black](2,2){0.12}
\pscircle[linecolor=black,fillstyle=solid,fillcolor=black](3,2){0.12}
\pscircle[linecolor=black,fillstyle=solid,fillcolor=black](3,3){0.12}
\pscircle[linecolor=black,fillstyle=solid,fillcolor=black](4,3){0.12}
\endpspicture
\hspace{1.5cm}
\pspicture(0,0)(4,3.5)
\rput(2,2){\pspolygon[linewidth=0pt,linecolor=lightyellow,fillstyle=solid,fillcolor=lightyellow](0,0)(0,1)(1,1)(1,0)}
\rput(2.5,2.5){\small $x_2$}
\psline[linewidth=2.0pt,linecolor=blue](1,0)(1,1)
\psline[linewidth=2.0pt,linecolor=blue](2,1)(2,2)
\psline[linewidth=2.0pt,linecolor=blue](3,3)(4,3)
\pscircle[linecolor=black,fillstyle=solid,fillcolor=black](1,0){0.12}
\pscircle[linecolor=black,fillstyle=solid,fillcolor=black](1,1){0.12}
\pscircle[linecolor=black,fillstyle=solid,fillcolor=black](2,1){0.12}
\pscircle[linecolor=black,fillstyle=solid,fillcolor=black](2,2){0.12}
\pscircle[linecolor=black,fillstyle=solid,fillcolor=black](3,2){0.12}
\pscircle[linecolor=black,fillstyle=solid,fillcolor=black](3,3){0.12}
\pscircle[linecolor=black,fillstyle=solid,fillcolor=black](4,3){0.12}
\endpspicture
\hspace{1.5cm}
\pspicture(0,0)(4,3.5)
\rput(3,3){\pspolygon[linewidth=0pt,linecolor=lightyellow,fillstyle=solid,fillcolor=lightyellow](0,0)(0,1)(1,1)(1,0)}
\rput(3.5,3.5){\small $x_3$}
\psline[linewidth=2.0pt,linecolor=blue](1,0)(1,1)
\psline[linewidth=2.0pt,linecolor=blue](2,1)(2,2)
\psline[linewidth=2.0pt,linecolor=blue](3,2)(3,3)
\pscircle[linecolor=black,fillstyle=solid,fillcolor=black](1,0){0.12}
\pscircle[linecolor=black,fillstyle=solid,fillcolor=black](1,1){0.12}
\pscircle[linecolor=black,fillstyle=solid,fillcolor=black](2,1){0.12}
\pscircle[linecolor=black,fillstyle=solid,fillcolor=black](2,2){0.12}
\pscircle[linecolor=black,fillstyle=solid,fillcolor=black](3,2){0.12}
\pscircle[linecolor=black,fillstyle=solid,fillcolor=black](3,3){0.12}
\pscircle[linecolor=black,fillstyle=solid,fillcolor=black](4,3){0.12}
\endpspicture
\end{center}
\end{figure}

We define $T_{n,k}(a,b)$ as the weighted sum over the perfect matchings which have exactly $k$ vertical edges along the northwest boundary (and necessarily $n-k$ horizontal edges). As is apparent in the picture above, $T_{n,k}(a,b)$ is the coefficient of $x_k$ in $T_n(a,b|x_0,x_1,\ldots,x_n)$. The recurrence relation (\ref{recurF}) then yields
\be
T_{n,k}(a,b) \, T_{n-2}(a,b) = T_{n-1}(a,b) \, T_{n-1,k-1}(a,b) + T_{n-1}(b,a) \, T_{n-1,k}(b,a).
\ee
Noting that $T_{n,0}(a,b) = \{b^{-1} {\rm \ or\ } a^{-1}\} T_{n-1}(b,a)$ depending on whether $n$ is even or odd, and $T_{n,n}(a,b) = a^{-1}T_{n-1}(a,b)$, we check that the recurrence holds for $0 \le k \le n$ provided we set $T_{n,-1}(a,b) = T_{n,n+1}(a,b) = 0$.

In order to simplify the recurrence relation, we introduce the following ratios,
\be
S_{n,k}(a,b) \equiv a \, \frac{T_{n,k}(a,b)}{T_{n-1}(a,b)}.
\ee
The recurrence then becomes linear for the $S_{n,k}$,
\be
S_{n,k}(a,b) = S_{n-1,k-1}(a,b) + S_{n-1,k}(b,a) \times \begin{cases} 1 & {\rm if\ } n=0,1 \bmod 4,\\
\beta & {\rm if\ } n=2,3 \bmod 4,
\end{cases}
\qquad\qquad \beta = \frac{a^2}{b^2},
\ee
for which the explicit expression (\ref{Fn}) of $T_n(a,b)$ has been used to compute the ratios 
\be
\frac{T_n(b,a)}{T_n(a,b)} = 1, \, \frac ab, \, 1, \, \frac ba, \qquad {\rm resp. \ for\ } n = 0, 1,2,3 \bmod 4. 
\ee
Because $S_{1,0}(a,b) = S_{1,1}(a,b) = 1$, all $S_{n,k}(a,b)=S_{n,k}(\beta)$ are functions of the single variable $\beta$, so we can write,
\be
S_{n,k}(\beta) = S_{n-1,k-1}(\beta) + \left\{1 \atop \beta\right\} S_{n-1,k}(\textstyle\frac1\beta), \qquad \displaystyle {\rm for\ }n = \left\{0,1 \atop 2,3\right\} \bmod 4.
\label{recurS}
\ee
It nicely generalizes Pascal's triangular relation, a fact that is not completely surprising since for $a=b=1$ (uniform measure), the numbers $S_{n,k}(1)$ are binomial coefficients, $S_{n,k}(1) = {n \choose k}$. The above recurrence makes it clear that the functions $S_{n,k}(\beta)$ are Laurent polynomials in $\beta$.

%%%%%%%%%%%%%%%%%%%%%%%%%%%%%%%%%%%%%%%%%%%%%%%%%%%%%%%%%%%%%%%%%%%%%%%%%%

\vskip 0.5truecm
\noindent
{\bf 4.1 Generating functions and asymptotics}
\addcontentsline{toc}{subsubsection}{4.1 Generating functions and asymptotics}

\medskip
\noindent
To apply the tangent method, we need to know the asymptotics of the functions $S_{n,k}(\beta)$ when $n$ and $k$ are large. For this we consider the generating function, which we split into four pieces, according to the value of $n$ modulo 4,
\be
G(u,v;\beta) = \sum_{n,k \ge 0} S_{n,k}(\beta) \, u^n v^k = \sum_{a=0}^3 \: \sum_{n,k \ge 0 \atop n=a \bmod 4} S_{n,k}(\beta) \, u^n v^k = \sum_{a=0}^3 \:G^{(a)}(u,v;\beta).
\ee

By multiplying the recurrence relations (\ref{recurS}) by $u^nv^k$, summing over $n,k$ for $n$ within each the four classes modulo 4 and using $S_{n,-1}(\beta)=0$, one readily derives the following relations,
\be
G^{(a)}(u,v;\beta) = \begin{cases} 
\delta_{a,0} + u\,v\, G^{(a-1)}(u,v;\beta) + u \,G^{(a-1)}(u,v;\textstyle \frac1\beta) & {\rm for\ } a=0,1 \bmod 4,\\
\noalign{\smallskip}
u\,v\, G^{(a-1)}(u,v;\beta) + \beta\,u\, G^{(a-1)}(u,v;\textstyle \frac1\beta) & {\rm for\ } a=2,3 \bmod 4,
\end{cases}
\label{Ga}
\ee
and four more relations obtained by replacing $\beta$ by $\frac 1\beta$. Together they provide an inhomogeneous linear system for the eight functions $G^{(a)}(u,v;\beta),\,G^{(a)}(u,v;\frac 1\beta)$. One finds
\be
G^{(0)}(u,v;\beta) = \frac{1-u^4\big[1+2\,(1+\beta+\frac1\beta)\,v^2+v^4\big] + 2u^4v \big[1+\frac1\beta+(1+\beta)\,v^2\big]}{1 - 2u^4(1+v^2)^2 - 4u^4v^2(\beta + \frac 1\beta) + u^8(1-v^2)^4},
\ee
and similar functions for the other three, with in each case the same denominator. With the recurrence relations (\ref{Ga}), the full generating function can be written
\bea
G(u,v;\beta) \egal \big[1 + uv + u^2\,(\beta+v^2) + u^3v\,(1+2\beta+v^2)\big] \, G^{(0)}(u,v;\beta)\nonumber\\
&& \hspace{1cm} + \; \big[u + (1+\beta)\,u^2v + u^3 \, (1+(1+2\beta)v^2)\big] \, G^{(0)}(u,v;\textstyle \frac 1\beta).
\eea

The asymptotics of the coefficients of multivariate generating functions has been studied in \cite{PW08,PW13}. In the present case, the generating function is rational and somewhat simpler. Writing the coefficient $S_{n,k}$ as a double Cauchy integral, the asymptotics can be obtained by doing a saddle point analysis which only involves the denominator of $G(u,v;\beta)$, which we denote by $P(u,v)$. Let $k=r n$, for $0 \le r \le 1$. The asymptotic value of $S_{n,rn}(\beta)$ is given by\footnote{We refer the reader to \cite{DGR19} where the procedure required to compute the asymptotics, following \cite{PW08}, is summarized.}
\be
S_{n,rn}(\beta) \simeq {\rm e}^{n F_1(r)}, \quad {\rm with\ }F_1(r) = -\log u(r) - r \log v(r),
\label{Ltilde}
\ee
where the functions $u(r)$ and $v(r)$ are the positive solutions of the following algebraic system,
\be
\begin{cases}
P(u,v) = 1 - 2u^4(1+v^2)^2 - 4u^4v^2(\beta + \textstyle\frac 1\beta) + u^8(1-v^2)^4 = 0, & \\
r\, u \, \partial_u P(u,v) = v\,\partial_v P(u,v). &
\end{cases}
\ee
The derivative $F_1'(r)$ is the only quantity that will matter for the tangent method; the previous algebraic system implies that it is given by
\be
F_1'(r) = -\log v(r).
\label{L1prime}
\ee

The first condition $P(u,v)=0$ allows to express $u$ in terms of $v$,
\be
u^2(v) = \frac1{\sqrt{1+(\beta+\textstyle\frac1\beta)v^2+v^4} + v \, \frac{1+\beta}{\sqrt{\beta}}} = \frac1{(1-v^2)^2} \Big[\sqrt{1+(\beta+\textstyle\frac1\beta)v^2+v^4} - v \, \textstyle\frac{1+\beta}{\sqrt{\beta}}\Big].
\label{usq}
\ee
From the second condition and a few straightforward calculations, one obtains the following relation between $v$ and $r$,
\be
r(v) = \frac v{1-v^2} \: \Bigg\{\frac{1+\beta}{2\sqrt{\beta}} \: \frac{1+v^2}{\sqrt{1+(\beta+\textstyle\frac1\beta)v^2+v^4}} - v\Bigg\},
\label{quart}
\ee
which we leave as it is.

Summarizing, we have the following asymptotics,
\be
F_1(r) = \lim_{n \to \infty} \frac 1n \log S_{n,rn}(\beta), \qquad \frac{{\rm d}F_1(r)}{{\rm d}r} = \log\frac1{v(r)},
\ee
where $v(r)$ is found by solving (\ref{quart}), a quartic equation for $v^2$.

Let us finish this part with a few remarks. From (\ref{quart}), one can see that $r(v)$ is a strictly increasing function of $v$, varying from 0 to 1 when $v$ runs from 0 to infinity, with $r(1)=\frac 12$. The inverse function $v(r)$ is therefore also strictly increasing, implying that $F_1(r)$ is strictly concave since $F_1''(r) = -\frac{v'(r)}{v(r)} < 0$. From $F_1'(r)=-\log v(r)$, $F_1$ has a unique maximum at $v=1$, $r=\frac12$, where it is equal to $F_1(\frac12) = -\log u(r)\big|_{r=\frac12} = -\log u(v)\big|_{v=1} = \frac12 \log\frac{2(1+\beta)}{\sqrt{\beta}}$.

In preparation for what comes later, we reformulate the above results for the conventional weighting of the two-periodic diamonds. We will be interested in the asymptotic exponential rate of the ratios $\frac{Z_{n,k}(a,b)}{Z_{n-1}(a,b)}$ for $k=rn$ large, with $Z_{n,k}$ defined in the obvious way. From (\ref{ZF}), we readily obtain that the exponential rate,
\bea
\lim_{n \to \infty} \frac 1n \log\frac{Z_{n,rn}(a,b)}{Z_{n-1}(a,b)} \egal \lim_{n \to \infty} \frac 1n \log\Big[(ab)^n\, \frac{T_{n,rn}(\textstyle \frac1a,\frac1b)}{T_{n-1}(\frac1a,\frac1b)}\Big] = \log(ab) + \lim_{n \to \infty} \frac 1n \log S_{n,rn}(\textstyle\frac1\beta)\nonumber\\
\egal \log(ab) + F_1(r),
\label{lattice}
\eea
is the same for both distributions, up to an additive constant. Indeed, because the denominator of the generating function $G(u,v;\beta)$ is symmetric under $\beta \leftrightarrow \frac1\beta$, the asymptotics of $S_{n,rn}(\beta)$ and $S_{n,rn}(\frac1\beta)$ are identical. Thus up to the term $\log (ab)$, $F_1(r)$ appears as a one-refined relative free energy, in the thermodynamic limit. Later on, we will similarly define a two-refined relative free energy $F_2(r,s)$. These two functions depend on $a,b$ through the parameter $\beta$, which we do not display explicitly to keep the notation as simple as possible.

Let us also look at the probability $\P_{n,k}(a,b)$ that a two-periodic diamond tiling has $k$ vertical dominos along the NW boundary, equivalently that the uppermost path stays next to the NW boundary over a distance $k=rn$, for $n$ large. It is given by
\bea
\P_{n,k}(a,b) \egal \frac{Z_{n,k}(a,b)}{Z_n(a,b)} = \frac{T_{n,k}(\frac1a,\frac1b)}{T_n(\frac1a,\frac1b)} = a \, S_{n,rn}({\textstyle\frac1\beta}) \, \frac{T_{n-1}(\frac1a,\frac1b)}{T_n(\frac1a,\frac1b)} \simeq \exp\Big\{n\Big[F_1(r) - \frac12 \log\frac{2(1+\beta)}{\sqrt{\beta}}\Big]\Big\},\nonumber\\
\eea
where we have used (\ref{Fn}) to find the asymptotic value of $\frac{T_{n-1}(a,b)}{T_n(a,b)}$. From the remark made above, the exponential rate, inside the square brackets, has its maximum at $r=\frac 12$, where it vanishes, so that the rate is negative for $r \neq \frac12$. Thus for $n$ large, the probability $\P_{n,k}(a,b)$ vanishes exponentially for all $k$ not in the scaling  neighbourhood of $k=\frac n2$: in the scaling limit, the uppermost path follows, with probability 1, the NW boundary until the midpoint $r=\frac12$, independently of the values of $a,b$. Moreover, expanding $F_1(r)$ to second order around $r=\frac 12$, one finds that the rescaled random variable $(rn - \frac n2)/\sqrt{n}$ converges to a normal distribution, with zero mean and a variance given by $\frac\beta{(1+\beta)^2}$. %In terms of the particle description of domino tilings of Aztec diamonds, this is the two-periodic extension of the distribution for the single particle on the first line; the question remains open as to whether the full GUE minor process found in the standard Aztec diamonds \cite{JN06,FF11} essentially remains in the two-periodic case.

%%%%%%%%%%%%%%%%%%%%%%%%%%%%%%%%%%%%%%%%%%%%%%%%%%%%%%%%%%%%%%%%%%%%%%%%%%

\vskip 0.5truecm
\noindent
{\bf 4.2 Path generating functions and asymptotics}
\addcontentsline{toc}{subsubsection}{4.2 Path generating functions and asymptotics}

\medskip
\noindent
We now turn to the path description of the Aztec diamond tilings, the other main ingredient for the tangent method. Being simpler for this description, we use the conventional measure (\ref{conv}).  

\begin{figure}[t]
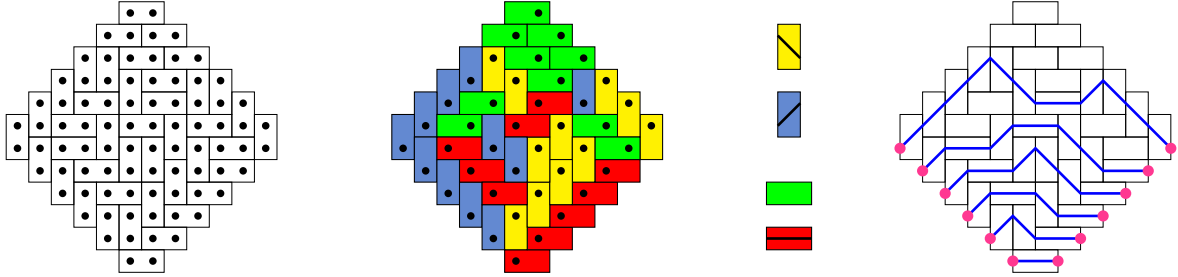

\begin{center}
\psset{unit=.3cm}
\pspicture(0,0)(10,12)
\rput(5,10){\hdimerw}
\rput(4,9){\hdimerw}
\rput(6,9){\hdimerw}
\rput(4,7){\vdimerw}
\rput(5,8){\hdimerw}
\rput(7,8){\hdimerw}
\rput(2,6){\vdimerw}
\rput(1,5){\vdimerw}
\rput(0,4){\vdimerw}
\rput(1,3){\vdimerw}
\rput(2,2){\vdimerw}
\rput(3,1){\vdimerw}
\rput(4,0){\vdimerw}
\rput(5,-1){\hdimerw}
\rput(6,0){\hdimerw}
\rput(7,1){\hdimerw}
\rput(8,2){\hdimerw}
\rput(9,3){\hdimerw}
\rput(11,4){\vdimerw}
\rput(10,5){\vdimerw}
\rput(9,6){\vdimerw}
\rput(8,6){\vdimerw}
\rput(6,7){\hdimerw}
\rput(5,6){\vdimerw}
\rput(3,7){\vdimerw}
\rput(8,5){\hdimerw}
\rput(9,4){\hdimerw}
\rput(7,4){\vdimerw}
\rput(8,3){\vdimerw}
\rput(6,6){\hdimerw}
\rput(5,0){\vdimerw}
\rput(6,1){\vdimerw}
\rput(7,2){\vdimerw}
\rput(6,3){\vdimerw}
\rput(5,5){\hdimerw}
\rput(2,5){\hdimerw}
\rput(2,4){\hdimerw}
\rput(3,6){\hdimerw}
\rput(4,4){\vdimerw}
\rput(3,3){\hdimerw}
\rput(4,2){\hdimerw}
\rput(5,3){\vdimerw}
\multido{\nt=4+1}{2}{\rput(0,\nt){\pscircle[linecolor=black,fillstyle=solid,fillcolor=black](0.5,0.5){0.12}}}
\multido{\nt=3+1}{4}{\rput(1,\nt){\pscircle[linecolor=black,fillstyle=solid,fillcolor=black](0.5,0.5){0.12}}}
\multido{\nt=2+1}{6}{\rput(2,\nt){\pscircle[linecolor=black,fillstyle=solid,fillcolor=black](0.5,0.5){0.12}}}
\multido{\nt=1+1}{8}{\rput(3,\nt){\pscircle[linecolor=black,fillstyle=solid,fillcolor=black](0.5,0.5){0.12}}}
\multido{\nt=0+1}{10}{\rput(4,\nt){\pscircle[linecolor=black,fillstyle=solid,fillcolor=black](0.5,0.5){0.12}}}
\multido{\nt=-1+1}{12}{\rput(5,\nt){\pscircle[linecolor=black,fillstyle=solid,fillcolor=black](0.5,0.5){0.12}}}
\multido{\nt=-1+1}{12}{\rput(6,\nt){\pscircle[linecolor=black,fillstyle=solid,fillcolor=black](0.5,0.5){0.12}}}
\multido{\nt=0+1}{10}{\rput(7,\nt){\pscircle[linecolor=black,fillstyle=solid,fillcolor=black](0.5,0.5){0.12}}}
\multido{\nt=1+1}{8}{\rput(8,\nt){\pscircle[linecolor=black,fillstyle=solid,fillcolor=black](0.5,0.5){0.12}}}
\multido{\nt=2+1}{6}{\rput(9,\nt){\pscircle[linecolor=black,fillstyle=solid,fillcolor=black](0.5,0.5){0.12}}}
\multido{\nt=3+1}{4}{\rput(10,\nt){\pscircle[linecolor=black,fillstyle=solid,fillcolor=black](0.5,0.5){0.12}}}
\multido{\nt=4+1}{2}{\rput(11,\nt){\pscircle[linecolor=black,fillstyle=solid,fillcolor=black](0.5,0.5){0.12}}}
\endpspicture
%%%%%%%%%
\hspace{2cm}
\pspicture(0,0)(10,10)
\rput(5,10){\hdimerg}
\rput(4,9){\hdimerg}
\rput(6,9){\hdimerg}
\rput(4,7){\vdimery}
\rput(5,8){\hdimerg}
\rput(7,8){\hdimerg}
\rput(2,6){\vdimerb}
\rput(1,5){\vdimerb}
\rput(0,4){\vdimerb}
\rput(1,3){\vdimerb}
\rput(2,2){\vdimerb}
\rput(3,1){\vdimerb}
\rput(4,0){\vdimerb}
\rput(5,-1){\hdimerr}
\rput(6,0){\hdimerr}
\rput(7,1){\hdimerr}
\rput(8,2){\hdimerr}
\rput(9,3){\hdimerr}
\rput(11,4){\vdimery}
\rput(10,5){\vdimery}
\rput(9,6){\vdimery}
\rput(8,6){\vdimerb}
\rput(6,7){\hdimerg}
\rput(5,6){\vdimery}
\rput(3,7){\vdimerb}
\rput(8,5){\hdimerg}
\rput(9,4){\hdimerg}
\rput(7,4){\vdimery}
\rput(8,3){\vdimery}
\rput(6,6){\hdimerr}
\rput(5,0){\vdimery}
\rput(6,1){\vdimery}
\rput(7,2){\vdimery}
\rput(6,3){\vdimery}
\rput(5,5){\hdimerr}
\rput(2,5){\hdimerg}
\rput(2,4){\hdimerr}
\rput(3,6){\hdimerg}
\rput(4,4){\vdimerb}
\rput(3,3){\hdimerr}
\rput(4,2){\hdimerr}
\rput(5,3){\vdimerb}
\multido{\nt=4+2}{1}{\rput(0,\nt){\pscircle[linecolor=black,fillstyle=solid,fillcolor=black](0.5,0.5){0.12}}}
\multido{\nt=3+2}{2}{\rput(1,\nt){\pscircle[linecolor=black,fillstyle=solid,fillcolor=black](0.5,0.5){0.12}}}
\multido{\nt=2+2}{3}{\rput(2,\nt){\pscircle[linecolor=black,fillstyle=solid,fillcolor=black](0.5,0.5){0.12}}}
\multido{\nt=1+2}{4}{\rput(3,\nt){\pscircle[linecolor=black,fillstyle=solid,fillcolor=black](0.5,0.5){0.12}}}
\multido{\nt=0+2}{5}{\rput(4,\nt){\pscircle[linecolor=black,fillstyle=solid,fillcolor=black](0.5,0.5){0.12}}}
\multido{\nt=-1+2}{6}{\rput(5,\nt){\pscircle[linecolor=black,fillstyle=solid,fillcolor=black](0.5,0.5){0.12}}}
\multido{\nt=0+2}{6}{\rput(6,\nt){\pscircle[linecolor=black,fillstyle=solid,fillcolor=black](0.5,0.5){0.12}}}
\multido{\nt=1+2}{5}{\rput(7,\nt){\pscircle[linecolor=black,fillstyle=solid,fillcolor=black](0.5,0.5){0.12}}}
\multido{\nt=2+2}{4}{\rput(8,\nt){\pscircle[linecolor=black,fillstyle=solid,fillcolor=black](0.5,0.5){0.12}}}
\multido{\nt=3+2}{3}{\rput(9,\nt){\pscircle[linecolor=black,fillstyle=solid,fillcolor=black](0.5,0.5){0.12}}}
\multido{\nt=4+2}{2}{\rput(10,\nt){\pscircle[linecolor=black,fillstyle=solid,fillcolor=black](0.5,0.5){0.12}}}
\multido{\nt=5+2}{1}{\rput(11,\nt){\pscircle[linecolor=black,fillstyle=solid,fillcolor=black](0.5,0.5){0.12}}}
\endpspicture
%%%%%%%%
\hspace{2cm}
\pspicture(0,0)(0,10)
\rput(-0.5,0){\hdimerr \psline[linewidth=1pt](0,.5)(2,0.5)}
\rput(-0.5,2){\hdimerg}
\rput(0,5){\vdimerb \psline[linewidth=1pt](0,.5)(1,1.5)}
\rput(0,8){\vdimery \psline[linewidth=1pt](0,1.5)(1,0.5)}
\endpspicture
%%%%%%%%
\hspace{1.5cm}
\pspicture(0,0)(10,10)
\rput(5,10){\hdimerw}
\rput(4,9){\hdimerw}
\rput(6,9){\hdimerw}
\rput(4,7){\vdimerw}
\rput(5,8){\hdimerw}
\rput(7,8){\hdimerw}
\rput(2,6){\vdimerw}
\rput(1,5){\vdimerw}
\rput(0,4){\vdimerw}
\rput(1,3){\vdimerw}
\rput(2,2){\vdimerw}
\rput(3,1){\vdimerw}
\rput(4,0){\vdimerw}
\rput(5,-1){\hdimerw}
\rput(6,0){\hdimerw}
\rput(7,1){\hdimerw}
\rput(8,2){\hdimerw}
\rput(9,3){\hdimerw}
\rput(11,4){\vdimerw}
\rput(10,5){\vdimerw}
\rput(9,6){\vdimerw}
\rput(8,6){\vdimerw}
\rput(6,7){\hdimerw}
\rput(5,6){\vdimerw}
\rput(3,7){\vdimerw}
\rput(8,5){\hdimerw}
\rput(9,4){\hdimerw}
\rput(7,4){\vdimerw}
\rput(8,3){\vdimerw}
\rput(6,6){\hdimerw}
\rput(5,0){\vdimerw}
\rput(6,1){\vdimerw}
\rput(7,2){\vdimerw}
\rput(6,3){\vdimerw}
\rput(5,5){\hdimerw}
\rput(2,5){\hdimerw}
\rput(2,4){\hdimerw}
\rput(3,6){\hdimerw}
\rput(4,4){\vdimerw}
\rput(3,3){\hdimerw}
\rput(4,2){\hdimerw}
\rput(5,3){\vdimerw}
\psline[linewidth=1.0pt,linecolor=blue](5,-0.5)(7,-0.5)
\psline[linewidth=1.0pt,linecolor=blue](4,0.5)(5,1.5)(6,0.5)(8,0.5)
\psline[linewidth=1.0pt,linecolor=blue](3,1.5)(4,2.5)(6,2.5)(7,1.5)(9,1.5)
\psline[linewidth=1.0pt,linecolor=blue](2,2.5)(3,3.5)(5,3.5)(6,4.5)(8,2.5)(10,2.5)
\psline[linewidth=1.0pt,linecolor=blue](1,3.5)(2,4.5)(4,4.5)(5,5.5)(7,5.5)(9,3.5)(11,3.5)
\psline[linewidth=1.0pt,linecolor=blue](0,4.5)(4,8.5)(5,7.5)(6,6.5)(8,6.5)(9,7.5)(12,4.5)
\pscircle[linecolor=mypink,fillstyle=solid,fillcolor=mypink](5,-0.5){0.2}
\pscircle[linecolor=mypink,fillstyle=solid,fillcolor=mypink](4,0.5){0.2}
\pscircle[linecolor=mypink,fillstyle=solid,fillcolor=mypink](3,1.5){0.2}
\pscircle[linecolor=mypink,fillstyle=solid,fillcolor=mypink](2,2.5){0.2}
\pscircle[linecolor=mypink,fillstyle=solid,fillcolor=mypink](1,3.5){0.2}
\pscircle[linecolor=mypink,fillstyle=solid,fillcolor=mypink](0,4.5){0.2}
\pscircle[linecolor=mypink,fillstyle=solid,fillcolor=mypink](7,-0.5){0.2}
\pscircle[linecolor=mypink,fillstyle=solid,fillcolor=mypink](8,0.5){0.2}
\pscircle[linecolor=mypink,fillstyle=solid,fillcolor=mypink](9,1.5){0.2}
\pscircle[linecolor=mypink,fillstyle=solid,fillcolor=mypink](10,2.5){0.2}
\pscircle[linecolor=mypink,fillstyle=solid,fillcolor=mypink](11,3.5){0.2}
\pscircle[linecolor=mypink,fillstyle=solid,fillcolor=mypink](12,4.5){0.2}
\endpspicture
\end{center}
\caption{Illustration of the bijection between Aztec tilings by dominos and non-intersecting paths.}
\label{fig2bis}
\end{figure}

The path description of the Aztec diamond tiling is well-known \cite{Jo02}, and is illustrated in Figure \ref{fig2bis}. The left figure shows a domino tiling of order 6, along with the vertices of the dual graph. In the central figure, half of the vertices are retained, so that each domino covers exactly one such vertex. The dominos are then colored according to which half of the domino contains the vertex: left (red), right (green), bottom (blue) or top (yellow). Finally a path segment is placed on the red, blue and yellow dominos as shown, whereas the green dominos do not carry any segment. As one can see on the right figure, all segments connect and form $n$ non-intersecting paths, each one starting and ending at the same height in the lower half of the diamond.

Each of the $n$ paths is Schr\"oder-like, made of the three elementary steps $(1,1), (1,-1)$ and (2,0), and never going below the baseline joining the starting and ending points. The full set of $2^{n(n+1)/2}$ domino tilings of the Aztec diamond of order $n$ is in one-to-one correspondence with the set of non-intersecting Schr\"oder paths contained in the diamond, starting and ending in the way described.

For the two-periodic measure, the paths must be appropriately weighted so that each set of $n$ paths and its corresponding tiling have the same weight. In turn, appropriate weights can be assigned individually to the three elementary steps but will depend on where the steps are taken. Let us first discuss the case of a single path.

\begin{figure}[t]
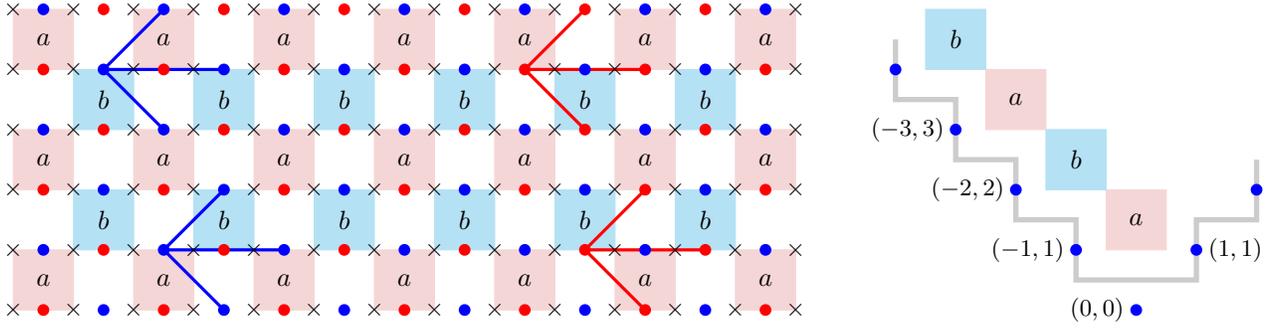

\begin{center}
\psset{unit=0.8cm}
\pspicture(0,0)(10,5)
\rput(-0.5,0){\multido{\nx=0+2}{7}{\rput(\nx,0){\pspolygon[linewidth=0pt,linecolor=lightred,fillstyle=solid,fillcolor=lightred](0,0)(0,1)(1,1)(1,0)
\rput(0.5,0.5){\small $a$}}}}
\rput(0.5,1){\multido{\nx=0+2}{6}{\rput(\nx,0){\pspolygon[linewidth=0pt,linecolor=lightblue,fillstyle=solid,fillcolor=lightblue](0,0)(0,1)(1,1)(1,0)
\rput(0.5,0.5){\small $b$}}}}
\rput(-0.5,2){\multido{\nx=0+2}{7}{\rput(\nx,0){\pspolygon[linewidth=0pt,linecolor=lightred,fillstyle=solid,fillcolor=lightred](0,0)(0,1)(1,1)(1,0)
\rput(0.5,0.5){\small $a$}}}}
\rput(0.5,3){\multido{\nx=0+2}{6}{\rput(\nx,0){\pspolygon[linewidth=0pt,linecolor=lightblue,fillstyle=solid,fillcolor=lightblue](0,0)(0,1)(1,1)(1,0)
\rput(0.5,0.5){\small $b$}}}}
\rput(-0.5,4){\multido{\nx=0+2}{7}{\rput(\nx,0){\pspolygon[linewidth=0pt,linecolor=lightred,fillstyle=solid,fillcolor=lightred](0,0)(0,1)(1,1)(1,0)
\rput(0.5,0.5){\small $a$}}}}
\rput(2,1){\psline[linewidth=1.2pt,linecolor=blue](0,0)(1,1)
\psline[linewidth=1.2pt,linecolor=blue](0,0)(2,0)
\psline[linewidth=1.2pt,linecolor=blue](0,0)(1,-1)}
\rput(1,4){\psline[linewidth=1.2pt,linecolor=blue](0,0)(1,1)
\psline[linewidth=1.2pt,linecolor=blue](0,0)(2,0)
\psline[linewidth=1.2pt,linecolor=blue](0,0)(1,-1)}
\rput(9,1){\psline[linewidth=1.2pt,linecolor=red](0,0)(1,1)
\psline[linewidth=1.2pt,linecolor=red](0,0)(2,0)
\psline[linewidth=1.2pt,linecolor=red](0,0)(1,-1)}
\rput(8,4){\psline[linewidth=1.2pt,linecolor=red](0,0)(1,1)
\psline[linewidth=1.2pt,linecolor=red](0,0)(2,0)
\psline[linewidth=1.2pt,linecolor=red](0,0)(1,-1)}
%
%\multido{\nx=0+2}{5}{\rput(\nx,0){\pscircle[linecolor=black,fillstyle=solid,fillcolor=black](0,0){0.04}}}
%\multido{\nx=1+2}{4}{\rput(\nx,1){\pscircle[linecolor=black,fillstyle=solid,fillcolor=black](0,0){0.04}}}
%\multido{\nx=0+2}{5}{\rput(\nx,2){\pscircle[linecolor=black,fillstyle=solid,fillcolor=black](0,0){0.04}}}
%\multido{\nx=1+2}{4}{\rput(\nx,3){\pscircle[linecolor=black,fillstyle=solid,fillcolor=black](0,0){0.04}}}
%\multido{\nx=0+2}{5}{\rput(\nx,4){\pscircle[linecolor=black,fillstyle=solid,fillcolor=black](0,0){0.04}}}
%
\multido{\nx=1+2}{6}{\rput(\nx,0){\pscircle[linecolor=blue,fillstyle=solid,fillcolor=blue](0,0){0.08}}}
\multido{\nx=0+2}{7}{\rput(\nx,1){\pscircle[linecolor=blue,fillstyle=solid,fillcolor=blue](0,0){0.08}}}
\multido{\nx=1+2}{6}{\rput(\nx,2){\pscircle[linecolor=blue,fillstyle=solid,fillcolor=blue](0,0){0.08}}}
\multido{\nx=0+2}{7}{\rput(\nx,3){\pscircle[linecolor=blue,fillstyle=solid,fillcolor=blue](0,0){0.08}}}
\multido{\nx=1+2}{6}{\rput(\nx,4){\pscircle[linecolor=blue,fillstyle=solid,fillcolor=blue](0,0){0.08}}}
\multido{\nx=0+2}{7}{\rput(\nx,5){\pscircle[linecolor=blue,fillstyle=solid,fillcolor=blue](0,0){0.08}}}
\multido{\nx=0+2}{7}{\rput(\nx,0){\pscircle[linecolor=red,fillstyle=solid,fillcolor=red](0,0){0.08}}}
\multido{\nx=1+2}{6}{\rput(\nx,1){\pscircle[linecolor=red,fillstyle=solid,fillcolor=red](0,0){0.08}}}
\multido{\nx=0+2}{7}{\rput(\nx,2){\pscircle[linecolor=red,fillstyle=solid,fillcolor=red](0,0){0.08}}}
\multido{\nx=1+2}{6}{\rput(\nx,3){\pscircle[linecolor=red,fillstyle=solid,fillcolor=red](0,0){0.08}}}
\multido{\nx=0+2}{7}{\rput(\nx,4){\pscircle[linecolor=red,fillstyle=solid,fillcolor=red](0,0){0.08}}}
\multido{\nx=1+2}{6}{\rput(\nx,5){\pscircle[linecolor=red,fillstyle=solid,fillcolor=red](0,0){0.08}}}
\rput(-0.5,0){\multido{\nx=0+1}{14}{\rput(\nx,0){\footnotesize $\times$}}}
\rput(-0.5,1){\multido{\nx=0+1}{14}{\rput(\nx,0){\footnotesize $\times$}}}
\rput(-0.5,2){\multido{\nx=0+1}{14}{\rput(\nx,0){\footnotesize $\times$}}}
\rput(-0.5,3){\multido{\nx=0+1}{14}{\rput(\nx,0){\footnotesize $\times$}}}
\rput(-0.5,4){\multido{\nx=0+1}{14}{\rput(\nx,0){\footnotesize $\times$}}}
\rput(-0.5,5){\multido{\nx=0+1}{14}{\rput(\nx,0){\footnotesize $\times$}}}
%
%\pscircle[linecolor=mypink,fillstyle=solid,fillcolor=mypink](1,2){0.1}
%
\endpspicture
\hspace{4cm}
%\vspace{3mm}
\pspicture(0,0)(5,5)
\psline[linewidth=2pt,linecolor=mygray](5,2.5)(5,1.5)(4,1.5)(4,0.5)(3,0.5)(2,0.5)(2,1.5)(1,1.5)(1,2.5)(0,2.5)(0,3.5)(-1,3.5)(-1,4.5)
\rput(2.5,1){\pspolygon[linewidth=0pt,linecolor=lightred,fillstyle=solid,fillcolor=lightred](0,0)(0,1)(1,1)(1,0)}
\rput(1.5,2){\pspolygon[linewidth=0pt,linecolor=lightblue,fillstyle=solid,fillcolor=lightblue](0,0)(0,1)(1,1)(1,0)}
\rput(0.5,3){\pspolygon[linewidth=0pt,linecolor=lightred,fillstyle=solid,fillcolor=lightred](0,0)(0,1)(1,1)(1,0)}
\rput(-0.5,4){\pspolygon[linewidth=0pt,linecolor=lightblue,fillstyle=solid,fillcolor=lightblue](0,0)(0,1)(1,1)(1,0)}
\pscircle[linecolor=blue,fillstyle=solid,fillcolor=blue](3,0){0.08}
\pscircle[linecolor=blue,fillstyle=solid,fillcolor=blue](2,1){0.08}
\pscircle[linecolor=blue,fillstyle=solid,fillcolor=blue](1,2){0.08}
\pscircle[linecolor=blue,fillstyle=solid,fillcolor=blue](0,3){0.08}
\pscircle[linecolor=blue,fillstyle=solid,fillcolor=blue](-1,4){0.08}
\pscircle[linecolor=blue,fillstyle=solid,fillcolor=blue](4,1){0.08}
\pscircle[linecolor=blue,fillstyle=solid,fillcolor=blue](5,2){0.08}
\rput(4.65,1){\footnotesize $(1,1)$}
\rput(2.35,0){\footnotesize $(0,0)$}
\rput(1.2,1){\footnotesize $(-1,1)$}
\rput(0.2,2){\footnotesize $(-2,2)$}
\rput(-0.8,3){\footnotesize $(-3,3)$}
\rput(3,1.5){\small $a$}
\rput(1,3.5){\small $a$}
\rput(2,2.5){\small $b$}
\rput(0,4.5){\small $b$}
\endpspicture
\end{center}
\caption{The left figure shows part of the dual graph (with vertices marked by crosses and faces colored according to their weight) along with part of the lattice $\Z^2$ on which the random walk takes place (blue and red dots). The blue resp. red dots belong to the even resp. odd sublattice. Depending on whether the starting point of the walk is even or odd, the walk stays on the blue or on the red dots. The three elementary steps are shown in the four different cases, and help to figure out what their weight is. The right figure shows the location of the starting points of the paths in the Aztec diamond, assumed to be even, relative to the faces.}
\label{fig3}
\end{figure}

Figure \ref{fig3} depicts the geometric setting. As indicated in the right figure, the starting points are assumed to be even, so that we may focus on the blue steps in the left picture. We then see that a diagonal step upwards gets a weight $a$ if it enters a red face, and a weight $b$ if it enters a blue face. Likewise, a diagonal step downwards gets a weight $a$ if it leaves a red face, and a weight $b$ if it leaves a blue face. As to the horizontal steps, carried by the red dominos, the figure suggests that they get weight $a$ if they go from a blue face to a blue face, weight $b$ if they connect two red faces. 

These are indeed the correct weights for elementary steps carried by dominos, but the green dominos do not carry any and yet contribute to the weight of a tiling. To resolve this problem, one may observe that there are as many red dominos as green ones, but more is actually true, namely the numbers of $a$--weighted red dominos and of $a$--weighted green dominos are equal, and the same for the $b$--weighted ones. This simply follows from the two-periodicity of the weights and from the way flips\footnote{A flip acts by rotation of a pair of parallel dominos (fully contained in a square of side 2), transforming two vertical dominos into two horizontal dominos, or vice-versa. Any domino tiling can be obtained from any other by a sequence of flips \cite{EKLP92a}.}  act on domino tilings. Thus the weights of the green dominos can simply be transfered to the red ones by squaring the weight that the latter would normally have. It follows that a horizontal step (attached to a red domino) gets a weight $a^2$ or $b^2$, as determined above.

In order to write the weights, we assume that among the even sites, the odd-odd ones are located on the top boundary of a blue face and the even-even ones on the top boundary of a red face, in agreement with the parity of the coordinates, $(-i,i)$ for $i \ge 0$, of the starting points of the paths in the Aztec diamond. We then find the following weights for the three steps, as functions of the site where the step is taken,
\begin{subequations}
\vspace{-3mm}
\bea
&& w[(i,j) \rightarrow (i,j)+(1,1)] = \begin{cases}
b & {\rm if\ } i {\rm\ is\ even,}\\
a & {\rm if\ } i {\rm\ is\ odd,}
\end{cases} \\
&& w[(i,j) \rightarrow (i,j)+(2,0)] = \begin{cases}
b^2 & {\rm if\ } i {\rm\ is\ even,}\\
a^2 & {\rm if\ } i {\rm\ is\ odd,}
\end{cases}\\
&& w[(i,j) \rightarrow (i,j)+(1,-1)] = \begin{cases}
a & {\rm if\ } i {\rm\ is\ even,}\\
b & {\rm if\ } i {\rm\ is\ odd.}
\end{cases}
\eea
\label{pathw}
\end{subequations}
%If the starting point is odd-odd, we simply interchange $a$ and $b$ in the expressions (\ref{pathw}).

Let us denote by $Z_{i,j}(a,b)$ the partition function for a single path going from an even-even site $P$ to $P+(i,j)$. It is the weighted sum over all such paths, the weight of each path being the product of the weights of its elementary steps, as given in (\ref{pathw}). The sum  $i+j$ is necessarily even, so we define $Z_{i,j}^{\rm e}(a,b)$ and $Z_{i,j}^{\rm o}(a,b)$ according to whether $i,j$ are both even or both odd. From (\ref{pathw}), we obtain the following recurrence relations, for $i \ge 0$ and $j \in \Z$,
\begin{subequations}
\bea
Z_{i,j}^{\rm e}(a,b) \egal \delta_{i,0}\, \delta_{j,0} + b^2 \, Z_{i-2,j}^{\rm e}(a,b) + b \, Z_{i-1,j+1}^{\rm o}(a,b) + a \, Z_{i-1,j-1}^{\rm o}(a,b),\\
\noalign{\smallskip}
Z_{i,j}^{\rm o}(a,b) \egal a^2 \, Z_{i-2,j}^{\rm o}(a,b) + a \, Z_{i-1,j+1}^{\rm e}(a,b) + b \, Z_{i-1,j-1}^{\rm e}(a,b).
\eea
\end{subequations}
The associated generating functions $G^{\rm e,o}(x,y) = \sum_{i,j} Z^{\rm e,o}_{i,j}(a,b) \,x^i \, y^j
$ satisfy
\begin{subequations}
\bea
G^{\rm e}(x,y) \egal 1 + b^2  x^2 \, G^{\rm e}(x,y) + x \, \Big(ay + \frac by\Big) \, G^{\rm o}(x,y),\\
G^{\rm o}(x,y) \egal a^2  x^2 \, G^{\rm o}(x,y) + x \, \Big(by + \frac ay\Big) \, G^{\rm e}(x,y).
\eea
\end{subequations}
This linear system is easily solved and yields the full generating function $G(x,y) = G^{\rm e}(x,y) + G^{\rm o}(x,y)$,
\be
G(x,y) = \frac{1-a^2x^2+x(by + a/y)}{(1-a^2x^2)(1-b^2x^2) - x^2(ay+b/y)(by+a/y)}. \qquad \text{(even-even starting point)}
\ee
If the starting point $P$ is odd-odd, the weights $a$ and $b$ get simply interchanged.

The generating function can be used to compute the finite $n$ partition functions of the two-periodic Aztec diamonds, by using the Lindstr\"om-Gessel-Viennot lemma \cite{Li73,GV85}. The only thing one has to remember is that, as indicated above, the starting points of the paths alternate between even-even and odd-odd, implying that the weights $a$ and $b$ must be accordingly interchanged. We can then write
\be
Z_n(a,b) = \det\big(A_{i,j}\big)_{0 \le i,j \le n}, \qquad {\rm with\ } A_{i,j} = \begin{cases}
\;Z_{i+j,j-i}(a,b) & {\rm for\ } i {\rm \ even},\\
\;Z_{i+j,j-i}(b,a) & {\rm for\ } i {\rm \ odd}.
\end{cases}
\ee
For $n=3$, we obtain
\bea
Z_3(a,b) \egal \small \det \begin{bmatrix} 
1 & b & ab & ab^2 \\
b & 2a^2 + b^2 & 2a^3+3ab^2 & 4a^3b + 3ab^3 \\
ab & 2a^3+3ab^2 & 2a^4 + 7a^2b^2 + 4b^4 & 8a^4b + 13a^2b^3 + 4b^5 \\
ab^2 & 4a^3b + 3ab^3 & 8a^4b + 13a^2b^3 + 4b^5 & 8a^6 + 28a^4b^2 + 23a^2b^4 + 4b^6
\end{bmatrix} \nonumber\\
\noalign{\medskip}
\egal 16a^4b^4\,(a^2+b^2)^2,
\eea
and recover the result found in (\ref{Zn}).

More importantly the generating function may be used to evaluate the asymptotics of the coefficients $Z_{i,j}(a,b)$ for large $i,j$ of order $n$. The computation is similar to the one in the previous section. For large $n$, and setting $i=rn$ and $j=sn$, one has
\be
Z_{rn,sn}(a,b) \simeq {\rm e}^{nrL(t)}, \qquad {\rm for\ }t = \textstyle\frac sr, \quad L(t) = -\log x(t) - t \log y(t), \qquad -1 \le t \le 1,
\label{L}
\ee
where $x(t)$ and $y(t)$ are the positive solutions of the algebraic system,
\be
\begin{cases}
Q(x,y) = (1-a^2x^2)(1-b^2x^2) - x^2(ay+b/y)(by+a/y) = 0, & \\
t\, x \, \partial_x Q(x,y) = y\,\partial_y Q(x,y). &
\end{cases}
\label{Qsystem}
\ee
Similarly to the function $F_1(r)$ of the previous section, $L(t)$ satisfies $L'(t) = -\log y(t)$.

We will see in the next section that the tangent method only requires to know the product $x(t)y(t)$. So, instead of looking for $x$ and $y$ individually, we solve for $\frac xy$ and $xy$. Together the two equations above yield a quadratic equation for $\frac xy$, with the following positive solution,
\be
\frac xy = \sqrt{\frac{t/ab - (xy)^2}{ab \, t (xy)^2-1}} \equiv \frac 1{\sqrt{ab}} \; \sqrt{\frac{t - p^2}{tp^2-1}},
\label{fracxy}
\ee
where we have set $p \equiv p(t) = \sqrt{ab} \, x(t) y(t)$. Inserting $\frac xy$ in the first equation allows to compute $t$ as a function of $p$,
\be
t(p) = \frac 1{(1-p^2)^4 - 4\alpha^2p^4}\:\Bigg\{2\alpha\, p(1-p^4) \sqrt{p^4+(\alpha^2-2)p^2+1} - 2\alpha^2 p^2(1+p^4) - (1-p^2)^4\Bigg\},
\label{tp}
\ee
with $\alpha=\frac{a^2+b^2}{ab}=\frac{1+\beta}{\sqrt{\beta}}$. Again, we do not invert this relation to get $p$, equivalently $xy$, as a function of $t$; the result would be fairly complicated and is in any case not useful.

To finish, we give the values of the function $L$ at $t=\pm1$, as they will be used in the next section. From $t(p=1)=1$, we obtain $x(1)y(1)=\frac1{\sqrt{ab}}$ and thus $L(1) = -\log\big[x(1)y(1)\big] = \log \sqrt{ab}$. Likewise since $t(p=+\infty)=-1$, we have $p(t=-1)=+\infty$ and therefore $\frac{x(-1)}{y(-1)}=\frac1{\sqrt{ab}}$ by using (\ref{fracxy}). It yields $L(-1) = -\log\frac{x(-1)}{y(-1)} = \log \sqrt{ab}$, equal to $L(1)$ as expected.

%%%%%%%%%%%%%%%%%%%%%%%%%%%%%%%%%%%%%%%%%%%%%%%%%%%%%%%%%%%%%%%%%%%%%%%%%%

\vskip 0.5truecm
\noindent
{\bf 4.3 Geometric tangent method}
\addcontentsline{toc}{subsubsection}{4.3 Geometric tangent method}

\medskip
\noindent
The tangent method, as originally proposed in \cite{CS16}, required an extension of the domain of interest and a saddle point analysis to determine the likeliest entry point in the domain. It has been slightly reformulated in \cite{DR21} in a way that does not require an extension of the domain but instead uses directly the one-point boundary function. We briefly review it. 

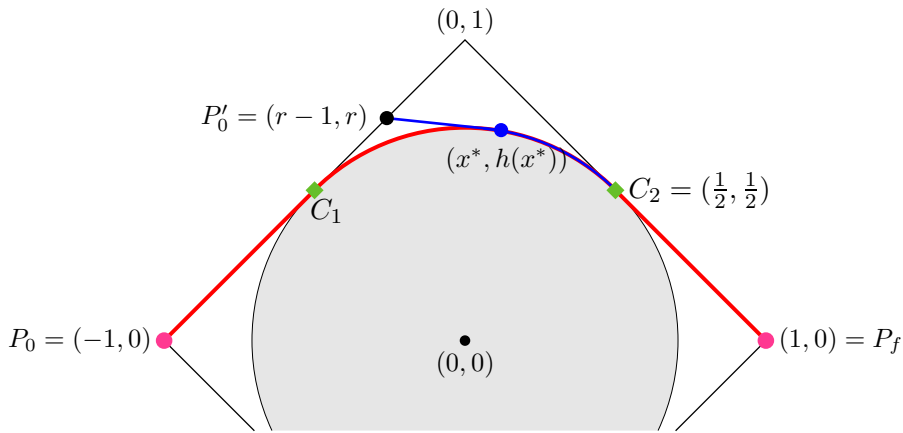
\begin{figure}[h]
\begin{center}
\psset{unit=.8cm}
%\setlength{\unitlength}{.7cm}
%
%\hspace{-1.5cm}
\begin{pspicture}(-5,-1)(5,5.7)
\psarc[linewidth=0.1pt,linecolor=black,fillstyle=solid,fillcolor=mygrey](0,0){3.54}{-25}{205}
\pscircle[linecolor=black,fillstyle=solid,fillcolor=black](0,0){0.07}
\rput(0,-0.4){\small $(0,0)$}
\psline[linewidth=0.5pt,linecolor=black](-3.5,-1.5)(-5,0)(0,5)(5,0)(3.5,-1.5)
\psline[linewidth=1.5pt,linecolor=red](-5,0)(-2.5,2.5)
%\psline[linewidth=2pt,linecolor=blue](2.5,2.5)(5,0)
\psline[linewidth=1.5pt,linecolor=red](2.5,2.5)(5,0)
\psarc[linewidth=1.5pt,linecolor=red](0,0){3.54}{45}{135}
\psarc[linewidth=1pt,linecolor=blue](0,0){3.54}{45}{85}
%\psellipticarc[linewidth=1.5pt,linecolor=red](0,0)(4.11,2.93){26}{152}
%\psellipticarc[linewidth=1pt,linestyle=dashed,dash=3pt 2pt,linecolor=gray](0,0)(4.11,2.93){-20}{26}
%\psellipticarc[linewidth=1pt,linestyle=dashed,dash=3pt 2pt,linecolor=gray](0,0)(4.11,2.93){152}{200}
%\psellipse[linewidth=1pt,linecolor=red](0,0)(4.11,2.93)
%\psellipticarc[linewidth=1pt,linecolor=blue](0,0)(4.09,2.93){26}{80}
\pscircle[linecolor=mypink,fillstyle=solid,fillcolor=mypink](-5,0){0.12}
\pscircle[linecolor=mypink,fillstyle=solid,fillcolor=mypink](5,0){0.12}
\psdiamond[linecolor=mygreen,fillstyle=solid,fillcolor=mygreen](-2.5,2.5)(0.15,0.15)
\psdiamond[linecolor=mygreen,fillstyle=solid,fillcolor=mygreen](2.5,2.5)(0.15,0.15)
\rput(-2.3,2.15){$C_1$}
\rput(3.9,2.5){$C_2=(\frac12,\frac12)$}
\rput(-6.4,0){\small $P_0=(-1,0)$}
%\psline[linewidth=0.5pt,linecolor=black](-2.5,3.35)(3,2.5)
\psline[linewidth=1pt,linecolor=blue](-1.3,3.7)(0.6,3.5)
\pscircle[linecolor=black,fillstyle=solid,fillcolor=black](-1.3,3.7){0.10}
\rput(-3.,3.7){\small $P'_0=(r-1,r)$}
\pscircle[linecolor=blue,fillstyle=solid,fillcolor=blue](0.6,3.5){0.10}
\rput(0.7,3){\small $(x^*,h(x^*))$}
\rput(6.25,0){\small $(1,0)=P_f$}
\rput(0,5.3){\small $(0,1)$}
\end{pspicture}
\end{center}
\caption{One-refinement for the Aztec diamond: the uppermost path is forced to leave the NW boundary at $P'_0$. The shaded region represents the non-frozen phase.}
\label{fig4}
\end{figure}

\medskip
Figure \ref{fig4} depicts, after rescaling all distances by $n$, the path description of a generic tiling of the Aztec diamond of order $n$ going to infinity, when the uppermost path is conditioned to leave the NW boundary at the point $P'_0$, parametrized by $r$. In the scaling limit, the uppermost path starts from $P_0$, follows the NW boundary till $P'_0$, then reaches the contact point $C_2$ by following the blue path, and eventually merges with the NE boundary from $C_2$ to $P_f$. The red line depicts the second uppermost path, starting from $P_0$ and ending at $P_f$. In actual facts, a fraction $n^{\varepsilon<1}$ of the uppermost paths will condense on the red curve in the scaling limit; the part of the red curve between the contact points $C_1,C_2$, is a portion of the arctic curve. The assumptions of the tangent method are, in the scaling limit, (1) that the shape of the arctic curve is not affected at all by the fact that the uppermost path has been conditioned to leave the NW boundary at $P_0'$, and (2) that, with probability 1, the blue path leaves $P'_0$ on a straight line which touches the arctic curve tangentially, and from there on, merges with the arctic curve till the contact point $C_2$, and eventually reaches $P_f$ on a straight line. Although random for finite $n$, the uppermost path and the fraction $n^\varepsilon$ below it become deterministic in the limit $n \to \infty$. Though reasonable, the hypothesis (1) is stronger and hard to control in general; it has however been made rigorous in the context of the six-vertex model \cite{Ag20}. Under the hypothesis (1), assumption (2) has been proved in a fairly general framework \cite{DGR19}.

In concrete terms, the tangent method works as follows. The point $P'_0$ is fixed in terms of a parameter $r$. With probability 1, the blue path is a straight line that hits the arctic curve tangentially at the blue point, whose coordinates are $(x^*,h(x^*))$, $h(x)$ being the functional description of the arctic curve. It follows that the equation of the straight line is $Y = t^* (X+1-r) + r$, where the slope $t^*=h'(x^*)$ is seen as a function of $r$. Varying $r$ provides a family of straight lines tangent to the arctic curve, which can then be retrieved as its envelope, 
\be
\begin{cases}
X(r) = \displaystyle \Big(\frac{{\rm d}t^*}{{\rm d}r}\Big)^{-1} (t^*-1) + r - 1, & \\
\noalign{\medskip}
Y(r) = t^* \big(X(r)-r+1\big) + r, &
\end{cases}
\label{recti}
\ee
The main problem is therefore to compute the function $t^*=t^*(r)$; once this is done, the equations (\ref{recti}) provide an explicit parametrization of the arctic curve. Because the contact points are at $y=\frac 12$ for the two-periodic Aztec diamonds (from the discussion at the end of Section 4.1), the parameter $r$ will vary in $[\frac12,1]$. We have $t^*(\frac 12)=1$ and $t^*(1)=-1$.

In its original form \cite{CS16}, the starting point $Q_0$ of the blue path was some (suitably chosen) point outside the domain (see Section 6.1). The point of entry in the domain $P'_0$ was then determined as the one with the highest probability, and becomes deterministic in the scaling limit. The slope $t^*$ was then obtained from $Q_0$ (given) and $P'_0$ (computed from $Q_0$). With the reformulation proposed in \cite{DR21}, the extension of the domain is no longer necessary.

The reformulation is based on a factorization property: since the uppermost path has no effect, in the scaling limit, on what the other paths do, in particular on the arctic curve itself, the partition function of the complete set of $n$ paths should factorize, at dominant order, into that of the smaller system with $n-1$ paths and the partition function $Z_1^{\rm up}$ for the single, uppermost path, constrained not to cross the arctic curve and to stay in the domain. Applied to two-periodic diamonds for which the uppermost path is conditioned to leave the NW boundary at the point $P'_0$, the factorization leads to \cite{DR21}
\be
\lim_{n \to \infty} \frac 1n \, \log\frac{Z_{n,rn}(a,b)}{Z_{n-1}(a,b)} = \lim_{n \to \infty} \frac 1n \, \log Z_1^{\rm up}.
\label{main}
\ee
The left-hand side constitutes the (in general hard) combinatorial input, specific to the lattice model considered; for the two-periodic diamonds, it has been determined in Section 4.1. The right-hand side has been discussed in more generality in \cite{DGR19}.

$Z_1^{\rm up}$ is the partition function for the uppermost path, made of the same weighted elementary steps as all the other paths, with starting and ending points $P_0$ and $P_f$, conditioned to follow the NW boundary till $P_0'$, and confined in the region bordered by the boundary of the domain and the arctic curve. It is the sum of the weights of all paths satisfying these conditions. It has been shown that when the starting and ending points are separated by a large distance, say proportional to a large number $n$, this sum is exponentially dominated by the lattice paths which collapse in the scaling limit onto a unique continuous trajectory $f(x)$, solution of a variational problem and given by the shortest path from the initial to final points and wholy contained in the allowed region. In addition, the partition function is, for large $n$, explicitly given by \cite{DGR19}
\be
Z_1^{\rm up}[f] \simeq \exp\{n S[f]\}, \qquad S[f] = \int_{P_0}^{P_f} {\rm d}x \, L\big(f'(x)\big),
\ee
where $L$ is a computable function that only depends on the nature of the lattice paths (f.i. the set of elementary steps and their weights): in fact $L(t)$ is precisely the function that controls the asymptotics of the partition function of a single path going from the origin to a distant point $(rn,sn)$ with no other constraints, $Z_{rn,sn} \simeq \exp{[nr L(t)]}$, for $t=\frac rs$. For two-periodic Aztec diamonds, that function has been discussed and (implicitly) determined in Section 4.2 in terms of the functions $x(t)$ and $y(t)$.

For the one-refinement shown in Figure \ref{fig4}, $f$ is made of three pieces: a straight portion, from $P_0$ to $P'_0$ with slope $+1$, the blue curve between $P'_0$ and $C_2$, and another straight portion from $C_2$ to $P_f$ with slope $-1$; the whole trajectory is indeed the shortest one between $P_0$ and $P_f$ under the constraint to pass through $P'_0$. The exponential rate $S[f]$ is then equal to
\bea
S[f] \egal \int_{-1}^{r-1} {\rm d}x \, L(1) + \int_{r-1}^{x^*} {\rm d}x \, L(t^*) + \int_{x^*}^{\frac12} {\rm d}x \, L\big(h'(x)\big) + \int_{\frac12}^1 {\rm d}x \, L(-1) \nonumber\\
\egal r \log\sqrt{ab} + (x^* - r + 1) L(t^*) + \int_{x^*}^{\frac12} {\rm d}x \, L\big(h'(x)\big) + \frac 12 \log\sqrt{ab},
\eea
where the values $L(\pm 1) = \log{\sqrt{ab}}$, computed at the end of Section 4.2, have been used.

Differentiating with respect to $r$ yields
\be
\frac{{\rm d}S[f]}{{\rm d}r} = \log\sqrt{ab} - L(t^*) + (x^*-r+1) \, L'(t^*) \, h''(x^*) \,\frac{{\rm d}x^*}{{\rm d}r}.
\ee
Since the rectilinear part of the blue line, whose equation is given in (\ref{recti}), passes through the point $(x^*,h(x^*))$, we obtain
\be
r = \frac{h(x^*) - t^*(x^*+1)}{1-t^*},
\ee
from which we readily compute
\be
\frac{{\rm d}r}{{\rm d}x^*} = h''(x^*) \, \frac{r-1-x^*}{1-t^*}.
\ee
From this, we find
\be
\frac{{\rm d}S[f]}{{\rm d}r} = \log\sqrt{ab} - L(t^*) - (1-t^*) L'(t^*) = \log\big[\sqrt{ab}\,x(t^*)y(t^*)\big] = \log{p(t^*)},
\ee
where we have used $L(t) = -\log x(t) - t \log y(t)$, and $L'(t) = -\log y(t)$, see Section 4.2.

Going back to (\ref{main}) and inserting the result (\ref{lattice}) obtained in Section 4.1 for the left-hand side, we can write
\be
\log (ab) + F_1(r) = S[f],
\ee
and upon differentiating with respect to $r$,
\be
-\log v(r) = \log{p(t^*)}, \qquad {\rm or\ } \quad p(t^*) = \frac 1{v(r)}. 
\ee
This in principle solves our problem, which was to compute $t^*(r)$ as a function of $r$. 

%%%%%%%%%%%%%%%%%%%%%%%%%%%%%%%%%%%%%%%%%%%%%%%%%%%%%%%%%%%%%%%%%%%%%%%%%%

\vskip 0.5truecm
\noindent
{\bf 4.4 Two-periodic arctic curves}
\addcontentsline{toc}{subsubsection}{4.4 Two-periodic arctic curves}

\medskip
\noindent
Let us summarize what we have done so far. The equations (\ref{recti}) provide a parametric form $\big(X(r),Y(r)\big)$ of the arctic curve where $t^*(r)$ is obtained in terms of $r$ as the unique solution of $p(t^*) = \frac1{v(r)}$. The inverse function of $p$ is therefore required and this is precisely what we have computed in (\ref{tp}). Since $t^*$ is the value of $t$ for which $p = \frac 1v$, we obtain directly from (\ref{tp}),
\be
t^* = t\Big(p=\frac 1v\Big) = \frac 1{(v^2-1)^4 - 4\alpha^2v^4}\:\Bigg\{2\alpha\, v(v^4-1) \sqrt{v^4+(\alpha^2-2)v^2+1} - 2\alpha^2 v^2(v^4+1) - (v^2-1)^4\Bigg\},
\ee
with $\alpha=\frac{1+\beta}{\sqrt{\beta}}$. We still have to find $v=v(r)$ if we insist that $r$ is the right parameter, but we can instead change and choose $v$ as a new parameter. From the discussion at the end of Section 4.1, we know that $v$ varies from 1 to $+\infty$ when $r$ ranges from $\frac12$ to 1.

In terms of $v$, the parametrization of $X$ and $Y$ can now be written
\be
\begin{cases}
X(v) = \displaystyle \Big(\frac{{\rm d}t^*}{{\rm d}r}\Big)^{-1} \big(t^*(v)-1\big) + r(v) - 1, & \\
\noalign{\medskip}
Y(v) = t^*(v) \Big(X(v)-r(v)+1\Big) + r(v), &
\end{cases}
\ee
and made fully explicit in view of
\be
\frac{{\rm d}t^*}{{\rm d}r}(v) = \frac{{\rm d}t^*}{{\rm d}v} \, \Big(\frac{{\rm d}r}{{\rm d}v}\Big)^{-1},
\ee
and the fact that $t^*(v)$ and $r(v)$ are explicitly known. The remaining calculations are straightforward.

\begin{figure}[t]
\begin{center}
%\psset{unit=1cm}
\setlength{\unitlength}{1cm}
%
%\hspace{-1.5cm}
\begin{pspicture}(2,3.5)
\rput[bl](-0.1,0.4){\includegraphics[scale=0.32]{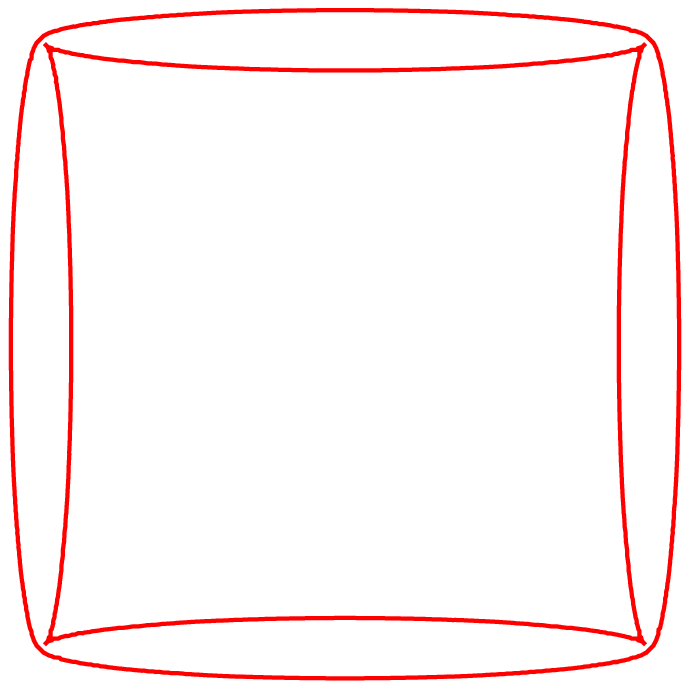}}
\rput(0,0.5){\psline[linewidth=0.5pt,linecolor=black](1,-1)(3,1)(1,3)(-1,1)(1,-1)}
\end{pspicture}
\hspace{2.3cm}
\begin{pspicture}(2,3.5)
\rput[bl](-0.26,0.24){\includegraphics[scale=0.32]{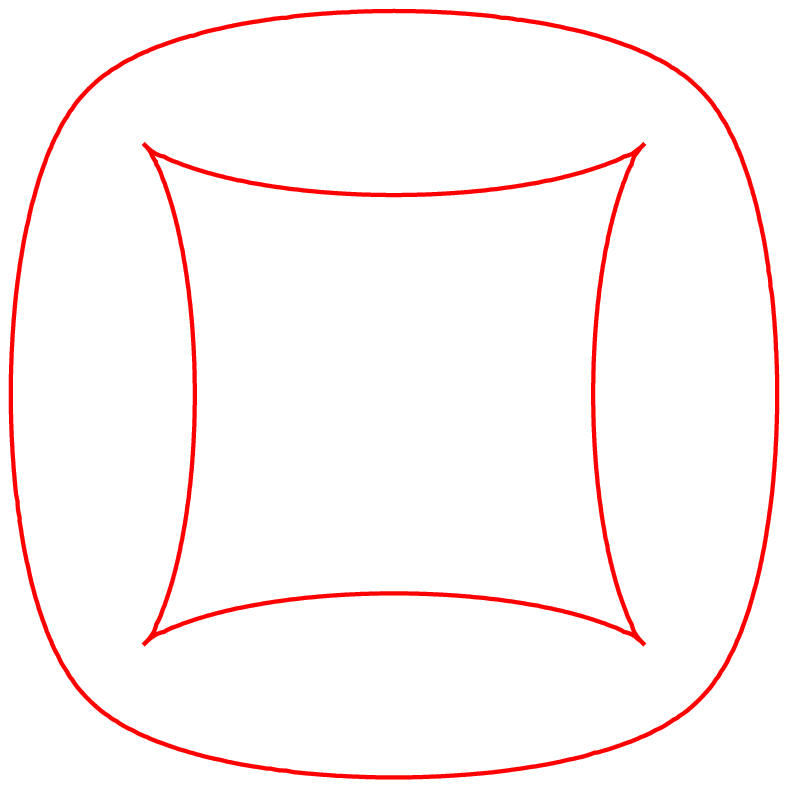}}
\rput(0,0.5){\psline[linewidth=0.5pt,linecolor=black](1,-1)(3,1)(1,3)(-1,1)(1,-1)}
\end{pspicture}
\hspace{2.3cm}
\begin{pspicture}(2,3.5)
\rput[bl](-0.40,0.11){\includegraphics[scale=0.32]{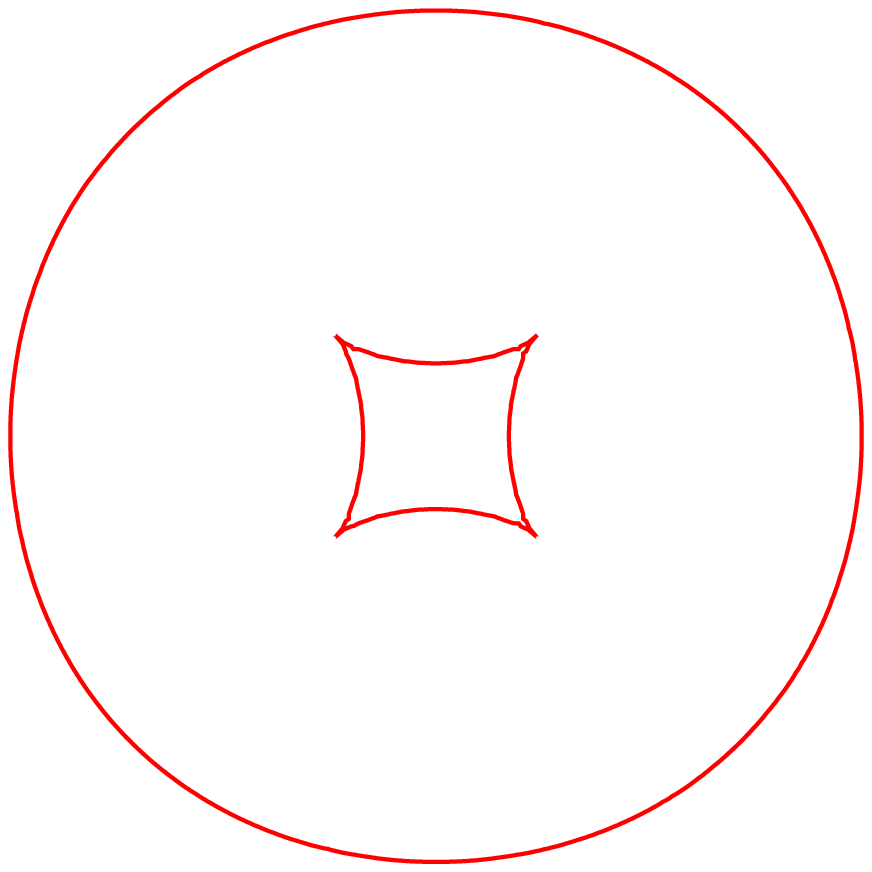}}
\rput(0,0.5){\psline[linewidth=0.5pt,linecolor=black](1,-1)(3,1)(1,3)(-1,1)(1,-1)}
\end{pspicture}
\hspace{2.3cm}
\begin{pspicture}(2,3.5)
\rput[bl](-0.415,0.085){\includegraphics[scale=0.32]{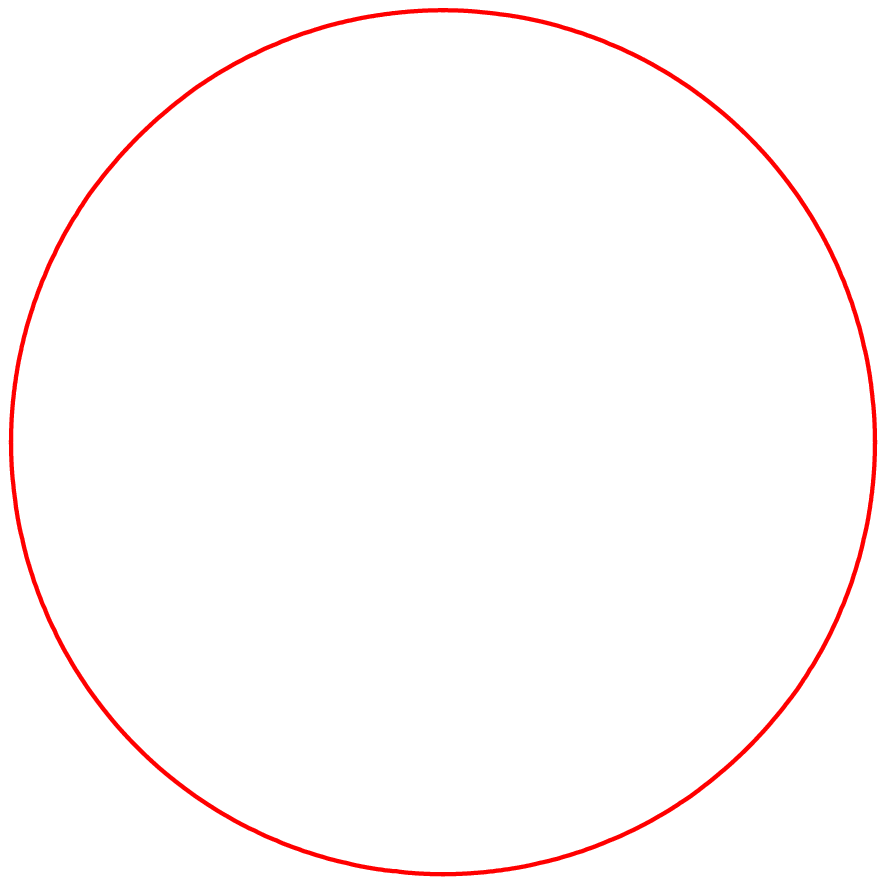}}
\rput(0,0.5){\psline[linewidth=0.5pt,linecolor=black](1,-1)(3,1)(1,3)(-1,1)(1,-1)}
\end{pspicture}
\end{center}
\caption{Arctic curves for the two-periodic Aztec diamonds with parameter values $\beta=0.01$, $\beta=0.1$, $\beta=0.5$ and $\beta=1$ from left to right.}
\label{fig5}
\end{figure}

Our final result for the parametric form of the arctic curves reads,
\be
\left\{ X(v) \atop Y(v) \right\} = \frac1{(v^2+1)^4 + 4\frac{(\beta-1)^2}\beta\,v^4}\:\left\{\frac12 (v^4-1)(v^2+1)^2 \mp 2\frac{\sqrt{\beta}}{\beta+1}\,v \Big[v^4+\big(\beta+\frac1\beta\big)v^2+1\Big]^{\frac32}\right\}.
\label{param}
\ee
It is symmetric under $\beta \leftrightarrow \frac1\beta$, equivalently under $a \leftrightarrow b$. 

For any $\beta$, the arctic curve is known to be algebraic of degree 8. We can indeed check that the following equation is satisfied by $U(v)=X(v)+Y(v),\,V(v)=Y(v)-X(v)$, for any complex $v$,
\bea
&& \hspace{-5mm}(\beta+1)^6 \, (U^8+V^8) - 4(\beta+1)^4(\beta^2-6\beta+1) \,U^2V^2(U^4+V^4)  \nonumber\\
&& \; + \: 2(\beta+1)^2(3\beta^4-20\beta^3+82\beta^2-20\beta+3) \, U^4V^4 - 4(\beta+1)^4(\beta^2-\beta+1) \, (U^6+V^6) \nonumber\\
&& \; + \: 4(\beta+1)^2(\beta^4+17\beta^3-48\beta^2+17\beta+1)\,U^2V^2(U^2+V^2) \nonumber\\
&& \; + \: 6(\beta^4-1) (\beta^2-1) \, (U^4+V^4) + 4(\beta-1)^2(\beta^4-22\beta^3-42\beta^2-22\beta+1) \, U^2V^2 \nonumber\\
&& \; - \: 4(\beta-1)^4(\beta^2+\beta+1)\, (U^2+V^2) + (\beta-1)^6 = 0.
\eea
The real section of this algebraic curve has two components (sometimes called arctic and antarctic), except for $\beta=1$ and $\beta=0$, in which cases it degenerates into a circle and a square respectively, both inscribed in the Aztec diamond. As is well-known, the two components separate three distinct phases, a frozen phase outside of the outer component, a liquid phase inbetween the two components and a gaseous phase inside the inner component, with its characteristic cusps. As one can see in Figure \ref{fig5}, for $0 \le \beta \le 1$, the gaseous phase shrinks as $\beta$ increases and disappears for $\beta=1$; likewise the liquid phase gets squeezed when $\beta$ decreases and vanishes for $\beta=0$. The parametrization (\ref{param}) yields the north portion of the outer component when $v$ varies from 1 to $+\infty$ by real values; the complete outer component is obtained by letting  $v$ take all real values. The inner component cannot be obtained from the parametrization (\ref{param}).

To the best of our knowledge, the above algebraic equation has been first found in the context of two-periodic diamonds by Di Francesco and Soto-Garrido \cite{DFSG14} who used the octahedron recurrence to compute a density profile throughout the Aztec domain; it has been rederived by Chhita and Johansson \cite{CJ16} by using an improved formula for the inverse Kasteleyn matrix (i.e. the correlation kernel); it has also been obtained by Duits and Kuijlaars \cite{DK21} who relied on the non-intersecting path description and used matrix-valued orthogonal polynomials to obtain an alternative expression for the inverse Kasteleyn matrix.

%%%%%%%%%%%%%%%%%%%%%%%%%%%%%%%%%%%%%%%%%%%%%%%%%%%%%%%%%%%%%%%%%%%%%%%%%%

\vskip 0.5truecm
\noindent
{\bf \large 5. Two-refined partition functions}
\addcontentsline{toc}{subsection}{5. Two-refined partition functions}
\setcounter{section}{5}
\setcounter{equation}{0}

\medskip
\noindent
The previous calculations can be extended to two-refined partition functions, for which we fix the numbers of vertical edges along the NW and the NE boundaries. In terms of paths, it means that the uppermost path is forced to follow the NW and NE boundaries over fixed distances. Sportiello \cite{Sp19} has proposed the two-refinement tangent method as an alternative mean to compute arctic curves. Based on two-refined partition functions, this method relies on a 0-to-1 discontinuity to locate the arctic curve; the mechanism behind the discontinuity is well understood from the factorization properties put forward in \cite{DR21}.  

As in Section 4, we assign weights to the outer faces along the NW boundary, and also to those along the NE boundary. We keep the notation $x_0,x_1,\ldots,x_n$ for the NW faces and denote those of the NE faces by $\bar x_0,\bar x_1,\ldots,\bar x_n$, so that the weights $x_i,\bar x_i$ refer to pairs of faces which are located at the same height. As there is a unique face at the top of the graph, we set $\bar x_n=x_n$.

We denote by $T_n\big(a,b\Big|{x_0,x_1,\cdots,x_{n} \atop \bar x_0,\bar x_1,\cdots,\bar x_{n}}\big)$ the partition function, with respect to the octahedron measure, of the two-periodic Aztec diamond of order $n$ with the boundary face weights as above; the two functions $T_n(a,b|x_0,x_1,\cdots,x_n)$ and $T_n(a,b)$ are specializations of the former, and have the same meaning as before, see Section 4. By symmetry, one needs not distinguish whether the face weights in $T_n(a,b|x_0,x_1,\cdots,x_n)$ are those along the NW boundary or the NE one. 

The recurrence (\ref{octa}) yields
\bea
T_n\big(a,b\Big|{x_0,x_1,\cdots,x_{n} \atop \bar x_0,\bar x_1,\cdots,\bar x_{n}}\big) \, T_{n-2}(a,b) \egal T_{n-1}(a,b) \, T_{n-1}\big(a,b\Big|{x_1,x_2,\cdots,x_n \atop \bar x_1,\bar x_2,\cdots,\bar x_n}\big) \nonumber\\
\noalign{\medskip}
&& \hspace{-1cm} + \: T_{n-1}(b,a|x_0,x_1,\cdots,x_{n-1}) \, T_{n-1}(b,a|\bar x_0,\bar x_1,\cdots,\bar x_{n-1}),
\label{octa2}
\eea
along with the initial conditions,
\be
T_0 = 1, \qquad T_1\big(a,b\Big|{x_0,x_1 \atop \bar x_0,x_1}\big) = \frac 1a \, (x_1 + x_0 \bar x_0).
\ee
One finds for instance,
\be
T_2\big(a,b\Big|{x_0,x_1,x_2 \atop \bar x_0,\bar x_1,x_2}\big) = \frac2{a^2}\,(x_2 + x_1 \bar x_1) + \frac1{b^2}\,(x_1 \bar x_1 + x_1 \bar x_0 + x_0 \bar x_1 + x_0 \bar x_0),
\ee
where the eight terms correspond to the following eight configurations, shown in the same order.

\medskip
\begin{figure}[h]
\begin{center}
\psset{unit=.4cm}
\hspace{.5cm}
\pspicture(0,0)(3,3)
\psline[linewidth=2.0pt,linecolor=blue](0,0)(0,1)
\psline[linewidth=2.0pt,linecolor=blue](1,0)(1,1)
\psline[linewidth=2.0pt,linecolor=blue](-1,1)(-1,2)
\psline[linewidth=2.0pt,linecolor=blue](2,1)(2,2)
\psline[linewidth=2.0pt,linecolor=blue](0,2)(0,3)
\psline[linewidth=2.0pt,linecolor=blue](1,2)(1,3)
\pscircle[linecolor=black,fillstyle=solid,fillcolor=black](0,0){0.12}
\pscircle[linecolor=black,fillstyle=solid,fillcolor=black](1,0){0.12}
\pscircle[linecolor=black,fillstyle=solid,fillcolor=black](-1,1){0.12}
\pscircle[linecolor=black,fillstyle=solid,fillcolor=black](0,1){0.12}
\pscircle[linecolor=black,fillstyle=solid,fillcolor=black](1,1){0.12}
\pscircle[linecolor=black,fillstyle=solid,fillcolor=black](2,1){0.12}
\pscircle[linecolor=black,fillstyle=solid,fillcolor=black](-1,2){0.12}
\pscircle[linecolor=black,fillstyle=solid,fillcolor=black](0,2){0.12}
\pscircle[linecolor=black,fillstyle=solid,fillcolor=black](1,2){0.12}
\pscircle[linecolor=black,fillstyle=solid,fillcolor=black](2,2){0.12}
\pscircle[linecolor=black,fillstyle=solid,fillcolor=black](0,3){0.12}
\pscircle[linecolor=black,fillstyle=solid,fillcolor=black](1,3){0.12}
\endpspicture
\hspace{.8cm}
\pspicture(0,0)(3,3)
\psline[linewidth=2.0pt,linecolor=blue](0,0)(1,0)
\psline[linewidth=2.0pt,linecolor=blue](0,1)(1,1)
\psline[linewidth=2.0pt,linecolor=blue](-1,1)(-1,2)
\psline[linewidth=2.0pt,linecolor=blue](2,1)(2,2)
\psline[linewidth=2.0pt,linecolor=blue](0,2)(0,3)
\psline[linewidth=2.0pt,linecolor=blue](1,2)(1,3)
\pscircle[linecolor=black,fillstyle=solid,fillcolor=black](0,0){0.12}
\pscircle[linecolor=black,fillstyle=solid,fillcolor=black](1,0){0.12}
\pscircle[linecolor=black,fillstyle=solid,fillcolor=black](-1,1){0.12}
\pscircle[linecolor=black,fillstyle=solid,fillcolor=black](0,1){0.12}
\pscircle[linecolor=black,fillstyle=solid,fillcolor=black](1,1){0.12}
\pscircle[linecolor=black,fillstyle=solid,fillcolor=black](2,1){0.12}
\pscircle[linecolor=black,fillstyle=solid,fillcolor=black](-1,2){0.12}
\pscircle[linecolor=black,fillstyle=solid,fillcolor=black](0,2){0.12}
\pscircle[linecolor=black,fillstyle=solid,fillcolor=black](1,2){0.12}
\pscircle[linecolor=black,fillstyle=solid,fillcolor=black](2,2){0.12}
\pscircle[linecolor=black,fillstyle=solid,fillcolor=black](0,3){0.12}
\pscircle[linecolor=black,fillstyle=solid,fillcolor=black](1,3){0.12}
\endpspicture
\hspace{.8cm}
\pspicture(0,0)(3,3)
\psline[linewidth=2.0pt,linecolor=blue](0,0)(1,0)
\psline[linewidth=2.0pt,linecolor=blue](0,1)(1,1)
\psline[linewidth=2.0pt,linecolor=blue](-1,1)(-1,2)
\psline[linewidth=2.0pt,linecolor=blue](2,1)(2,2)
\psline[linewidth=2.0pt,linecolor=blue](0,2)(1,2)
\psline[linewidth=2.0pt,linecolor=blue](0,3)(1,3)
\pscircle[linecolor=black,fillstyle=solid,fillcolor=black](0,0){0.12}
\pscircle[linecolor=black,fillstyle=solid,fillcolor=black](1,0){0.12}
\pscircle[linecolor=black,fillstyle=solid,fillcolor=black](-1,1){0.12}
\pscircle[linecolor=black,fillstyle=solid,fillcolor=black](0,1){0.12}
\pscircle[linecolor=black,fillstyle=solid,fillcolor=black](1,1){0.12}
\pscircle[linecolor=black,fillstyle=solid,fillcolor=black](2,1){0.12}
\pscircle[linecolor=black,fillstyle=solid,fillcolor=black](-1,2){0.12}
\pscircle[linecolor=black,fillstyle=solid,fillcolor=black](0,2){0.12}
\pscircle[linecolor=black,fillstyle=solid,fillcolor=black](1,2){0.12}
\pscircle[linecolor=black,fillstyle=solid,fillcolor=black](2,2){0.12}
\pscircle[linecolor=black,fillstyle=solid,fillcolor=black](0,3){0.12}
\pscircle[linecolor=black,fillstyle=solid,fillcolor=black](1,3){0.12}
\endpspicture
\hspace{.8cm}
\pspicture(0,0)(3,3)
\psline[linewidth=2.0pt,linecolor=blue](0,0)(0,1)
\psline[linewidth=2.0pt,linecolor=blue](1,0)(1,1)
\psline[linewidth=2.0pt,linecolor=blue](-1,1)(-1,2)
\psline[linewidth=2.0pt,linecolor=blue](2,1)(2,2)
\psline[linewidth=2.0pt,linecolor=blue](0,2)(1,2)
\psline[linewidth=2.0pt,linecolor=blue](0,3)(1,3)
\pscircle[linecolor=black,fillstyle=solid,fillcolor=black](0,0){0.12}
\pscircle[linecolor=black,fillstyle=solid,fillcolor=black](1,0){0.12}
\pscircle[linecolor=black,fillstyle=solid,fillcolor=black](-1,1){0.12}
\pscircle[linecolor=black,fillstyle=solid,fillcolor=black](0,1){0.12}
\pscircle[linecolor=black,fillstyle=solid,fillcolor=black](1,1){0.12}
\pscircle[linecolor=black,fillstyle=solid,fillcolor=black](2,1){0.12}
\pscircle[linecolor=black,fillstyle=solid,fillcolor=black](-1,2){0.12}
\pscircle[linecolor=black,fillstyle=solid,fillcolor=black](0,2){0.12}
\pscircle[linecolor=black,fillstyle=solid,fillcolor=black](1,2){0.12}
\pscircle[linecolor=black,fillstyle=solid,fillcolor=black](2,2){0.12}
\pscircle[linecolor=black,fillstyle=solid,fillcolor=black](0,3){0.12}
\pscircle[linecolor=black,fillstyle=solid,fillcolor=black](1,3){0.12}
\endpspicture
\hspace{.8cm}
\pspicture(0,0)(3,3)
\psline[linewidth=2.0pt,linecolor=blue](0,0)(1,0)
\psline[linewidth=2.0pt,linecolor=blue](0,1)(0,2)
\psline[linewidth=2.0pt,linecolor=blue](-1,1)(-1,2)
\psline[linewidth=2.0pt,linecolor=blue](2,1)(2,2)
\psline[linewidth=2.0pt,linecolor=blue](1,1)(1,2)
\psline[linewidth=2.0pt,linecolor=blue](0,3)(1,3)
\pscircle[linecolor=black,fillstyle=solid,fillcolor=black](0,0){0.12}
\pscircle[linecolor=black,fillstyle=solid,fillcolor=black](1,0){0.12}
\pscircle[linecolor=black,fillstyle=solid,fillcolor=black](-1,1){0.12}
\pscircle[linecolor=black,fillstyle=solid,fillcolor=black](0,1){0.12}
\pscircle[linecolor=black,fillstyle=solid,fillcolor=black](1,1){0.12}
\pscircle[linecolor=black,fillstyle=solid,fillcolor=black](2,1){0.12}
\pscircle[linecolor=black,fillstyle=solid,fillcolor=black](-1,2){0.12}
\pscircle[linecolor=black,fillstyle=solid,fillcolor=black](0,2){0.12}
\pscircle[linecolor=black,fillstyle=solid,fillcolor=black](1,2){0.12}
\pscircle[linecolor=black,fillstyle=solid,fillcolor=black](2,2){0.12}
\pscircle[linecolor=black,fillstyle=solid,fillcolor=black](0,3){0.12}
\pscircle[linecolor=black,fillstyle=solid,fillcolor=black](1,3){0.12}
\endpspicture
\hspace{.8cm}
\pspicture(0,0)(3,3)
\psline[linewidth=2.0pt,linecolor=blue](0,0)(1,0)
\psline[linewidth=2.0pt,linecolor=blue](0,1)(0,2)
\psline[linewidth=2.0pt,linecolor=blue](-1,1)(-1,2)
\psline[linewidth=2.0pt,linecolor=blue](1,2)(2,2)
\psline[linewidth=2.0pt,linecolor=blue](1,1)(2,1)
\psline[linewidth=2.0pt,linecolor=blue](0,3)(1,3)
\pscircle[linecolor=black,fillstyle=solid,fillcolor=black](0,0){0.12}
\pscircle[linecolor=black,fillstyle=solid,fillcolor=black](1,0){0.12}
\pscircle[linecolor=black,fillstyle=solid,fillcolor=black](-1,1){0.12}
\pscircle[linecolor=black,fillstyle=solid,fillcolor=black](0,1){0.12}
\pscircle[linecolor=black,fillstyle=solid,fillcolor=black](1,1){0.12}
\pscircle[linecolor=black,fillstyle=solid,fillcolor=black](2,1){0.12}
\pscircle[linecolor=black,fillstyle=solid,fillcolor=black](-1,2){0.12}
\pscircle[linecolor=black,fillstyle=solid,fillcolor=black](0,2){0.12}
\pscircle[linecolor=black,fillstyle=solid,fillcolor=black](1,2){0.12}
\pscircle[linecolor=black,fillstyle=solid,fillcolor=black](2,2){0.12}
\pscircle[linecolor=black,fillstyle=solid,fillcolor=black](0,3){0.12}
\pscircle[linecolor=black,fillstyle=solid,fillcolor=black](1,3){0.12}
\endpspicture
\hspace{.8cm}
\pspicture(0,0)(3,3)
\psline[linewidth=2.0pt,linecolor=blue](0,0)(1,0)
\psline[linewidth=2.0pt,linecolor=blue](-1,2)(0,2)
\psline[linewidth=2.0pt,linecolor=blue](-1,1)(0,1)
\psline[linewidth=2.0pt,linecolor=blue](2,1)(2,2)
\psline[linewidth=2.0pt,linecolor=blue](1,1)(1,2)
\psline[linewidth=2.0pt,linecolor=blue](0,3)(1,3)
\pscircle[linecolor=black,fillstyle=solid,fillcolor=black](0,0){0.12}
\pscircle[linecolor=black,fillstyle=solid,fillcolor=black](1,0){0.12}
\pscircle[linecolor=black,fillstyle=solid,fillcolor=black](-1,1){0.12}
\pscircle[linecolor=black,fillstyle=solid,fillcolor=black](0,1){0.12}
\pscircle[linecolor=black,fillstyle=solid,fillcolor=black](1,1){0.12}
\pscircle[linecolor=black,fillstyle=solid,fillcolor=black](2,1){0.12}
\pscircle[linecolor=black,fillstyle=solid,fillcolor=black](-1,2){0.12}
\pscircle[linecolor=black,fillstyle=solid,fillcolor=black](0,2){0.12}
\pscircle[linecolor=black,fillstyle=solid,fillcolor=black](1,2){0.12}
\pscircle[linecolor=black,fillstyle=solid,fillcolor=black](2,2){0.12}
\pscircle[linecolor=black,fillstyle=solid,fillcolor=black](0,3){0.12}
\pscircle[linecolor=black,fillstyle=solid,fillcolor=black](1,3){0.12}
\endpspicture
\hspace{.8cm}
\pspicture(0,0)(3,3)
\psline[linewidth=2.0pt,linecolor=blue](0,0)(1,0)
\psline[linewidth=2.0pt,linecolor=blue](-1,2)(0,2)
\psline[linewidth=2.0pt,linecolor=blue](-1,1)(0,1)
\psline[linewidth=2.0pt,linecolor=blue](1,2)(2,2)
\psline[linewidth=2.0pt,linecolor=blue](1,1)(2,1)
\psline[linewidth=2.0pt,linecolor=blue](0,3)(1,3)
\pscircle[linecolor=black,fillstyle=solid,fillcolor=black](0,0){0.12}
\pscircle[linecolor=black,fillstyle=solid,fillcolor=black](1,0){0.12}
\pscircle[linecolor=black,fillstyle=solid,fillcolor=black](-1,1){0.12}
\pscircle[linecolor=black,fillstyle=solid,fillcolor=black](0,1){0.12}
\pscircle[linecolor=black,fillstyle=solid,fillcolor=black](1,1){0.12}
\pscircle[linecolor=black,fillstyle=solid,fillcolor=black](2,1){0.12}
\pscircle[linecolor=black,fillstyle=solid,fillcolor=black](-1,2){0.12}
\pscircle[linecolor=black,fillstyle=solid,fillcolor=black](0,2){0.12}
\pscircle[linecolor=black,fillstyle=solid,fillcolor=black](1,2){0.12}
\pscircle[linecolor=black,fillstyle=solid,fillcolor=black](2,2){0.12}
\pscircle[linecolor=black,fillstyle=solid,fillcolor=black](0,3){0.12}
\pscircle[linecolor=black,fillstyle=solid,fillcolor=black](1,3){0.12}
\endpspicture
\end{center}
\end{figure}

%%%%%%%%%%%%%%%%%%%%%%%%%%%%%%%%%%%%%%%%%%%%%%%%%%%%%%%%%%%%%%%%%%%%%%%%%%

\eject
\noindent
{\bf 5.1 Asymptotics}
\addcontentsline{toc}{subsubsection}{5.1 Asymptotics}

\medskip
\noindent
Let us denote by $T_{n,k,\ell}(a,b)$ the weighted sum over the perfect matchings which have exactly $k$ resp. $\ell$  vertical edges along the NW resp. NE boundary. $T_{n,k,\ell}(a,b)$ is the coefficient of $x_k \bar x_\ell$ in $T_n\big(a,b\Big|{x_0,x_1,\cdots,x_{n} \atop \bar x_0,\bar x_1,\cdots,\bar x_{n}}\big)$ if $k,\ell \le n-1$, and the coefficient of $x_n$ if $k=\ell=n$. From (\ref{octa2}), it satisfies the following recurrence
\be
T_{n,k,\ell}(a,b) \, T_{n-2}(a,b) = T_{n-1}(a,b) \, T_{n-1,k-1,\ell-1}(a,b) + T_{n-1,k}(b,a) \, T_{n-1,\ell}(b,a).
\label{1rec}
\ee
By using the boundary conditions $T_{n,0,\ell}(a,b) = \{b^{-1} {\rm \ or\ } a^{-1}\} T_{n-1,\ell}(b,a)$ for $n$ even resp. odd, and $T_{n,n,\ell}(a,b) = a^{-1} \,\delta_{\ell,n} \,T_{n-1}(a,b) $, one easily checks that the above recurrence is valid for all values $0 \le k,\ell \le n$. As before we define the following ratio,
\be
S_{n,k,\ell}(a,b) \equiv a \, \frac{T_{n,k,\ell}(a,b)}{T_{n-1}(a,b)},
\ee
and obtain a new recurrence from (\ref{1rec}) by multiplying by $a$ and dividing by $T_{n-2}(a,b)T_{n-1}(a,b)$, 
\be
S_{n,k,\ell}(a,b) = S_{n-1,k-1,\ell-1}(a,b) + \frac a{b^2} \frac{T^2_{n-2}(b,a)}{T_{n-2}(a,b)T_{n-1}(a,b)} \, S_{n-1,k}(b,a)\,S_{n-1,\ell}(b,a),
\ee
with $S_{n,k}(a,b)$ related to the 1-refined partition function and defined in Section 4. By using once more the explicit expression of the unrefined partition functions $T_n(a,b)$ given in (\ref{Fn}), we compute that the ratio
\bea
\frac a{b^2} \frac{T^2_{n-2}(b,a)}{T_{n-2}(a,b)T_{n-1}(a,b)} \egal \Big[\frac{\sqrt{\beta}}{2(1+\beta)}\Big]^{\frac n2} \times  \Bigg\{1+\beta,\, \Big[\frac{2(1+\beta)}{\sqrt{\beta}}\Big]^{\frac12}, \,(1+\beta)\sqrt{\beta} , \, \Big[\frac{2(1+\beta)}{\sqrt{\beta}}\Big]^{\frac12} \beta^{\frac32}\Bigg\}, \nonumber\\
&& \hspace{5.5cm} {\rm resp.\ for\ } n=0,1,2,3 \bmod 4,
\label{inter}
\eea
is a function of $\beta=\frac{a^2}{b^2}$; since $S_{n,k}(a,b)$ is a function of $\beta$, so is $S_{n,k,\ell}(a,b)$.

The recurrence now reads
\be
S_{n,k,\ell}(\beta) = S_{n-1,k-1,\ell-1}(\beta) + \Big[\frac{\sqrt{\beta}}{2(1+\beta)}\Big]^{\frac n2} \, \varepsilon_n(\beta) \, S_{n-1,k}(\textstyle \frac 1\beta) \, S_{n-1,\ell}(\frac 1\beta),
\ee
where $\varepsilon_n(\beta)$ is the function in the curly brackets in (\ref{inter}), which depends on the residue of $n$ modulo 4. Interestingly, we have a simple linear recurrence of order 1 in terms of the known functions $S_{n,k}(\beta)$, from which we can derive a closed expression for $S_{n,k,\ell}(\beta)$, namely,
\be
S_{n,k,\ell}(\beta) = \delta_{k,n} \, \delta_{\ell,n} + \sum_{i=0}^{\min(k,\ell)} \: \Big[\frac{\sqrt{\beta}}{2(1+\beta)}\Big]^{\frac {n-i}2} \, \varepsilon_{n-i}(\beta) \, S_{n-1-i,k-i}(\textstyle \frac 1\beta) \, S_{n-1-i,\ell-i}(\frac 1\beta).
\ee
In particular, for uniformly weighted Aztec diamonds, we obtain, since $\varepsilon_n(\beta=1)=2$ for all $n$, the following exact expression for the boundary two-refined partition functions, 
\be
S_{n,k,\ell}(1) = \delta_{k,n} \, \delta_{\ell,n} + 2^{1-n} \sum_{i=0}^{\min(k,\ell)} 2^i \: {n-1-i \choose k-i} \, {n-1-i \choose \ell-i}.
\ee

As before, the asymptotic value of $S_{n,k,\ell}(\beta)$ is needed to apply the two-refined tangent method. It can be computed from that of $S_{n,k}$, obtained in Section 4.1. Assume $k \ge \ell$. Letting $k=rn$ and $\ell=sn$ ($r \ge s$), the sum over $i$ can be turned into an integral over $\xi=\frac in$. Using the asymptotic value of $S_{n,rn}(\beta) \simeq \exp{nF_1(r)}$, we obtain
\be
S_{n,rn,sn}(\beta) \simeq \int_0^s {\rm d}\xi \: \exp\Big\{n\Big[\frac{1-\xi}2 \log\frac{\sqrt{\beta}}{2(1+\beta)} + (1-\xi) \, F_1\Big(\frac{r-\xi}{1-\xi}\Big) + (1-\xi) \, F_1\Big(\frac{s-\xi}{1-\xi}\Big)\Big]\Big\}
\ee
Let us denote by $H(\xi;r,s)$ the function of $\xi$ within the square brackets. 

We first observe that $H$ is a strictly concave function of $\xi$. Indeed the first two derivatives yield 
\bea
H'(\xi;r,s) \egal \frac12 \log\frac{2(1+\beta)}{\sqrt{\beta}} - F_1\Big(\frac{r-\xi}{1-\xi}\Big) - F_1\Big(\frac{s-\xi}{1-\xi}\Big) - \frac{1-r}{1-\xi} \, F_1'\Big(\frac{r-\xi}{1-\xi}\Big) - \frac{1-s}{1-\xi}\, F_1'\Big(\frac{s-\xi}{1-\xi}\Big) \nonumber\\
\egal \frac12 \log\frac{2(1+\beta)}{\sqrt{\beta}} + \log uv\Big(\frac{r-\xi}{1-\xi}\Big) + \log uv\Big(\frac{s-\xi}{1-\xi}\Big),  \qquad (uv(z)=u(z)v(z))\\
H''(\xi;r,s) \egal \frac{(1-r)^2}{(1-\xi)^3} \: F_1''\Big(\frac{r-\xi}{1-\xi}\Big) + \frac{(1-s)^2}{(1-\xi)^3} \: F_1''\Big(\frac{s-\xi}{1-\xi}\Big),
\eea
where we have used $F_1(z) = -\log u(z) - z \log v(z)$ and $F_1'(z) = -\log v(z)$. The strict concavity of $H$ follows from the expression of $H''$ and the fact that $F_1$ itself is strictly concave, $F_1''(z)<0$, see Section 4.1.

In view of the strict concavity of $H$, it is instructive to evaluate its first derivative at the lower and upper bounds of the integration domain. One finds respectively,
\bea
H'(\xi=0;r,s) \egal \frac12 \log\frac{2(1+\beta)}{\sqrt{\beta}} + \log\!\big[uv(r)\big] + \log\!\big[uv(s)\big], \\
H'(\xi=s;r,s) \egal \frac12 \log\frac{2(1+\beta)}{\sqrt{\beta}} + \log uv\Big( \frac{r-s}{1-s}\Big) + \log\!\big[uv(0^+)\big] = -\infty.
\eea
The value at the upper bound follows from the explicit forms of $u$ and $v$ found in Section 4.1, from which we deduce $u(z) = 1 - z + \ldots$ and $v(z) = \frac{2\sqrt{\beta}}{1+\beta}z + \big[\frac{2\sqrt{\beta}}{1+\beta}\big]^3 z^2 + \ldots$ for small and positive $z$.
%For $u(x)$ and $v(x)$ as above, we obtain for small and positive $x$,
%\be
%x(v) = \frac{1+x}{2\sqrt{x}} \, v - v^2 + \ldots \quad \Longrightarrow \quad v(x) = \frac{1+x}{2\sqrt{x}} x + \Big(\frac{1+x}{2\sqrt{x}}\Big)^3 x^2 + \ldots, \quad u(x) = 1 - x + \ldots
%\ee

The result of the saddle point analysis therefore depends on the sign of $H'(0;r,s)$. If $H'(0;r,s) > 0$, the strict concavity of $H$ implies that it has a unique maximum at a point $\xi^*$ strictly inside the interval $[0,s]$, and satisfying
\be
H'(\xi^*;r,s) = 0 \quad \Longleftrightarrow \quad uv\Big(\frac{r-\xi^*}{1-\xi^*}\Big) \, uv\Big(\frac{s-\xi^*}{1-\xi^*}\Big) = \Big[\frac{\sqrt{\beta}}{2(1+\beta)}\Big]^{\frac12}.
\label{max}
\ee
If $H'(0;r,s) \le 0$, the function $H$ is monotonically decreasing on the interval $[0,s]$ and its maximum is at $\xi=0$. 

To summarize, the asymptotic value of the two-refined partition function is given by $S_{n,rn,sn}(\beta) \simeq \exp\{n F_2(r,s)\}$, where
\be
F_2(r,s) = 
\begin{cases}
\displaystyle H(\xi^*;r,s) = (1-r) \log v\Big(\frac{r-\xi^*}{1-\xi^*}\Big) + (1-s) \log v\Big(\frac{s-\xi^*}{1-\xi^*}\Big) & {\rm if\ } H'(0;r,s) \ge 0,\\
\noalign{\smallskip}
\displaystyle H(0;r,s) = \frac12 \log\frac{\sqrt{\beta}}{2(1+\beta)} + F_1(r) + F_1(s) & {\rm if\ } H'(0;r,s) \le 0,
\end{cases}
\label{Fprime}
\ee
where $\xi^*$ is the unique solution of (\ref{max}). For $H'(0;r,s)=0$, both expressions coincide on account of (\ref{max}) with $\xi^*=0$.

The rewriting of the above result for the conventional weighting of the two-periodic Aztec diamonds is the same as for the one-refined functions. If the conventional two-refined partition function is denoted by $Z_{n,k,\ell}(a,b)$, we have, for $k=rn$ and $\ell=s n$,
\bea
\lim_{n \to \infty} \frac 1n \log\frac{Z_{n,rn,sn}(a,b)}{Z_{n-1}(a,b)} \egal \lim_{n \to \infty} \frac 1n \log\Big[(ab)^n\, \frac{T_{n,rn,sn}(\textstyle \frac1a,\frac1b)}{T_{n-1}(\frac1a,\frac1b)}\Big] = \log(ab) + \lim_{n \to \infty} \frac 1n \log S_{n,rn,sn}(\textstyle\frac1\beta)\nonumber\\
\egal \log(ab) + F_2(r,s).
\label{2-lattice}
\eea

%%%%%%%%%%%%%%%%%%%%%%%%%%%%%%%%%%%%%%%%%%%%%%%%%%%%%%%%%%%%%%%%%%%%%%%%%%

\vskip 0.5truecm
\noindent
{\bf 5.2 Two-refined tangent method}
\addcontentsline{toc}{subsubsection}{5.2 Two-refined tangent method}

\begin{figure}[t]
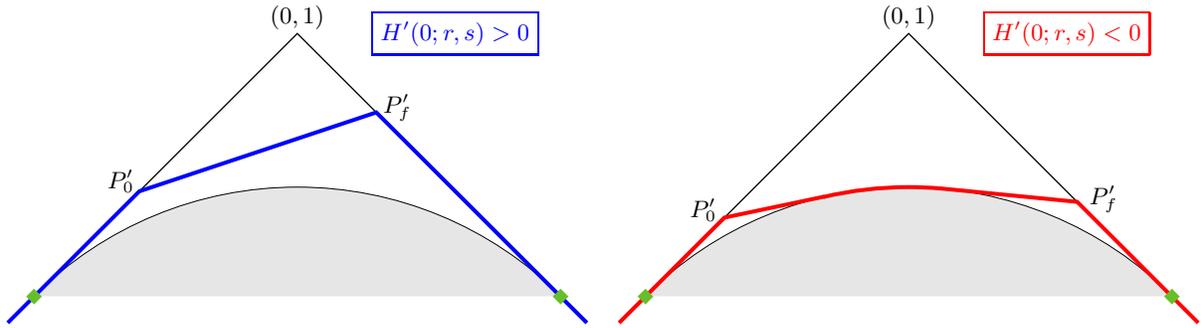

\begin{center}
\psset{unit=.7cm}
%\setlength{\unitlength}{.7cm}
%
%\hspace{-1.5cm}
\pspicture(-5,0)(5,5.5)
%\pscircle[linecolor=black,fillstyle=solid,fillcolor=black](0,0){0.07}
\psarc[linewidth=0.1pt,linecolor=black,fillstyle=solid,fillcolor=mygrey](0,-5){7.08}{45}{135}
\psline[linewidth=0.5pt,linecolor=black](-5,0)(0,5)(5,0)
%\rput(0.5,2.5){\footnotesize $h(x)$}
%\psellipticarc[linewidth=1pt,linestyle=dashed,dash=3pt 2pt,linecolor=red](0,0)(4.11,2.93){-20}{26}
%\psellipticarc[linewidth=1pt,linestyle=dashed,dash=3pt 2pt,linecolor=red](0,0)(4.11,2.93){152}{200}
%\pscircle[linecolor=red,fillstyle=solid,fillcolor=red](0,2.9){0.10}
%\rput(0,2.35){\small $(0,\frac1{\sqrt{1+w}})$}
%\psellipse[linewidth=1pt,linecolor=red](0,0)(4.11,2.93)
%\psline[linewidth=0.5pt,linecolor=black](-4.5,3.65)(3,2.5)
\psline[linewidth=1.5pt,linecolor=blue](-5.5,-.5)(-3.,2)(1.5,3.5)(5.5,-0.5)
\rput(-3.35,2.2){\footnotesize $P'_0$}
\rput(1.9,3.6){\footnotesize $P'_f$}
\rput(3,5){\blue \footnotesize \fbox{$H'(0;r,s)>0$}}
\rput(0,5.3){\footnotesize $(0,1)$}
\psdiamond[linecolor=mygreen,fillstyle=solid,fillcolor=mygreen](-5,0)(0.15,0.15)
\psdiamond[linecolor=mygreen,fillstyle=solid,fillcolor=mygreen](5,0)(0.15,0.15)
\endpspicture
\hspace{1cm}
\pspicture(-5,0)(5,5.5)
%\pscircle[linecolor=black,fillstyle=solid,fillcolor=black](0,0){0.07}
\psarc[linewidth=0.1pt,linecolor=black,fillstyle=solid,fillcolor=mygrey](0,-5){7.08}{45}{135}
\psarc[linewidth=1.5pt,linecolor=red](0,-5){7.08}{85}{103.2}
\rput(-3.9,1.65){\footnotesize $P'_0$}
\rput(3.7,1.85){\footnotesize $P'_f$}
\psline[linewidth=0.5pt,linecolor=black](-5,0)(0,5)(5,0)
\psline[linewidth=1.5pt,linecolor=red](-5.5,-0.5)(-3.5,1.5)(-1.6,1.9)
\psline[linewidth=1.5pt,linecolor=red](0.6,2.055)(3.2,1.8)(5.5,-0.5)
\rput(3,5){\red \footnotesize \fbox{$H'(0;r,s)<0$}}
%\psline[linewidth=0.5pt,linestyle=dashed,dash=3pt 2pt](2.1,2.513)(2.1,0)
%\rput(2.1,-0.35){\footnotesize $x_2(s)$}
%\psline[linewidth=0.5pt,linestyle=dashed,dash=3pt 2pt](-1.55,2.71)(-1.55,0)
%\rput(-1.55,-0.35){\footnotesize $x_1(r)$}
%\rput(0.5,2.5){\footnotesize $h(x)$}
%\psellipticarc[linewidth=1pt,linestyle=dashed,dash=3pt 2pt,linecolor=red](0,0)(4.11,2.93){-20}{26}
%\psellipticarc[linewidth=1pt,linestyle=dashed,dash=3pt 2pt,linecolor=red](0,0)(4.11,2.93){152}{200}
%\pscircle[linecolor=red,fillstyle=solid,fillcolor=red](0,2.9){0.10}
%\rput(0,2.35){\small $(0,\frac1{\sqrt{1+w}})$}
%\psellipse[linewidth=1pt,linecolor=red](0,0)(4.11,2.93)
%\psline[linewidth=0.5pt,linecolor=black](-4.5,3.65)(3,2.5)
\rput(0,5.3){\footnotesize $(0,1)$}
\psdiamond[linecolor=mygreen,fillstyle=solid,fillcolor=mygreen](-5,0)(0.15,0.15)
\psdiamond[linecolor=mygreen,fillstyle=solid,fillcolor=mygreen](5,0)(0.15,0.15)
\endpspicture
\end{center}
\caption{Geometric setting for the two-refined case: the (forced) passage points $P'_0=(r-1,r)$ and $P'_f=(1-s,s)$ are parametrized by $r,s \ge \frac 12$. In both cases, the paths start from $P_0=(-1,0)$ and end at $P_f=(1,0)$, not shown on the pictures.}
\label{fig6}
\end{figure}

\smallskip
\noindent
Let us first recall the principle underlying the two-refined tangent method, as proposed in \cite{Sp19}. The two-refined case is in fact very similar to the one-refined method, the only difference being that the uppermost path is forced to go through two points instead of one, say $P'_0$ and $P'_f$, located on the NW and NE boundaries respectively, as shown in Figure \ref{fig6}. In the scaling limit, the uppermost path becomes a deterministic trajectory $f$, which, as before, is the shortest one going from $P_0$ to $P_f$ under the constraint to pass through $P'_0$ and $P'_f$ and not to cross the arctic curve. The optimal solution will have two straight segments, namely from $P_0$ to $P'_0$ and from $P'_f$ to $P_f$, which are easy to understand. The part between $P'_0$ and $P'_f$ is more interesting and will depend on the relative locations of these two points, that is, on the values of $r$ and $s$.

There are essentially two different regimes. The following definition will be useful for the discussion: we denote by $Z_{0,rn,sn,f}(a,b)$ the partition function for a single path from $P_0$ to $P_f$ passing through $P'_0$ and $P'_f$, and {\it no other constraint}.
\begin{compactenum}
\item If $r$ and $s$ are large enough, say close to 1, the uppermost path will be almost surely a straight line between $P'_0$ and $P'_f$; in the scaling limit the deterministic $f$ is then piecewise rectilinear and comprise three linear segments (left of Figure \ref{fig6}). Since the uppermost path barely interferes with the other $n-1$ paths, one expects a factorization $Z_{n,rn,sn}(a,b) \simeq Z_{n-1}(a,b) Z_{0,rn,sn,f}(a,b)$.
\item If on the other hand $r$ and $s$ are close enough to $\frac12$, the straight line between $P'_0$ and $P'_f$ would cross the arctic curve and this forces the uppermost path to take a longer way to go `around' the bulk of the other paths forming the arctic circle. This introduces frustration and lowers considerably the entropy contribution of the uppermost path hence of the whole system. Thus one would expect in this case a strong inequality $Z_{n,rn,sn}(a,b) \ll Z_{n-1}(a,b) Z_{0,rn,sn,f}(a,b)$. It is also expected \cite{DGR19}, although this is {\it not} an assumption of the two-refined tangent method, that in the scaling limit, the uppermost path between $P'_0$ and $P'_f$ will follow a deterministic trajectory which includes part of the arctic curve and two straight segments tangent to the arctic curve and attached to $P'_0$ and $P'_f$ respectively (right of Figure \ref{fig6}). 
\end{compactenum}

It follows from this discussion that one should observe a sharp 1-to-0 transition in the $(r,s)$ plane when one crosses the curve ${\cal C}_{r,s}$ which expresses the condition that the straight segment from $P'_0$ and $P'_f$ is exactly tangent to the arctic curve, namely
\be
\frac{Z_{n,rn,sn}(a,b)}{Z_{n-1}(a,b) Z_{0,rn,sn,f}(a,b)} \simeq  
\begin{cases}
1 & {\rm if\ } (r,s) {\rm \ is\ above\ or\ on\ } {\cal C}_{r,s},\\
0 & {\rm if\ } (r,s) {\rm \ is\ below\ } {\cal C}_{r,s}.
\end{cases}
\label{ratio}
\ee
The two cases refer respectively to the straight line from $P'_0$ and $P'_f$ not crossing or crossing the arctic curve. The whole problem therefore reduces to the computation of ${\cal C}_{r,s}$. Once this curve is known, the family of straight segments joining $P'_0$ to $P'_f$ with $r,s$ on ${\cal C}_{r,s}$, provides a one-parameter family of tangents to the arctic curve, which can then be determined as before. This is the content of the two-refined tangent method \cite{Sp19}. Unlike the previous, original method, no tangency assumption is made. In a sense the above ratio serves as an indicator of the location of the arctic curve.

\begin{figure}[t]
\begin{center}
\includegraphics[scale=0.6]{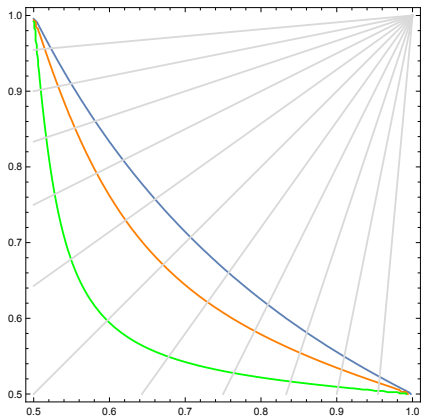}
\end{center}
\caption{Plot of the critical curve $H'(0;r,s)=0$ across which the 1-to-0 transition takes place, for different values of $\beta$: $\beta=1$ (blue), $\beta=4$ (orange) and $\beta=20$ (green). The sheaf of confluent straight lines are the curves along which  $t=\frac{s-r}{2-r-s}$ is constant; the lines shown correspond to $t$ ranging from $\frac56$ (smallest slope) down to $-\frac56$ (largest slope).} 
\label{fig7}
\end{figure}

In view of the analysis of the previous section, it is natural to guess that the two regimes described above correspond to the two cases in (\ref{Fprime}), and this is indeed what we will show: the curve ${\cal C}_{r,s}$ is precisely determined by the condition $H'(0;r,s)=0$ separating the two cases in (\ref{Fprime}),
\be
{\cal C}_{r,s}\::\; H'(0;r,s) = 0 \quad \Longleftrightarrow \quad uv(r) \, uv(s) = \Big[\frac{\sqrt{\beta}}{2(1+\beta)}\Big]^{\frac12}. \qquad \qquad {\rm (tangency\ condition)}
\ee
A plot of this curve in the $(r,s)$ plane, for different values of $\beta$, is shown in Figure \ref{fig7}.

Before further analyzing this relation and its consequences, we compute the limit in (\ref{ratio}) and show that the curve ${\cal C}_{r,s}$ is indeed characterized by the condition $H'(0;r,s)=0$. The ratio $Z_{n,rn,sn}(a,b)/Z_{n-1}(a,b)$ has been computed in Section 5.1 and given in (\ref{2-lattice}), and so the single path partition function $Z_{0,rn,sn,f}(a,b)$ is all what remains to compute. Because the path is constrained to pass though $P'_0$ and $P'_f$, it factorizes as $Z_{0,rn,sn,f}(a,b) = Z_{rn,rn}(a,b) \, Z_{(2-r-s)n,(s-r)n}(a,b) \, Z_{sn,-sn}(a,b)$ in the notations of Section 4.2, where $Z_{i,j}(a,b)$ is the partition function of a single path going from the origin to the site $(i,j)$. For distant sites, this has been computed in terms of the $L$ function. Using (\ref{L}), we obtain for large $n$,
\bea
Z_{0,rn,sn,f}(a,b) &\!\!\simeq\!\!& \exp\Big\{nr \, L(1) + n(2-r-s) \, L(t) + ns \, L(-1)\Big\}, \nonumber\\
&\!\!\simeq\!\!& \exp\Big\{n(r+s) \log\sqrt{ab} + n(2-r-s) \, L(t)\Big\}, \qquad t = \frac{s-r}{2-r-s}.
\eea
Using (\ref{2-lattice}), we can reformulate the limit (\ref{ratio}) as,
\be
\lim_{n \to \infty} \, \frac1n \, \log\frac{Z_{n,rn,sn}(a,b)}{Z_{n-1}(a,b) Z_{0,rn,sn,f}(a,b)} = F_2(r,s) - (2-r-s) \big[L(t) - \log\sqrt{ab}\big],
\label{lim}
\ee
where $F_2(r,s)$ is given in (\ref{Fprime}). In this form, the 1-0 transition in (\ref{ratio}) amounts to prove that the previous limit either vanishes or is strictly negative, depending on the values of $r,s$: it vanishes when $H'(0;r,s) \ge 0$ and is strictly negative when $H'(0;r,s) < 0$. 

We start with the case $H'(0;r,s) \ge 0$ and show that the limit (\ref{lim}) vanishes. Using the explicit expression of $F_2(r,s)$ in (\ref{Fprime}), 
\bea
F_2(r,s) + (2-r-s) \log\sqrt{ab} \egal (1-r) \log v\Big(\frac{r-\xi^*}{1-\xi^*}\Big) + (1-s) \log v\Big(\frac{s-\xi^*}{1-\xi^*}\Big) + (2-r-s) \log\sqrt{ab} \nonumber\\
&& \hspace{-5cm} = \; (2-r-s) \log\left[ab \, v\Big(\frac{r-\xi^*}{1-\xi^*}\Big)v\Big(\frac{s-\xi^*}{1-\xi^*}\Big)\right]^{\frac12} + (s-r) \log\left[v\Big(\frac{r-\xi^*}{1-\xi^*}\Big)\Big/v\Big(\frac{s-\xi^*}{1-\xi^*}\Big)\right]^{\frac12} \nonumber\\
\noalign{\smallskip}
&& \hspace{-5cm} \stackrel{?}{=} \; (2-r-s)\,L(t) = -(2-r-s) \log x(t) - (s-r) \log y(t),
\label{=?}
\eea
we have to check that the equality on the third line holds. The function $L(t)$ has been written in terms of the functions $x(t)$ and $y(t)$ discussed in Section 4.2, as the solutions of the algebraic system (\ref{Qsystem}). In order to establish (\ref{=?}), it is therefore sufficient to check that the arguments of the two logarithms in the second line of (\ref{=?}) satisfy the same algebraic equations as $1/x(t)$ and $1/y(t)$ respectively. In view of the fact that $\xi^*$ satisfies the non-linear relation (\ref{max}), this looks like a non-trivial task; however the explicit expressions we have of the functions $u$ and $v$ from Section 4.1 allow for a fairly easy check. Details are given in Appendix A.

When $H'(0;r,s) < 0$, we have to show that the limit (\ref{lim}) is strictly negative, namely,
\be
\Phi(r,s) \equiv \frac12 \log\frac{\sqrt{\beta}}{2(1+\beta)} + F_1(r) + F_1(s) - (2-r-s) \big[L(t) - \log\sqrt{ab}\big] < 0.
\ee
As a first remark, we note that $\Phi(r,s)$ vanishes on the curve ${\cal C}_{r,s}$, namely $\Phi(r,s)\big|_{H'(0;r,s)=0}=0$, because $\Phi(r,s)=0$ is equivalent to the identity (\ref{=?}) when $H'(0;r,s)=0$. To prove that $\Phi(r,s)<0$ below the curve ${\cal C}_{r,s}$, we examine the derivative of $\Phi$ along the lines of constant values of $t$. These lines are shown in Figure \ref{fig7} as straight lines pointing towards the upper right corner. We show that in the region below the curve ${\cal C}_{r,s}$, the directional derivative of $\Phi$ is strictly positive in the upwards direction and vanishes exactly on ${\cal C}_{r,s}$. Because $\Phi$ itself vanishes on ${\cal C}_{r,s}$, as noted above, this implies that $\Phi$ is strictly negative throughout the region below that curve. Details are again given in Appendix A.

It remains to actually compute the arctic curve. The straight line connecting $P'_0=(r-1,r)$ and $P'_f=(1-s,s)$ has the equation
\be
Y = \frac{s-r}{2-r-s} \, X + \frac{r+s-2rs}{2-r-s},
\label{lines}
\ee
and is tangent to the arctic curve when $(r,s)$ is on the curve ${\cal C}_{r,s}$, namely when it satisfies the tangency condition $[uv(r) \, uv(s)]^2 = \frac{\sqrt{\beta}}{2(1+\beta)}$. If we can solve it and find $s=s(r)$, we will have a family of straight lines tangent to the arctic curve.

Like in Section 4.4 when the one-refined tangent was worked out, the choice of $r$ as parameter is not optimal. Here too we will use the invertible function $v : [0,1[ \to [0,+\infty[$, to trade $r$ and $s$ for $v_r=v(r)$ and $v_s=v(s)$ by using (\ref{quart}). In addition the tangency condition is much easier to solve for $v_s$ in terms of $v_r$, and eventually yields $s = v^{-1}(v_s) = v^{-1}\big(v_s(v_r)\big)$ as a function of $v_r$, taken as the new parameter.

Our last task is thus to solve the tangency condition $[uv(r) \, uv(s)]^2 = \frac{\sqrt{\beta}}{2(1+\beta)}$ and find $v_s$ in terms of $v_r$. Altough this equation looks complicated, the calculations of Appendix A (with $\xi^*=0$) show that it is equivalent to the quadratic equation $(v_r^2 - 1) \, (v_s^2 - 1) = 2 \textstyle \frac{1+\beta}{\sqrt{\beta}} \, v_r v_s$, from which we easily determine $v_s$ as a function of $v_r$. In turn this yields $s$ as a function of $v_r$, while $r$ is already known from $v_r$. Altogether we find
\begin{subequations}
\bea
r \egal v^{-1}(v_r) = \frac {v_r}{1-v_r^2} \: \Bigg\{\frac{1+\beta}{2\sqrt{\beta}} \: \frac{1+v_r^2}{\sqrt{1+(\beta+\textstyle\frac1\beta)v_r^2+v_r^4}} - v_r\Bigg\}, \\
\noalign{\smallskip}
s \egal v^{-1}\big(v_s(v_r)\big) = \frac12 + \frac{\sqrt{\beta}}{1+\beta} \frac{\sqrt{1+(\beta+\textstyle\frac1\beta)v_r^2+v_r^4}}{v_r(1+v_r^2)}.
\eea
\end{subequations}

Substituting in (\ref{lines}) yields a family of straight lines $Y(v_r) = a(v_r)\, X(v_r) + b(v_r)$ tangent to the arctic curve with parameter $v_r \ge 1$ for $r \ge \frac 12$. One finds the following functions $a(v)$ and $b(v)$,
\begin{subequations}
\bea
a(v) \egal \frac{\frac{1+\beta}{\sqrt{\beta}}v(v^2+1) - (v^2-1) \sqrt{1+(\beta+\textstyle\frac1\beta)v^2+v^4}}{\frac{1+\beta}{\sqrt{\beta}}v(v^2+1) + (v^2-1) \sqrt{1+(\beta+\textstyle\frac1\beta)v^2+v^4}},\\
\noalign{\smallskip}
b(v) \egal \left[\frac{v^2-1}{v^2+1} + \frac{1+\beta}{\sqrt{\beta}} \frac{v}{\sqrt{1+(\beta+\textstyle\frac1\beta)v^2+v^4}}\right]^{-1}.
\eea
\end{subequations}
We check the expected slope values $a(1)=+1$ and $a(+\infty)=-1$, corresponding respectively to $r=\frac12$ and $r=1$.

The parametric form of the arctic curve itself is then given by
\be
X(v_r) = -\frac{b'(v_r)}{a'(v_r)}, \qquad Y(v_r) = a(v_r) \, X(v_r) + b(v_r).
\ee
It turns out to be exactly identical to the parametrization (\ref{param}) we found in Section 4.4.

%%%%%%%%%%%%%%%%%%%%%%%%%%%%%%%%%%%%%%%%%%%%%%%%%%%%%%%%%%%%%%%%%%%%%%%%%%

\vskip 0.5truecm
\noindent 
{\bf \large 6. General remarks}
\addcontentsline{toc}{subsection}{6. General remarks}
\setcounter{section}{6}
\setcounter{equation}{0}

\medskip
\noindent
We would like to make a few general remarks both on the results obtained in the last two sections and on the method used. Let us recall, as explained in Section 4.3, that the factorization argued in \cite{DR21} is that in the scaling limit, the ratio of partition functions $Z_n/Z_{n-1}$ where $Z_n$ may or may not be refined, is equal to the contribution of the uppermost path $Z^{\rm up}_1$. Moreover the uppermost path being a deterministic trajectory $f$ in that limit, $Z^{\rm up}_1$ can be given an explicit form in terms of the function $L(t)$ associated with the lattice paths (see Section 4.2). Altogether the arguments of \cite{DR21} leads to the following asymptotics,
\be
\lim_{n \to \infty} \frac 1n \, \log\frac{Z^{\rm ref}_{n}}{Z_{n-1}} = \lim_{n \to \infty} \frac 1n \, \log Z_1^{\rm up} = S[f] = \int_{P_0}^{P_f}  {\rm d}x \, L\big(f'(x)\big).
\label{pred0}
\ee

%%%%%%%%%%%%%%%%%%%%%%%%%%%%%%%

\vskip 0.5truecm
\noindent 
{\bf 6.1 The tangent method implies factorization}
\addcontentsline{toc}{subsubsection}{6.1 The tangent method implies factorization}

\begin{figure}[t]
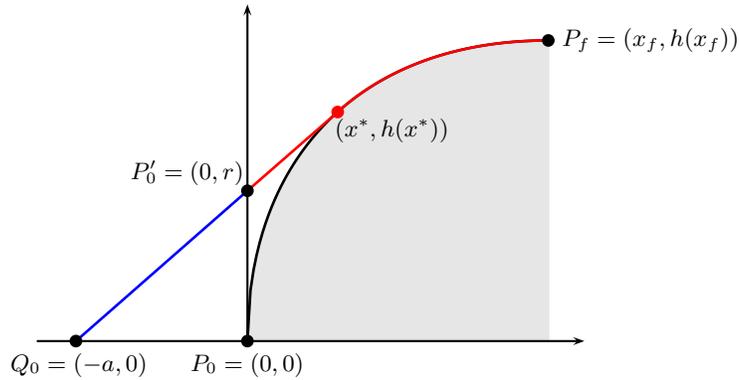

\begin{center}
\pspicture(-4,0)(5,5)
\psset{xunit=8cm}
\psset{yunit=8cm}
\rput(-0.0,-0.04){\footnotesize $P_0=(0,0)$}
\rput(-0.28,-0.04){\footnotesize $Q_0=(-a,0)$}
\rput(0.67,0.5){\footnotesize $P_f=(x_f,h(x_f))$}
\rput(-0.1,0.28){\footnotesize $P'_0=(0,r)$}
\psplot[plotpoints=100,linecolor=black,linewidth=1pt,fillstyle=solid,fillcolor=mygrey]{0}{0.5}{2 x sub x mul 3 mul sqrt 2 div x 2 div sub}
\psline[linecolor=mygrey,fillstyle=solid,fillcolor=mygrey](0,0)(0.5,0.5)(0.5,0)
\psline[linecolor=blue,linewidth=1pt](-0.285,0)(0,0.25)
\psline[linecolor=red,linewidth=1pt](0,0.25)(0.15,0.381)
\psplot[plotpoints=100,linecolor=red,linewidth=1pt]{0.15}{0.5}{2 x sub x mul 3 mul sqrt 2 div x 2 div sub}
%\psline[linecolor=red,linewidth=1.5pt](0.218,0.431)(0.33,0.5)
\pscircle[linecolor=black,fillstyle=solid,fillcolor=black](0.5,0.5){0.07}
\pscircle[linecolor=black,fillstyle=solid,fillcolor=black](0,0){0.07}
\pscircle[linecolor=black,fillstyle=solid,fillcolor=black](-0.285,0){0.07}
\pscircle[linecolor=black,fillstyle=solid,fillcolor=black](0,0.25){0.07}
%\rput(-0.03,0.328){\footnotesize $x_2$}
%\rput(0.1,0.53){\footnotesize $h(x_2)$}
%\psline[linewidth=0.5pt,linestyle=dashed,dash=3pt 2pt](0.1,0.328)(0,0.328)
%\psline[linewidth=0.5pt,linestyle=dashed,dash=3pt 2pt](0.1,0.328)(0.1,0.5)
%\rput(-0.03,0.431){\footnotesize $x_1$}
%\rput(0.22,0.53){\footnotesize $h(x_1)$}
%\psline[linewidth=0.5pt,linestyle=dashed,dash=3pt 2pt](0.22,0.431)(0,0.431)
%\psline[linewidth=0.5pt,linestyle=dashed,dash=3pt 2pt](0.22,0.431)(0.22,0.5)
\pscircle[linecolor=red,fillstyle=solid,fillcolor=red](0.15,0.381){0.07}
%\pscircle[linecolor=red,fillstyle=solid,fillcolor=red](0.22,0.431){0.07}
\rput(0.24,0.35){\footnotesize $(x^*,h(x^*))$}
%\rput(0.2,0.326){\small $(x_2,h(x_2))$}
%\psline[linecolor=blue,linewidth=1.5pt](0,0.35)(0.2,0.5)
\psline{->}(-0.35,0)(0.56,0)
\psline{->}(0,0)(0,0.56)
\endpspicture
\end{center}
\caption{Geometric setting to apply the tangent method. The portion of arctic curve from $P_0$ to $P_f$, in black and bordering the entropic region (shaded), is functionally given by $h(x)$. The vertical axis is assumed to be part of the boundary of the domain.}
\label{fig8}
\end{figure}

\smallskip
\noindent
In Section 4, we have recalled the arguments underlying the factorization property, summarized in (\ref{pred0}). Here we would like to take an alternative point of view leading to the same property, and which is based directly on the tangent method itself, worked out in the way it was originally conceived \cite{CS16}, namely with an extension of the domain. We keep the discussion as general as possible and consider a generic situation in the neighbourhood of a contact point $P_0$ between the arctic curve and the boundary of the domain, as shown in Figure \ref{fig8}.

The part of the arctic curve we focus on stretches between $P_0$ and $P_f$. In the scaling limit, the uppermost path, starting at $P_0$ and ending at $P_f$, is deterministic and condenses on the arctic curve. The original tangent method requires to move the starting point of that path from $P_0$ to some other point $Q_0$, here chosen on a horizontal line and parametrized by $a>0$. It also argues that, in the scaling limit, the displaced uppermost path is a straight line entering the domain at some point $P'_0$ and hitting the arctic curve tangentially, say at $(x^*,h(x^*))$, before following the arctic curve itself to reach $P_f$ (which has not been moved). The slope $t^*$ of the straight line can be computed from $P'_0$, and is equal to $t^* = h'(x^*)$ from the tangency condition. 

Given $Q_0$, the point $P'_0$ itself is determined as the entry point with the highest probability, in the following way. Let $Z_n^{\rm dis}(a)$ be the finite size partition function for the original model with however the uppermost path displaced as explained. At finite size, this path is random as is its entry point $P'_0$ into the domain, so that we may write
\be
Z_n^{\rm dis}(a) = \int  {\rm d}r \: Z_1(Q_0 \to P'_0) \,\times Z_n^{\rm ref}(r),
\ee
where $Z_1(Q_0 \to P'_0)$ is the partition function of a single path from $Q_0$ to $P'_0$ (in blue), and $Z_n^{\rm ref}(r)$ is the one-refined partition function for the original model in which the starting point of the uppermost path has been moved from $P_0$ to $P'_0$, on the boundary of the domain. 

Writing $Z_1(Q_0 \to P'_0) \simeq \exp\{na L(t)\}$ and $Z_n^{\rm ref}(r)/Z_{n-1} \simeq \exp\{nF_1(r)\}$ as we did before for the two-periodic Aztec diamond, we obtain
\be
\frac{Z_n^{\rm dis}(a)}{Z_{n-1}} \simeq \int  {\rm d}r \: \exp\big\{n [a L(t) + F_1(r)] \big\}, \qquad t = \frac ra.
\ee
The entry point with the highest probability is parametrized by the value of $r$ which solves the saddle point equation,
\be
F'_1(r) + L'(t) = 0.
\label{saddle}
\ee 
The principle of the tangent method is that this equation can be solved for $r$ in terms of $a$ if one can compute the proper function $L$ and the (logarithm of the) one-refined partition function $F_1(r)$ from lattice calculations; the solution then yields $t^*$ in terms of the parameter $a$. 

If however we view $t=t^*$ as the input, the saddle point equation gives a differential equation for $F_1(r)$. The main observation here is that its solution is precisely the integral form derived from the factorization assumption. In the present setting, this integral form reads
\be
S[f] = \int_{P'_0}^{P_f} {\rm d}x \, L\big(f'(x)\big) = \int_0^{x^*} {\rm d}x \, L(t^*) + \int_{x^*}^{P_f} {\rm d}x \, L\big(h'(x)\big) = x^* \, L(t^*) + \int_{x^*}^{P_f} {\rm d}x \, L\big(h'(x)\big),
\label{integral}
\ee
where $f(x)$ is the red trajectory in Figure \ref{fig8}, which coincides with the arctic curve from $x^*$ to $P_f$. To see this, we check that $S[f]$ given in the previous equation satisfies the saddle point equation (\ref{saddle}), and takes the correct value at some specific value of $r$. 

The dependence of $S[f]$ on $r$ is all contained in $x^*$ and $t^*$ so that the first derivative is
\be
\frac{{\rm d}S[f]}{{\rm d}r} = L(t^*) \frac{{\rm d}x^*}{{\rm d}r} + x^* \, L'(t^*) \frac{{\rm d}t^*}{{\rm d}r} - L(t^*) \frac{{\rm d}x^*}{{\rm d}r} = -L'(t^*).
\ee
The last equality follows from the fact that the red straight line from $P'_0=(0,r)$ to $(x^*,h(x^*))$ has slope $t^* = h'(x^*)$, implying
\be
h(x^*) = t^* x^* + r \quad \stackrel{\frac{{\rm d}}{{\rm d}r}}{\Longrightarrow} \quad h'(x^*) \frac{{\rm d}x^*}{{\rm d}r} = t^* \frac{{\rm d}x^*}{{\rm d}r} + \frac{{\rm d}t^*}{{\rm d}r} x^* + 1 \quad \Longrightarrow \quad x^* \frac{{\rm d}t^*}{{\rm d}r} = -1.
\ee
Moreover the integral form (\ref{integral}) yields the boundary value $S[f]\big|_{r_{\rm max}} = x_f L(t_f)$ at the largest possible value of $r$, the one for which the tangency point is $P_f$, with slope $t_f$. This is the correct value since in this case, $Z_n^{\rm ref}(r)/Z_{n-1}$ reduces, at dominant order (and sometimes exactly), to the partition function for a single path going from $(0,r_{\rm max})$ to $P_f$. As recalled above, it is given asymptotically by $\exp\{nx_f L(t_f)\}$ and indeed yields the value $F_1(r_{\rm max}) = x_fL(t_f)$. We thus conclude that $F_1(r) = S[f]$.

Therefore the factorization property and the integral form for the contribution of the uppermost path are both recovered in the scaling limit, from a direct use of the geometric tangent method, with no further assumption. By factorizing the uppermost path in systems of decrasing size $n$, $n-1$, $n-2$, ..., one can factorize $m = {\cal O}(1)$ uppermost paths. The resulting multiple contribution is simply additive with respect to the individual paths,
\be
\lim_{n \to \infty} \frac 1n \, \log\frac{Z^{\rm ref}_{n}(r_1,r_2,\ldots,r_m)}{Z_{n-m}} = F_1(r_1) + F_1(r_2) + \ldots + F_1(r_m) = S[f_1] + \ldots + S[f_m].
\label{add}
\ee
Rather than factorizing one uppermost path at a time, one may consider the $m$-refined free energy by moving the starting points of the $m$ uppermost paths. The saddle point analysis in this case readily yields a system of partial differential equations whose solution is the additive formula (\ref{add}).

The above result extends to cases where the paths are $q$-weighted, that is, when a path $p$, in addition to the weights of the elementary steps it is made of, receives a weight equal to $q^{A(p)}$, with $A(p)$ the area below $p$ \cite{DFG19}. Provided $q$ is sufficiently close to 1, the proper scaling rate being $q={\rm e}^{\lambda/n}$, the tangent method works in exactly the same way as when $q=1$, the only difference being that straight lines are to be replaced by curved geodesics $g(x)$ whose shape depends on $\lambda$: in the scaling limit, a free path constrained to start at $P_0$ and end at $P_f$ will almost surely follow the geodesic passing by $P_0$ and $P_f$. The geodesic equation is explicitly given by
\be
\frac{{\rm d}}{{\rm d}x} L'\big(g'(x)\big) = \lambda,
\ee
whose general solution depends on two integration constants $C_1$ and $C_2$ \cite{DGR19}
\be
g(x) = C_2 + \int_0^x {\rm d}u \, y^{-1}\big({\rm e}^{-\lambda u - C_1}\big), \qquad L'(t) = -\log y(t).
\label{geo}
\ee
It is worth noting that in the above expressions, the dependence in $\lambda$ has been made completely explicit. The function $L(t)$ and the associated function $y(t)$, discussed in Section 4.2 in the two-periodic Aztec diamonds, only depend on the elementary steps making the paths and their weights, and do not depend at all on $\lambda$. 

The arguments showing the tangency property in the tangent method have also been extended to this case \cite{DGR19}. Therefore Figure \ref{fig8} remains fully valid if the straight segment from $(-a,0)$ to $(x^*,h(x^*))$ is replaced by a curved geodesic, concave for $\lambda > 0$ and convex for $\lambda < 0$ (large area paths are favoured or disfavoured respectively). The computation of the entry point $P'_0$ is done in a similar way and simply requires the expression of the partition function for a single path. This has been computed in \cite{DGR19}, where it was shown that the contributions of all lattice paths which condense in the scaling limit onto a given trajectory $f$ starting at $P_0$ and ending at $P_f$ is given by
\be
Z_1[f] \simeq \exp\Big\{n \int_{P_0}^{P_f}  {\rm d}x \, L\big(f'(x)\big) + \lambda n \int_{P_0}^{P_f} {\rm d}x \, f(x)\Big\}.
\label{Zq}
\ee

In the present case, we need the partition function $Z_1(Q_0 \to P'_0)$ for a free path going from $Q_0$ to $P'_0$. Since such paths condense almost surely onto the geodesic $g_r$ passing through these two points, we obtain
\be
Z_1(Q_0 \to P'_0) \simeq \exp\Big\{n \int_{-a}^0  {\rm d}x \, L\big(g'_r(x)\big) + \lambda n \int_{-a}^0 {\rm d}x \, g_r(x)\Big\}.
\ee

This is all we need to analyze the saddle point. Letting $Z_n^{\rm ref}(r)/Z_{n-1} \simeq \exp\{nF_1(r)\}$ as before (implicit dependence on $\lambda$), the saddle point equation reads
\be 
F'_1(r) + \int_{-a}^0  {\rm d}x \, L'\big(g'_r(x)\big) \, \partial_r g'_r(x) + \lambda \int_{-a}^0 {\rm d}x \, \partial_r g_r(x) = 0.
\ee
The first integral may be simplified by writing $\partial_r g'_r(x) = \partial_x \partial_r g_r(x)$ and integrating by parts. Using the geodesic equation satisfied by $g_r(x)$, the remaining integrals cancel out leaving only the boundary terms,
\be
F'_1(r) + L'\big(g'_r(0)\big) \, \partial_r g_r(0) - L'\big(g'_r(-a)\big) \, \partial_r g_r(-a) = 0.
\ee
The second term at $x=-a$ can be seen to vanish by using the explicit form (\ref{geo}) of $g_r$. Indeed taking into account the two conditions $g_r(-a)=0$ and $g_r(0)=r$, one finds
\be
g_r(x) = \int_{-a}^x {\rm d}u \, y^{-1}\big({\rm e}^{-\lambda u - C_1(r)}\big).
\ee
The derivative with respect to $r$ is then
\bea
\partial_r g_r(x) \egal - C'_1(r) \int_{-a}^x {\rm d}u \, \big(y^{-1}\big)'\big({\rm e}^{-\lambda u - C_1(r)}\big) \, {\rm e}^{-\lambda u - C_1(r)} \nonumber\\
\egal \frac{C'_1(r)}\lambda  \int_{-a}^x {\rm d}u \, \frac{{\rm d}}{{\rm d}u} y^{-1}\big({\rm e}^{-\lambda u - C_1(r)}\big) = \frac{C'_1(r)}\lambda  \Big[y^{-1}\big({\rm e}^{-\lambda u - C_1(r)}\big)\Big]_{-a}^x.
\eea
It implies $\partial_rg_r(-a)=0$ so that the saddle point equation takes the final form,
\be
F'_1(r) + L'\big(g'_r(0)\big) \, \partial_r g_r(0) = 0.
\label{saddleq}
\ee

For $q$-weighted paths, the uppermost path, constrained to start at $P'_0$ and to end at $P_f$, will almost surely condense. Therefore its contribution, by applying (\ref{Zq}), ought to be given by
\be
S[f] \equiv \lim_{n \to \infty} \frac 1n \, \log\frac{Z^{\rm ref}_{n}}{Z_{n-1}} = \int_{P'_0}^{P_f}  {\rm d}x \, L\big(f'(x)\big) + \lambda \int_{P'_0}^{P_f} {\rm d}x \, f(x),
\label{predq}
\ee
where $f(x)$ is the ${\cal C}^1$ deterministic trajectory followed by the uppermost path in the scaling limit, made of a portion of geodesic and a piece of arctic curve, both being $\lambda$-dependent.

In the setting of Figure \ref{fig8}, the previous expression becomes
\be
S[f] = \int_0^{x^*}  {\rm d}x \, L\big(g'_r(x)\big) + \lambda \int_0^{x^*} {\rm d}x \, g_r(x) + \int_{x^*}^{P_f}  {\rm d}x \, L\big(h'(x)\big) + \lambda \int_{x^*}^{P_f} {\rm d}x \, h(x),
\label{integralq}
\ee
with $g_r$ the geodesic passing through $P'_0=(0,r)$ and tangent to the arctic curve at $x^*$, and $h$ the arctic curve. Similarly to what we did for $\lambda=0$, we want to show that $S[f] = F_1(r)$, and in particular that $S[f]$ satisfies the saddle point equation (\ref{saddleq}).

The first derivative yields
\bea
\frac{{\rm d}S[f]}{{\rm d}r} \egal L\big(g'_r(x^*)\big) \frac{{\rm d}x^*}{{\rm d}r} + \int_0^{x^*}  {\rm d}x \, L'\big(g'_r(x)\big) \partial_rg'_r(x) + \lambda g_r(x^* ) \frac{{\rm d}x^*}{{\rm d}r} + \lambda \int_0^{x^*} {\rm d}x \, \partial_r g_r(x) \nonumber\\
&& \hspace{0mm} - \: L\big(h'(x^*)\big) \frac{{\rm d}x^*}{{\rm d}r} - \lambda h(x^*) \frac{{\rm d}x^*}{{\rm d}r}.
\eea
Because the geodesic $g_r$ and the arctic curve merge tangentially at $x^*$, the third and sixth terms cancel each other, as do the first and fifth terms. The second term can be integrated by parts, as we did above, cancelling the fourth term upon using the geodesic equation, and only leaves two boundary terms,
\be
\frac{{\rm d}S[f]}{{\rm d}r} = L'\big(g'_r(x^*)\big) \, \partial_r g_r(x^*) - L'\big(g'_r(0)\big) \, \partial_r g_r(0).
\ee
Differentiating the merging condition $g_r(x^*) = h(x^*)$ with respect to $r$ and using once more the tangency condition gives $\partial_r g_r(x^*) = 0$. Therefore $S[f]$ satisfies the saddle point equation. 

Finally for the maximal value $r_{\rm max}$ of $r$, such that $x^*=x_f$, the value 
\be
S[f]\Big|_{r_{\rm max}} = \int_0^{x_f}  {\rm d}x \, L\big(g'_r(x)\big) + \lambda \int_0^{x_f} {\rm d}x \, g_r(x)
\ee
is the correct asymptotic value of $\lim_{n \to \infty} \frac 1n \, \log\frac{Z^{\rm ref}_{n}}{Z_{n-1}}(r_{\rm max})$. This proves the identity $F_1(r) = S[f]$.

%%%%%%%%%%%%%%%%%%%%%%%%%%%%%%%

\vskip 0.5truecm
\noindent 
{\bf 6.2 Application to the two-refined tangent method}
\addcontentsline{toc}{subsubsection}{6.2 Application to the two-refined tangent method}

\smallskip
\noindent
We note here that the 1-to-0 transition at the heart of the two-refined tangent method used in Section 5 readily follows from (\ref{pred0}), because it implies the following identity,
\be
\lim_{n \to \infty} \frac 1n \frac{Z_{n,rn,sn}(a,b)}{Z_{n-1}(a,b)} \stackrel{\footnotesize (\ref{2-lattice})}{=} F_2(r,s) + \log(ab) = \int_{P_0}^{P_f}  {\rm d}x \, L\big(f'(x)\big).
\label{pred}
\ee

Indeed if $r,s$ are sufficiently large so that $H'(0;r,s) \ge 0$, the uppermost trajectory $f$ is the blue line in the left panel of Figure \ref{fig6}, made of three rectilinear sections. The integral of $L(f')$ is trivial,
\bea
\int_{P_0}^{P_f}  {\rm d}x \, L\big(f'(x)\big) \egal \int_{-1}^1 {\rm d}x \, L\big(f'(x)\big) = \int_{-1}^{r-1} {\rm d}x \, L(1) + \int_{r-1}^{1-s} {\rm d}x \, L(t) + \int_{1-s}^1 {\rm d}x \, L(-1) \nonumber\\
\egal (r+s) \log\sqrt{ab} + (2-r-s) L(t), \qquad \quad t=\frac{s-r}{2-r-s},
\eea
and immediately yields the identity (\ref{=?}), which we have proved in Section 5 without using (\ref{pred}). %This identity (\ref{pred}) is non-trivial and somewhat surprising. We nonetheless proved it in the previous section and Appendix A. 

When $r,s$ are such that $H'(0;r,s)<0$, the uppermost trajectory $f$ is the red curve in the right panel of Figure \ref{fig6}, and includes a portion of the arctic curve. This makes the part of the integral of $L(f')$ from $P'_0$ to $P'_f$ difficult to evaluate. However it directly follows from the variational analysis in \cite{DGR19} that the integral 
\be
\int_{P'_0}^{P'_f}  {\rm d}x \, L\big(f'(x)\big)
\ee 
has a unique global maximum $f_{\rm max}$ in the set of continuous and piecewise $C^1$ functions, attained for the straight line connecting $P'_0$ and $P'_f$. In the case at hand, $f$ is not the straight line (which would cross the arctic curve), so we immediately get 
\be
\int_{P'_0}^{P'_f}  {\rm d}x \, L\big(f'(x)\big) < \int_{P'_0}^{P'_f}  {\rm d}x \, L\big(f_{\rm max}'(x)\big) < \int_{r-1}^{1-s}  {\rm d}x \, L(t) = (2-r-s) L(t).
\ee
The inequality, proved in Appendix A without using (\ref{pred}),
\bea
F_2(r,s) + \log(ab) \egal \int_{P_0}^{P_f}  {\rm d}x \, L\big(f'(x)\big) = (r+s) \log\sqrt{ab} + \int_{P'_0}^{P'_f}  {\rm d}x \, L\big(f'(x)\big) \nonumber\\
&\!\!\!<\!\!\!& \,(r+s) \log\sqrt{ab} + (2-r-s) L(t),
\eea
therefore readily follows form the factorization property.

%%%%%%%%%%%%%%%%%%%%%%%%%%%%%%%

\vskip 0.5truecm
\noindent 
{\bf 6.3 More multirefinements}
\addcontentsline{toc}{subsubsection}{6.3 More multirefinements}

\smallskip
\noindent
Our last remark concerns the value of $F_2(r,s)$, found in (\ref{Fprime}) in the case $H'(0;r,s) < 0$, namely
\be
F_2(r,s) = \frac12 \log\frac{\sqrt{\beta}}{2(1+\beta)} + F_1(r) + F_1(s), \qquad H'(0;r,s) < 0.
\label{F2}
\ee
It is surprisingly much simpler than in the case $H'(0;r,s)>0$, and strangely is the sum of two decoupled terms, one in $r$ and one in $s$. Moreover that function $F_1(z)$ is exactly the one controlling the asymptotic value of the one-refined partition function (this was also observed in other models \cite{DR21}). This observation is again fully consistent with the factorization discussed above and the specific value of $Z_1^{\rm up}$, as the integral of $L(f')$, see (\ref{pred0}).

To see this, one can symbolically decompose the integral into three terms as in the following picture, where each term corresponds to the integral of $L($derivative of the red curve$)$. 

\begin{figure}[!h]
\begin{center}
\psset{unit=.33cm}
%\setlength{\unitlength}{.7cm}
%
%\hspace{-1.5cm}
\pspicture(-5,0)(5,5.5)
%\pscircle[linecolor=black,fillstyle=solid,fillcolor=black](0,0){0.07}
\psarc[linewidth=0.1pt,linecolor=black,fillstyle=solid,fillcolor=mygrey](0,-5){7.08}{45}{135}
\psarc[linewidth=1.2pt,linecolor=red](0,-5){7.08}{85}{103.2}
%\rput(-3.9,1.65){\footnotesize $P'_0$}
%\rput(3.7,1.85){\footnotesize $P'_f$}
\psline[linewidth=0.5pt,linecolor=black](-5,0)(0,5)(5,0)
\psline[linewidth=1.2pt,linecolor=red](-5.5,-0.5)(-3.5,1.5)(-1.6,1.9)
\psline[linewidth=1.2pt,linecolor=red](0.6,2.055)(3.2,1.8)(5.5,-0.5)
\rput(6.5,2){\small $=$}
\endpspicture
\hspace{1cm}
\pspicture(-5,0)(5,5.5)
%\pscircle[linecolor=black,fillstyle=solid,fillcolor=black](0,0){0.07}
\psarc[linewidth=0.1pt,linecolor=black,fillstyle=solid,fillcolor=mygrey](0,-5){7.08}{45}{135}
\psarc[linewidth=1.2pt,linecolor=red](0,-5){7.08}{45}{103.2}
%\rput(-3.9,1.65){\footnotesize $P'_0$}
%\rput(3.7,1.85){\footnotesize $P'_f$}
\psline[linewidth=0.5pt,linecolor=black](-5,0)(0,5)(5,0)
\psline[linewidth=1.2pt,linecolor=red](-5.5,-0.5)(-3.5,1.5)(-1.6,1.9)
\psline[linewidth=1.2pt,linecolor=red](5,0)(5.5,-0.5)
\rput(6.5,2){\small $+$}
\endpspicture
\hspace{1cm}
\pspicture(-5,0)(5,5.5)
%\pscircle[linecolor=black,fillstyle=solid,fillcolor=black](0,0){0.07}
\psarc[linewidth=0.1pt,linecolor=black,fillstyle=solid,fillcolor=mygrey](0,-5){7.08}{45}{135}
\psarc[linewidth=1.2pt,linecolor=red](0,-5){7.08}{85}{135}
%\rput(-3.9,1.65){\footnotesize $P'_0$}
%\rput(3.7,1.85){\footnotesize $P'_f$}
\psline[linewidth=0.5pt,linecolor=black](-5,0)(0,5)(5,0)
\psline[linewidth=1.2pt,linecolor=red](-5.5,-0.5)(-5,0)
\psline[linewidth=1.2pt,linecolor=red](0.6,2.055)(3.2,1.8)(5.5,-0.5)
\rput(6.5,2){\small $-$}
\endpspicture
\hspace{1cm}
\pspicture(-5,0)(5,5.5)
%\pscircle[linecolor=black,fillstyle=solid,fillcolor=black](0,0){0.07}
\psarc[linewidth=0.1pt,linecolor=black,fillstyle=solid,fillcolor=mygrey](0,-5){7.08}{45}{135}
\psarc[linewidth=1.2pt,linecolor=red](0,-5){7.08}{45}{135}
%\rput(-3.9,1.65){\footnotesize $P'_0$}
%\rput(3.7,1.85){\footnotesize $P'_f$}
\psline[linewidth=0.5pt,linecolor=black](-5,0)(0,5)(5,0)
\psline[linewidth=1.2pt,linecolor=red](-5.5,-0.5)(-5,0)
\psline[linewidth=1.2pt,linecolor=red](5,0)(5.5,-0.5)
\endpspicture
\end{center}
\end{figure}

The red curve in the l.h.s. is the trajectory we called $f$ in the above discussion of the two-refined case. The integral of $L(f')$ from $P_0=(-1,0)$ to $P_f=(1,0)$ yields $F_2(r,s)+\log(ab)$ from (\ref{pred}). The first two terms in the r.h.s. correspond to one-refined situations, parametrized by $r$ on the NW boundary in the first term, and by $s$ on the NE boundary in the second one. According to (\ref{pred0}), these two terms yield respectively $F_1(r)+\log(ab)$ and $F_1(s)+\log(ab)$, from (\ref{lattice}). The last term corresponds to the unrefined case; using (\ref{pred0}) once more, the integral in that case equals the asymptotic value of $\frac1n \log Z_n(a,b)/Z_{n-1}(a,b)$, equivalently the value of the one-refined case at $r=\frac12$,
\be
\int_{-1}^{-\frac12} {\rm d}x \, L(1) + \int_{-\frac12}^{\frac12} {\rm d}x \, L\big(h'(x)\big) + \int_{\frac12}^1 {\rm d}x \, L(-1) = \log(ab) + F_1(\textstyle \frac12) = \frac12 \log\frac{2(1+\beta)}{\sqrt{\beta}} + \log(ab),
\label{integ}
\ee             
where $h(x)$ is the functional form of the arctic curve. The value of $F_1(\frac12)$ was computed at the end of Section 4.1. 

Altogether the pictorial equation above yields
\be
F_2(r,s) + \log(ab) = \Big[F_1(r) + \log(ab)\Big] + \Big[F_1(s) + \log(ab)\Big] - \Big[\textstyle \frac12 \log\frac{2(1+\beta)}{\sqrt{\beta}} + \log(ab)\Big],
\ee
and is precisely the form given in (\ref{F2}). We also note that, from (\ref{integ}), we get the value of the integral along the arctic curve, between the two contact points,
\be
\int_{-\frac12}^{\frac12} {\rm d}x \, L\big(h'(x)\big) = \frac12 \log\frac{2ab(1+\beta)}{\sqrt{\beta}},
\ee
a result which is remarkably simple and not manifestly easy to prove directly, given the complexity of the $L$ function and of the arctic curve itself. 

In a more general context, whenever the uppermost trajectory touches the arctic curve, the above relation between the two- and one-refined partition functions should be valid, as long as the identity (\ref{pred0}) holds. 

Altogether the identity (\ref{pred0}) provides a proper basis to believe that, {\it asymptotically, multirefined partition functions all reduce to one-refined ones}. For those trajectories which do not touch the arctic curve, the situation is combinatorially much simpler since it amounts to compute the appropriate $L$ function.

%%%%%%%%%%%%%%%%%%%%%%%%%%%%%%%%%%%%%%%%%%%%%%%%%%%%%%%%%%%%%%%%%%%%%%%%%%

\eject
\vskip 0.5truecm
\noindent 
{\bf \large 7. Discussion}
\addcontentsline{toc}{subsection}{7. Discussion}
\setcounter{section}{7}
\setcounter{equation}{0}

\medskip
\noindent
As mentioned earlier, two-periodic Aztec diamonds have been considered in \cite{DFSG14} (called $2 \times 2$ periodic), but the authors generalized this setting to measures with higher periodicity, called $m$-toroidal periodic or $2 \times 2m$ periodic Aztec diamonds. These have been revisited in \cite{Be19} where, relying on the formulation in terms of non-intersecting lattice paths, the kernels of the underlying determinantal point processes were computed and analyzed. In this last section, %we argue that no conceptual obstacle should prevent the application of the tangent methods to these higher periodic cases; at worse, when $k$ gets large, one could encounter difficulties in asymptotic computations due to the presence of high degree polynomials in the denominators of the generating functions. In fact the tangent methods should be as amenable to exact calculations as the methods used in \cite{DFSG14}.
we examine the applicability of the tangent method to the Aztec diamonds with higher periodicity; as explained below, we see no major obstacle to its application.

\begin{figure}[t]
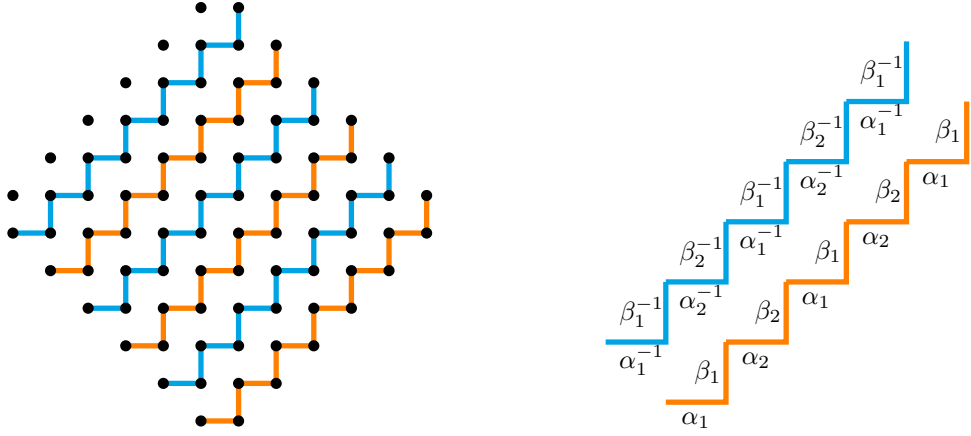

\begin{center}
\psset{unit=.5cm}
\pspicture(-2,0)(10,12)
\multido{\nt=0+1}{6}{\rput(\nt,\nt){\psline[linewidth=2pt,linecolor=cerulean](0.5,4.5)(1.5,4.5)(1.5,5.5)}}
\multido{\nt=1+1}{6}{\rput(\nt,\nt){\psline[linewidth=2pt,linecolor=orange](0.5,2.5)(1.5,2.5)(1.5,3.5)}}
\multido{\nt=2+1}{6}{\rput(\nt,\nt){\psline[linewidth=2pt,linecolor=cerulean](0.5,0.5)(1.5,0.5)(1.5,1.5)}}
\multido{\nt=3+1}{6}{\rput(\nt,\nt){\psline[linewidth=2pt,linecolor=orange](0.5,-1.5)(1.5,-1.5)(1.5,-0.5)}}
\multido{\nt=4+1}{6}{\rput(\nt,\nt){\psline[linewidth=2pt,linecolor=cerulean](0.5,-3.5)(1.5,-3.5)(1.5,-2.5)}}
\multido{\nt=5+1}{6}{\rput(\nt,\nt){\psline[linewidth=2pt,linecolor=orange](0.5,-5.5)(1.5,-5.5)(1.5,-4.5)}}
\multido{\nt=4+1}{2}{\rput(0,\nt){\pscircle[linecolor=black,fillstyle=solid,fillcolor=black](0.5,0.5){0.12}}}
\multido{\nt=3+1}{4}{\rput(1,\nt){\pscircle[linecolor=black,fillstyle=solid,fillcolor=black](0.5,0.5){0.12}}}
\multido{\nt=2+1}{6}{\rput(2,\nt){\pscircle[linecolor=black,fillstyle=solid,fillcolor=black](0.5,0.5){0.12}}}
\multido{\nt=1+1}{8}{\rput(3,\nt){\pscircle[linecolor=black,fillstyle=solid,fillcolor=black](0.5,0.5){0.12}}}
\multido{\nt=0+1}{10}{\rput(4,\nt){\pscircle[linecolor=black,fillstyle=solid,fillcolor=black](0.5,0.5){0.12}}}
\multido{\nt=-1+1}{12}{\rput(5,\nt){\pscircle[linecolor=black,fillstyle=solid,fillcolor=black](0.5,0.5){0.12}}}
\multido{\nt=-1+1}{12}{\rput(6,\nt){\pscircle[linecolor=black,fillstyle=solid,fillcolor=black](0.5,0.5){0.12}}}
\multido{\nt=0+1}{10}{\rput(7,\nt){\pscircle[linecolor=black,fillstyle=solid,fillcolor=black](0.5,0.5){0.12}}}
\multido{\nt=1+1}{8}{\rput(8,\nt){\pscircle[linecolor=black,fillstyle=solid,fillcolor=black](0.5,0.5){0.12}}}
\multido{\nt=2+1}{6}{\rput(9,\nt){\pscircle[linecolor=black,fillstyle=solid,fillcolor=black](0.5,0.5){0.12}}}
\multido{\nt=3+1}{4}{\rput(10,\nt){\pscircle[linecolor=black,fillstyle=solid,fillcolor=black](0.5,0.5){0.12}}}
\multido{\nt=4+1}{2}{\rput(11,\nt){\pscircle[linecolor=black,fillstyle=solid,fillcolor=black](0.5,0.5){0.12}}}
\endpspicture
%%%%%%
\hspace{1cm}
\psset{unit=.8cm}
\pspicture(0,0)(10,6)
\multido{\nt=2+1}{5}{\rput(\nt,\nt){\psline[linewidth=2pt,linecolor=cerulean](0.5,-1)(1.5,-1)(1.5,0)}}
\rput(-1.1,1){
\rput(4.2,-.3){\small $\alpha^{-1}_1$}
\rput(5.2,.7){\small $\alpha^{-1}_2$}
\rput(6.2,1.7){\small $\alpha^{-1}_1$}
\rput(7.2,2.7){\small $\alpha^{-1}_2$}
\rput(8.2,3.7){\small $\alpha^{-1}_1$}
\rput(4.2,.5){\small $\beta^{-1}_1$}
\rput(5.2,1.5){\small $\beta^{-1}_2$}
\rput(6.2,2.5){\small $\beta^{-1}_1$}
\rput(7.2,3.5){\small $\beta^{-1}_2$}
\rput(8.2,4.5){\small $\beta^{-1}_1$}}
\multido{\nt=3+1}{5}{\rput(\nt,\nt){\psline[linewidth=2pt,linecolor=orange](0.5,-3.)(1.5,-3.)(1.5,-2.)}}
\rput(4,-.3){\small $\alpha_1$}
\rput(5,.7){\small $\alpha_2$}
\rput(6,1.7){\small $\alpha_1$}
\rput(7,2.7){\small $\alpha_2$}
\rput(8,3.7){\small $\alpha_1$}
\rput(4.2,.5){\small $\beta_1$}
\rput(5.2,1.5){\small $\beta_2$}
\rput(6.2,2.5){\small $\beta_1$}
\rput(7.2,3.5){\small $\beta_2$}
\rput(8.2,4.5){\small $\beta_1$}
\endpspicture
\end{center}
\caption{Assignment of weights to the edges of the Aztec graph with $2 \times 2m$ periodicity. Along any broken orange line, the weights $\alpha_1,\beta_1,\alpha_2,\beta_2,\ldots ,\alpha_m,\beta_m,\alpha_1,\beta_1,\ldots$ are successively assigned to the edges starting with the lower horizontal edge and moving in the NE direction; in this way horizontal edges get an $\alpha$-weight, vertical edges a $\beta$-weight. Similarly on a broken blue line, the inverse weights are assigned in the same order $\alpha_1^{-1},\beta_1^{-1},\alpha_2^{-1},\beta_2^{-1},\ldots ,\alpha_m^{-1},\beta_m^{-1},\alpha_1^{-1},\beta_1^{-1},\ldots$ The right panel illustrates the basic weight assignment on two neighbouring diagonals of a $2\times 4$ periodic Aztec graph of order $n=5$ (to complete it, one just adds three more broken lines to the left).}
\label{fig9}
\end{figure}

As defined in \cite{DFSG14}, the most general $2 \times 2m$ periodic Aztec diamonds depend on $4m$ face parameters, but some of them can be gauged away: local transformations reduce this number to $2m-1$ while just affecting the measure by an irrelevant overall factor. A convenient gauging has been given in \cite{Be19} in terms of $2m$ edge weights $\alpha_1,\ldots,\alpha_m$ and $\beta_1, \ldots, \beta_m$, and their inverse, all related by the single relation $
\prod_i \alpha_i = \prod_i \beta_i$. These weights are periodically assigned to horizontal and vertical edges along parallel diagonals, and alternating with their inverse from one diagonal to the next one, see Figure \ref{fig9}. The uncolored edges get a weight 1. For $m=1$, the weights are all equal to $\alpha_1$ along the orange diagonals, and to $\alpha_1^{-1}$ along the blue ones, with the consequence that any flip of two parallel dimers is a weight preserving transformation; that case is therefore equivalent to a uniform distribution. The two-periodic case considered in this article, and shown in Figure \ref{fig1}, corresponds to $m=2$ and $(\alpha_1,\beta_1,\alpha_2,\beta_2)=(\frac ab,\frac ab,\frac ba,\frac ba)$.

If $T_n(\alpha_1,\beta_1,\ldots,\alpha_m,\beta_m)$ denotes the partition function for the $2 \times 2m$ periodic Aztec diamond of order $n$ with the edge weights defined above (and no face weights), the generalized octahedron recurrence \cite{Sp07} reads
\small
\bea
&& \!\!\!\!\!\!\! T_n(\alpha_1,\beta_1,\ldots,\alpha_m,\beta_m) \, T_{n-2}(\alpha_2^{-1},\beta_2^{-1},\ldots,\alpha_1^{-1},\beta_1^{-1}) = \alpha_1 \, T_{n-1}(\alpha_1^{-1},\beta_1^{-1},\ldots,\alpha_m^{-1},\beta_m^{-1}) \, T_{n-1}(\alpha_2,\beta_2,\ldots,\alpha_1,\beta_1) \nonumber\\ 
\noalign{\smallskip}
&& \hspace{2cm} + \; \beta_{n \bmod m} \: T_{n-1}(\alpha_1,\beta_1,\ldots,\alpha_m,\beta_m) \, T_{n-1}(\alpha_2^{-1},\beta_2^{-1},\ldots,\alpha_1^{-1},\beta_1^{-1}),
\label{mocta}
\eea
with initial conditions,
\be
T_0 = 1, \qquad T_1(\alpha_1,\beta_1) = \alpha_1 + \beta_1.
\ee

For what follows, it is convenient to extend periodically the finite sequences $(\alpha_i)_{1 \le i \le m}$ and $(\beta_i)_{1 \le i \le m}$ to any $i \in \N$, by letting $\alpha_{i+m}=\alpha_i$ and $\beta_{i+m}=\beta_i$. Also define the following infinite sequence $(X_j)_{j \ge 1}$ by setting
\be
X_{2i-1} = \alpha_i + \beta_i, \qquad X_{2i} = \alpha_{i+1}^{-1} + \beta_i^{-1}, \qquad (i \ge 1).
\ee
The sequence $(X_j)$ inherits a periodicity which is twice that of the $\alpha_i,\beta_i$, namely $X_j = X_{j+2m}$.

As noted in \cite{DFSG14}, the solution of the octahedron recurrence quite remarkably takes a maximally factorized form, for any $m$. In the present parametrization of the measure, it takes the following simple form,
\be
T_n(\alpha_1,\beta_1,\ldots) = \prod_{i=1}^n \: \prod_{j=0}^{i-1} \: X_{i+j} = X_n^{\lfloor \!\frac{n+1}2 \!\rfloor} \: \prod_{i=1}^{n-1} \: \big(X_i \, X_{2n-i}\big)^{\lfloor \!\frac{i+1}2 \!\rfloor}.
\label{solX}
\ee
Equivalently successive terms satisfy the simple recurrence
\be
\frac{T_n(\alpha_1,\beta_1,\ldots)}{T_{n-1}(\alpha_1,\beta_1,\ldots)} = X_n \, X_{n+1} \ldots X_{2n-1}, \qquad n \ge 1.
\ee
For $n=8$ and $m=3$ for instance, one obtains
\bea
T_8(\alpha_1,\beta_1,\alpha_2,\beta_2,\alpha_3,\beta_3) \egal (X_1X_2)\,(X_3X_4)^2 \, (X_5X_6)^3 \,(X_7X_8X_9)^4 \,(X_{10}X_{11})^3 \, (X_{12}X_{13})^2 \,(X_{14}X_{15}) \nonumber\\
\noalign{\smallskip}
&& \hspace{-3cm} = \: (\alpha_1\alpha_3\beta_2\beta_3)^{-5} \, (\alpha_2\beta_1)^{-6} \, (\alpha_1+\beta_1)^7 \, (\alpha_2+\beta_1)^6 \, (\alpha_2+\beta_2)^7 \, (\alpha_3+\beta_2)^5 \, (\alpha_3+\beta_3)^6 \, (\alpha_1+\beta_3)^5, 
\eea
where the second line is obtained by reduction modulo 6.

In order to apply the geometric tangent method, one should be able to compute (a) the one-refined partition functions, and (b) the partition function for a single path, weighted in such a way that the periodic weight of a perfect matching equals the total weight of the corresponding non-intersecting $n$ paths.

By assigning extra face variables\footnote{Alternatively, and since we are using here exclusively edge weights, we can assign a weight $x$ to each of the $n$ vertical edges along the NW boundary. The full partition function is then a polynomial in $x$ of degree $n$, $T_n(\alpha_1,\beta_1,\ldots|x) = \sum_{k=0}^n \: x^k \, T_{n,k}(\alpha_1,\beta_1,\ldots)$.} along the NW boundary as in Section 4, one readily obtain that the one-refined partition functions $T_{n,k}(\alpha_1,\beta_1,\ldots)$ (with $k$ the number of vertical dominos along the NW boundary) satisfy
\small
\bea
T_{n,k}(\alpha_1,\beta_1,\ldots) \, T_{n-2}(\alpha_2^{-1},\beta_2^{-1},\ldots) \egal \alpha_1 \, T_{n-1,k}(\alpha_1^{-1},\beta_1^{-1},\ldots) \, T_{n-1}(\alpha_2,\beta_2,\ldots) \nonumber\\ 
\noalign{\smallskip}
&& \hspace{1cm} + \; \beta_n \: T_{n-1}(\alpha_1,\beta_1,\ldots) \, T_{n-1,k-1}(\alpha_2^{-1},\beta_2^{-1},\ldots).
\eea

An important observation made in \cite{DFSG14} is that this recurrence can be linearized by taking appropriate quotients and reduced to a linear system with periodic coefficients, as in (\ref{recurS}). Defining $S_{n,k}(\alpha_1,\beta_1,\ldots) = T_{n,k}(\alpha_1,\beta_1,\ldots)/T_n(\alpha_1,\beta_1,\ldots)$, a slightly different quotient as the one used in Section 4, one obtains from the above recurrence that they satisfy
\be
S_{n,k}(\alpha_1,\beta_1,\ldots) = A_n \, S_{n-1,k}(\alpha_1^{-1},\beta_1^{-1},\ldots) + B_n \, S_{n-1,k-1}(\alpha_2^{-1},\beta_2^{-1},\ldots),
\ee
where, by using (\ref{solX}), the coefficients $A_n$ are explicitely given by
\be
A_n \equiv \alpha_1 \frac{T_{n-1}(\alpha_1^{-1},\beta_1^{-1},\ldots) \, T_{n-1}(\alpha_2,\beta_2,\ldots)}{T_n(\alpha_1,\beta_1,\ldots) \, T_{n-2}(\alpha_2^{-1},\beta_2^{-1},\ldots)} = \frac1{X_n} \times \begin{cases}
\alpha_{\frac{n+1}2} & \text{for $n$ odd},\\
\noalign{\smallskip}
\beta_{\frac n2}^{-1} & \text{for $n$ even},
\end{cases}
\ee
and $B_n = 1-A_n$ from (\ref{mocta}). The coefficients $A_n,B_n$ are indeed periodic with period $2m$. It implies that the generating functions $G^{(a)}(\alpha_1,\beta_1,\ldots)$ of the $S_{n,k}(\alpha_1,\beta_1,\ldots)$ for every class of $n$ modulo $2m$ satisfy coupled inhomogeneous linear equations. Following the procedure used in Section 4, this in principle allows to determine the asymptotic value of the one-refined partition functions $S_{n,k}$.

For the part (b), we need to fix the weights of the three elementary steps of the lattice paths so that the bijection with the sets of non-intersecting paths is weight-preserving. To do that, the problem we had in Section 4.2 concerned the green dominos, which carried no path segment but contributed to the total weight of a tiling. With the choice of parameters pictured in Figure \ref{fig9}, this problem does not arise because the horizontal dimers associated with green dominos will all be placed on uncoloured edges, and each of them therefore contributes a weight 1. The weight of the three elementary steps can thus be computed in a straightforward way: depending on their location, horizontal steps have weight $\alpha_i^{}$ or $\alpha_i^{-1}$, vertical steps have weight 1, $\beta_i^{}$ or $\beta_i^{-1}$. Following a procedure similar to that in Section 4.2, the appropriate partition functions and generating functions will also satisfy coupled linear equations. Their solutions then allow to compute the asymptotic value of the partition function for a single path in terms of a function $L$.

Filling up the details of parts (a) and (b) would then make the tangent method work.

%%%%%%%%%%%%%%%%%%%%%%%%%%%%%%%%%%%%%%%%%%%%%%%%%%%%%%%%%%%%%%%%%%%%%%%%%%

\vskip 0.5truecm
\noindent 
{\bf \large Acknowledgements}

\noindent
It is a pleasure to thank Bryan Debin, Jean-Fran\c cois de Kemmeter and Nicolas Robert for valuable discussions and suggestions. I also wish to thank the Galileo Galilei Institute (Florence, Italy) for its support and hospitality during the scientific program ``Randomness, Integrability, and Universality'', where the present project was finalized. This work was supported by the Fonds de la Recherche Scientifique\,--\,FNRS and the Fonds Wetenschappelijk Onderzoek\,--Vlaanderen (FWO) under EOS project no 30889451. The author is Senior Research Associate of FRS-FNRS (Belgian Fund for Scientific Research). 

%%%%%%%%%%%%%%%%%%%%%%%%%%%%%%%%%%%%%%%%%%%%%%%%%%%%%%%%%%%%%%%%%%%%%%%%%%

\vskip 0.5truecm
\appendix \noindent 
{\bf \large Appendix A. Details of proofs}
\addcontentsline{toc}{subsection}{Appendix A. Details of proofs}
\setcounter{section}{1}
\setcounter{equation}{0}

\medskip
\noindent
In this Appendix, we give the details of the proof that the limit (\ref{lim}) vanishes in the region of the $(r,s)$ plane above and on the curve $H'(0;r,s)=0$, and is strictly negative below that curve. 

When $H'(0;r,s) \ge 0$, we have to prove the identity (\ref{=?}), namely
\bea
&& \hspace{-7mm} (2-r-s) \log\left[ab \, v\Big(\frac{r-\xi^*}{1-\xi^*}\Big)v\Big(\frac{s-\xi^*}{1-\xi^*}\Big)\right]^{\frac12} + (s-r) \log\left[v\Big(\frac{r-\xi^*}{1-\xi^*}\Big)\Big/v\Big(\frac{s-\xi^*}{1-\xi^*}\Big)\right]^{\frac12} \nonumber\\
\noalign{\smallskip}
&& \hspace{1cm} = (2-r-s) L(t) = -\,(2-r-s) \log x(t) - (s-r) \log y(t),  \qquad\quad t = \frac{s-r}{2-r-s}.
\label{a1}
\eea
We recall that $\xi^*$ is the unique solution of 
\be
uv\Big(\frac{r-\xi^*}{1-\xi^*}\Big) \, uv\Big(\frac{s-\xi^*}{1-\xi^*}\Big) = \Big[\frac{\sqrt{\beta}}{2(1+\beta)}\Big]^{\frac12}.
\label{max2}
\ee

For convenience, we define $u_z \equiv u\big(\frac{z-\xi^*}{1-\xi^*}\big)$ and $v_z \equiv v\big(\frac{z-\xi^*}{1-\xi^*}\big)$ for $z=r,s$. The arguments of the two logarithms in the l.h.s. of (\ref{a1}) become $\sqrt{ab \, v_r v_s}$ and $\sqrt{v_r/v_s}$. (Note that for $\xi^*=0$, the $v_r$ and $v_s$ defined here reduce to those used in Section 5.2 where the tangency condition $H'(0;r,s)=0$ indeed implies $\xi^*=0$.)

Our strategy to prove (\ref{a1}) will be to show that their inverse, $1/\sqrt{ab \, v_r v_s}$ and $\sqrt{v_s/v_r}$, satisfy the same algebraic system as $x(t)$ and $y(t)$, given in (\ref{Qsystem}). Explicitly these equations read
\begin{subequations}
\bea
&& 1 - 2(a^2+b^2) \, x^2 + a^2b^2 \, x^4 - ab \, x^2 \big(y^2 + \frac1{y^2}\big) = 0, \\
&& t \, (a^2b^2 \, x^4 - 1) + ab \, x^2 \big(y^2 - \frac1{y^2}\big) = 0. 
\eea
\end{subequations}
Substituting $x$ and $y$ for $1/\sqrt{ab \, v_r v_s}$ and $\sqrt{v_s/v_r}$, and clearing negative powers yield
\begin{subequations}
\bea
&& (v_r^2 - 1) \, (v_s^2 - 1) = 2 \textstyle \frac{1+\beta}{\sqrt{\beta}} \, v_r v_s,  \label{first}\\
&& t \, (1 - v_r^2 \, v_s^2) = v_r^2 - v_s^2. \label{second}
\eea
\end{subequations}
We want to show that $v_r$ and $v_s$ satisfy these two equations as a consequence of (\ref{max2}).

In Section 4.1, we have not computed the functions $u$ and $v$ in terms of their argument $r$, but we have expressed $u^2$ as a function of $v$. This allows to write the product $(uv)^2$ as a function of $v$. Using (\ref{usq}), we obtain two possible forms,
\be
(uv)^2 = \frac{v^2}{\sqrt{1+(\beta+\textstyle\frac1\beta)v^2+v^4} + v \, \frac{1+\beta}{\sqrt{\beta}}} = \frac{v^2}{(1-v^2)^2} \Big[\sqrt{1+(\beta+\textstyle\frac1\beta)v^2+v^4} - v \, \textstyle\frac{1+\beta}{\sqrt{\beta}}\Big],
\label{A5}
\ee
From these, the equation (\ref{max2}), written as $(u_rv_r)^2 = \frac{\sqrt{\beta}}{2(1+\beta)} (u_sv_s)^{-2}$, yields the following relation between $v_r$ and $v_s$,
\be
\frac{v_r^2}{(1-v_r^2)^2} \Big[\sqrt{1+(\beta+\textstyle\frac1\beta)v_r^2+v_r^4} - v_r \, \textstyle\frac{1+\beta}{\sqrt{\beta}}\Big] = 
\frac{\sqrt{\beta}}{2(1+\beta)} \, \frac1{v_s^2} \, \Big[\sqrt{1+(\beta+\textstyle\frac1\beta)v_s^2+v_s^4} + v_s \, \frac{1+\beta}{\sqrt{\beta}}\Big].
\label{A6}
\ee
It can be turned into a polynomial relation by isolating the two radicals and squaring (twice), and then clearing denominators. The resulting relation is, with $\alpha = \frac{1+\beta}{\sqrt{\beta}}$,
\bea
&& \hspace{-6.5mm} \Big\{(v_r^2-1)^3 \, (v_s^2-1)^3 + 2 \alpha v_rv_s \, (v_r^2-1)^2 \, (v_s^2-1)^2 + (2 \alpha v_rv_s)^2 \,(v_r^2-1) \, (v_s^2-1) \nonumber\\
&& + \: (2 \alpha v_rv_s)^3 \, (v_r^2+v_s^2)\,(1+\frac{v_rv_s}\alpha) + (2 \alpha v_rv_s)^3 \, v_rv_s \, \big[(2\alpha-\frac1\alpha)\, v_r^2v_s^2 - 3v_rv_s - \frac1\alpha\big] \Big\} \nonumber\\
\noalign{\smallskip}
&& \times \big[(v_r^2-1) \, (v_s^2-1) - 2\alpha \, v_r v_s\big] = 0.
\eea
It can be shown that the first factor nowhere vanishes for $v_r,v_s \ge 0$, so that the second one must be zero, which is precisely the first equation (\ref{first}).

Alternatively, and more simply, one can solve (\ref{first}) for $v_s$ in terms of $v_r$, and check that the relation (\ref{A6}) is identically satisfied.

To check the second identity (\ref{second}), we use the inversion function of $v$, given in (\ref{quart}). For $v_z=v\big(\frac{z-\xi^*}{1-\xi^*}\big)$, we have
\be
\frac{z-\xi^*}{1-\xi^*} = v^{-1}(v_z) = \frac {v_z}{1-v_z^2} \: \Bigg\{\frac{1+\beta}{2\sqrt{\beta}} \: \frac{1+v_z^2}{\sqrt{1+(\beta+\textstyle\frac1\beta)v_z^2+v_z^4}} - v_z\Bigg\}.
\ee
We then write $t$ as a function of $v_r$ and $v_s$,
\be
t = \frac{s-r}{2-r-s} = \frac{\frac{s-\xi^*}{1-\xi^*} - \frac{r-\xi^*}{1-\xi^*}}{2 - \frac{r-\xi^*}{1-\xi^*} - \frac{s-\xi^*}{1-\xi^*}} = \frac{v^{-1}(v_s) - v^{-1}(v_r)}{2 - v^{-1}(v_r) - v^{-1}(v_s)},
\ee
and simply verify that 
\be
\frac{v^{-1}(v_s) - v^{-1}(v_r)}{2 - v^{-1}(v_s) - v^{-1}(v_r)} = \frac{v_r^2 - v_s^2}{1 - v_r^2v_s^2},
\ee
is indeed satisfied when one uses the positive solution $v_s$ in terms of $v_r$ as obtained from (\ref{first}).

By looking at (\ref{first}), one would suspect that the values $v_r=1$ or $v_s=1$ are special. However these values are ruled out by (\ref{max2}). If $v_r=1$ for instance, we obtain from (\ref{A5}) that $(u_rv_r)^2 = \frac{\sqrt{\beta}}{2(1+\beta)}$, which then implies $(u_sv_s)^2 = 1$ by (\ref{max2}). Then the first form of (\ref{A5}) shows that $v_s=+\infty$, equivalently $s=1$, a degenerate case.

Let us now discuss the second case, $H'(0;r,s) < 0$, for which we have to prove that the function $\Phi(r,s)$ is strictly negative in the region below the curve ${\cal C}_{r,s}$,
\be
\Phi(r,s) = \frac12 \log\frac{\sqrt{\beta}}{2(1+\beta)} + F_1(r) + F_1(s) - (2-r-s) \big[L(t) - \log\sqrt{ab}\big].
\ee

In order to show this, we compute the derivative of $\Phi(r,s)$ along the curves of constant values of $t=\frac{s-r}{2-r-s}$, see Figure \ref{fig7}. These level curves are straight lines passing through the point $(r,s)=(1,1)$ and having slope $\frac{1-t}{1+t}$. The corresponding differential operator, in the upwards direction (namely for increasing $r$ and $s$), is a positive multiple of $D = (1-r) \frac{\partial}{\partial r} + (1-s) \frac{\partial}{\partial s}$. Using $F_1'(z)=-\log v(z)$ as given in (\ref{L1prime}), we readily obtain
\be
D \Phi(r,s) = -(1-r) \log v(r) - (1-s) \log v(s) + (2-r-s) \big[L(t) - \log\sqrt{ab}\big].
\ee
Since $L(t)=L\big(\frac{s-r}{2-r-s}\big)$ is constant along a level curve, it is equal to $L(t)=L\big(\frac{s_0-r_0}{2-r_0-s_0}\big)$ where $(r_0,s_0)$ is the crossing point with the curve ${\cal C}_{r,s}$. From the first case examined in (\ref{a1}) (with $(r,s)=(r_0,s_0)$ and hence $\xi^*=0$), it also satisfies 
\be (2-r_0-s_0) \big[L(t) - \log\sqrt{ab}\big] = (1-r_0) \log v(r_0) + (1-s_0) \log v(s_0),
\ee
which readily shows that $D\Phi(r_0,s_0)=0$.

Substituting the last expression of $L(t)$ in the derivative of $\Phi(r,s)$ yields
\be
D\Phi(r,s) > (1-r_0) \log v(r_0) + (1-s_0) \log v(s_0) - (1-r) \log v(r) - (1-s) \log v(s),
\ee
where the inequality follows from $2-r-s > 2-r_0-s_0$ because $r<r_0$ and $s<s_0$. 

The conclusion would now follow if the function $f(z) \equiv (1-z) \log v(z)$ was increasing for $z \in [1/2,1]$, but this is not true when $z$ is close enough to 1. One may however note that $r,s$ satisfy $H'(0;r,s) < 0$ and therefore cannot be both large (even more so as $\beta$ increases, see Figure \ref{fig7}). Building on this remark, one can show that $f(r) + f(s)$ is indeed increasing along the straight lines of constant values of $t$. This concludes the proof: on a line of constant value of $t$, the derivative $D\Phi(r,s)$ is positive below the curve ${\cal C}_{r,s}$ and vanishes on the curve; because $\Phi$ itself vanishes on ${\cal C}_{r,s}$ ($\Phi(r_0,s_0)=0$), it is strictly negative below ${\cal C}_{r,s}$.

%%%%%%%%%%%%%%%%%%%%%%%%%%%%%%%%%%%%%%%%%%%%%%%%%%%%%%%%%%%%%%%%%%%%%%%%%%

\end{document}